%% file: paper.tex
\documentclass{JINST}
\pdfoutput=1 
\usepackage{graphicx} 
\usepackage{dcolumn}  
\usepackage{longtable}
\usepackage{bm}       
\usepackage{subfigure}
\usepackage{subeqn}
\usepackage{multirow} 
\title{How good are your fits? Unbinned multivariate goodness-of-fit tests in high energy physics.}
\author{Mike Williams\\ 
Imperial College London, London SW7 2AZ, UK
}

\abstract{ 
Multivariate analyses play an important role in high energy physics.  Such analyses often involve performing an unbinned maximum likelihood fit of a probability density function (p.d.f.) to the data.  This paper explores a variety of unbinned methods for determining the goodness of fit of the p.d.f.\ to the data.  The application and performance of each method is discussed in the context of a real-life high energy physics analysis (a Dalitz-plot analysis).  Several of the methods presented in this paper can also be used for the non-parametric determination of whether two samples originate from the same parent p.d.f.  This can be used, {\em e.g.}, to determine the quality of a detector Monte Carlo simulation without the need for a parametric expression of the efficiency.
}
\begin{document}

\input{intro.tex}

\input{toy.tex}

\input{methods.tex}

\input{conc.tex}

\acknowledgments
I want to thank Ulrik Egede, Jonas Rademacker and Ilya Narsky for their many useful comments on the content of this paper.  This work was supported by STFC grant number ST/H000992/1.

\input{bib.tex}
\input{appendix.tex}

\end{document}

%% file: intro.tex
\section{Introduction}

Multivariate analyses are playing an increasingly prominent role in high energy physics.  In such analyses a physicist will often employ an unbinned maximum likelihood fit of a probability density function (p.d.f.) to the data.  The fit p.d.f.\ is then used to extract the desired information ({\em e.g.}, some set of observables) from the data.  When performing this type of analysis it is important to determine the level of agreement between the fit p.d.f.\ and the data.  Unfortunately, the maximum likelihood value (m.l.v.) itself cannot be used to determine the goodness of fit (g.o.f.). 

A common practice in high energy physics is to instead bin the data and compute a $\chi^2$ value.  This statistic can be used to test the g.o.f.; however, it does have its limitations. In multivariate problems the available phase space is typically sparsely populated; this is known in the statistical literature as the {\em curse of dimensionality}~\cite{ref:bellman}. Employing a coarse binning scheme is often required in this situation to avoid having an abundance of low occupancy bins.  If the bin occupancies are too low, then the significance of any discrepancy between the data and the fit p.d.f.\ is often overestimated when using the $\chi^2$ method (see, {\em e.g.}, Ref.~\cite{ref:yates}). Of course, if the bin sizes are too large then it may not be possible to compare the finer structure of the fit p.d.f.\ with the data.  Apart from this, binning data always results in a loss of information; thus, one would expect unbinned g.o.f.\ methods to perform better in multivariate problems.

There are a large number of unbinned multivariate g.o.f.\ tests available in the statistical literature (see, {\em e.g.}, Ref.~\cite{ref:dagostino}); however, most of the high energy physics community appears to be unaware of their existence.  Because of this, many high energy physicists use the binned $\chi^2$ method even in analyses where its power is expected to be minimal.  Others employ g.o.f.\ tests that are not found in the statistical literature.  {\em E.g.}, consider a multivariate analysis where a p.d.f.\ has been fit to the data using an unbinned maximum likelihood fit.  Many high energy physics analyses 
have attempted to use the m.l.v., $\mathcal{L}_{\rm max}$, to determine the g.o.f.  An outline of the procedure used is as follows: the data is fit to obtain $\mathcal{L}_{\rm max}$; the fit p.d.f.\ is used to generate an ensemble of Monte Carlo data sets; the g.o.f.\ is determined using $\mathcal{L}_{\rm max}$ from the data and the distribution of m.l.v.'s obtained from the Monte Carlo.  This approach may sound reasonable, but it is fatally flawed and, in fact, often fails to provide any information regarding the g.o.f.~\cite{ref:heinrich} (see Appendix~A for a detailed discussion).  Rather than attempting to invent new unbinned multivariate g.o.f.\ tests, a more prudent approach for high energy physics would be to study the applicability and performance of the g.o.f.\ methods published in the statistical literature.  This paper carries out such a study. 

Even for one-dimensional data, there is no uniformly most powerful (u.m.p.) g.o.f.\ test; {\em i.e.}, no test is the most powerful in all situations.  The popularity of the $\chi^2$ test in high energy physics is a testament to its versatility and power but it does not mean that it is the u.m.p.\ g.o.f.\ test for one-dimensional data.  There are many situations where other tests are more powerful.  {\em E.g.}, the Kolmogorov-Smirnov test is typically better suited for comparing two samples (rather than a sample and a p.d.f.).  The situation for the unbinned multivariate case is the same; {\em i.e.}, there is no u.m.p.\ test.  Thus, it is vitally important to study the performance of the available unbinned multivariate g.o.f.\ methods in the context of real-world high energy physics analyses.  

This paper carries out a systematic study of the performance of a variety of unbinned multivariate g.o.f.\ methods in the context of a Dalitz-plot analysis.  For each method, the underlying concept used to test the g.o.f.\ is discussed first.  This is followed by an overview of the formalism with a strong emphasis on how to apply the method in a high energy physics analysis.  The performance of each method is then studied in detail, including examining the effects of test bias.  Guidelines for dealing with nuisance parameters (including, in some cases, explicit determination of the regions of validity) is also provided. Finally, a high energy physics multivariate g.o.f.\ road map is outlined in Section~4.
It is also worth noting that several of the methods discussed in this paper can be used for the non-parametric determination of whether two samples originate from the same parent p.d.f.  This could be used, {\em e.g.}, to determine the quality of a detector Monte Carlo simulation without the need for a parametric expression of the efficiency.

%% file: toy.tex
\section{Toy-Model Analysis}

A Dalitz-plot analysis provides an excellent testing ground for multivariate g.o.f.\ techniques.  It is often the case in these analyses that a p.d.f.\ with unknown parameters and of unknown quality is fit to the data in two (or more) dimensions.  Determining the g.o.f.\ of the p.d.f.\ to the data is crucial in these types of analyses. Calculating the g.o.f.\ is complicated by the fact that Dalitz-plot distributions are typically highly non-uniform and rapidly varying.  Because of this, even with moderate statistics binned g.o.f.\ tests are often inadequate.

In this paper I consider the decay $X \rightarrow a b c$, where $m_X = 1$ and $m_a = m_b = m_c = 0.1$ are the particle masses (in some units).  All four particles are pseudo-scalars; {\em i.e.}, they all have a spin-parity of $0^-$.  The model for the Dalitz-plot distribution of this decay is constructed using the isobar formalism in which the total amplitude is written as the coherent sum of contributions from resonant and nonresonant terms:
\begin{equation}
  \label{eq:isobar}
  {\cal M}(\vec{x}) = a_{\rm nr} e^{i\phi_{\rm nr}} + \sum_r a_r a^{i\phi_r} \mathcal{A}_r(\vec{x}). 
\end{equation}
In Eq.~\ref{eq:isobar}, $\vec{x} = (m^2_{ab},m^2_{ac})$ represents the position in the Dalitz plot and $a e^{i\phi}$ describes the complex amplitude for each component.  The terms $\mathcal{A}_r(\vec{x})$ denote the resonance amplitudes and contain contributions from Blatt-Weisskopf barrier form factors~\cite{Blatt}, relativistic Breit-Wigner line shapes to describe the propagators and spin factors obtained using the Zemach formalism~\cite{ref:zemach}.  All amplitudes are evaluated using the {\tt qft++} package~\cite{Williams:2008wu}. The properties of the resonances included in this model, along with their fit fractions, are shown in Table~\ref{tab:resonance-params}.  

\begin{table}
  \begin{center}
  \label{tab:resonance-params}
  \begin{tabular}{ccccc}
    \hline
    Daughters & $J^P$ & Mass & Width & Fit Fraction \\
    \hline
    $a,b$ & $0^+$ & 0.3 & 0.025 & 6\% \\
    $a,b$ & $2^+$ & 0.6 & 0.05 & 2\% \\
    $a,c$ & $1^-$ & 0.4 & 0.04 & 18\% \\
    $a,c$ & $0^+$ & 0.7 & 0.1 & 43\% \\
    $b,c$ & $1^-$ & 0.35 & 0.01 & 10\% \\
    $b,c$ & $0^+$ & 0.75 & 0.02 & 17\% \\
    $a,b,c$ & \multicolumn{3}{c}{non-resonant} & 1\% \\
    \hline
  \end{tabular}
  \caption{
    Resonances included in the Dalitz-plot model used in this paper.
  }
  \end{center}
\end{table}

\begin{figure}
  \centering
  \includegraphics[width=0.4\textwidth]{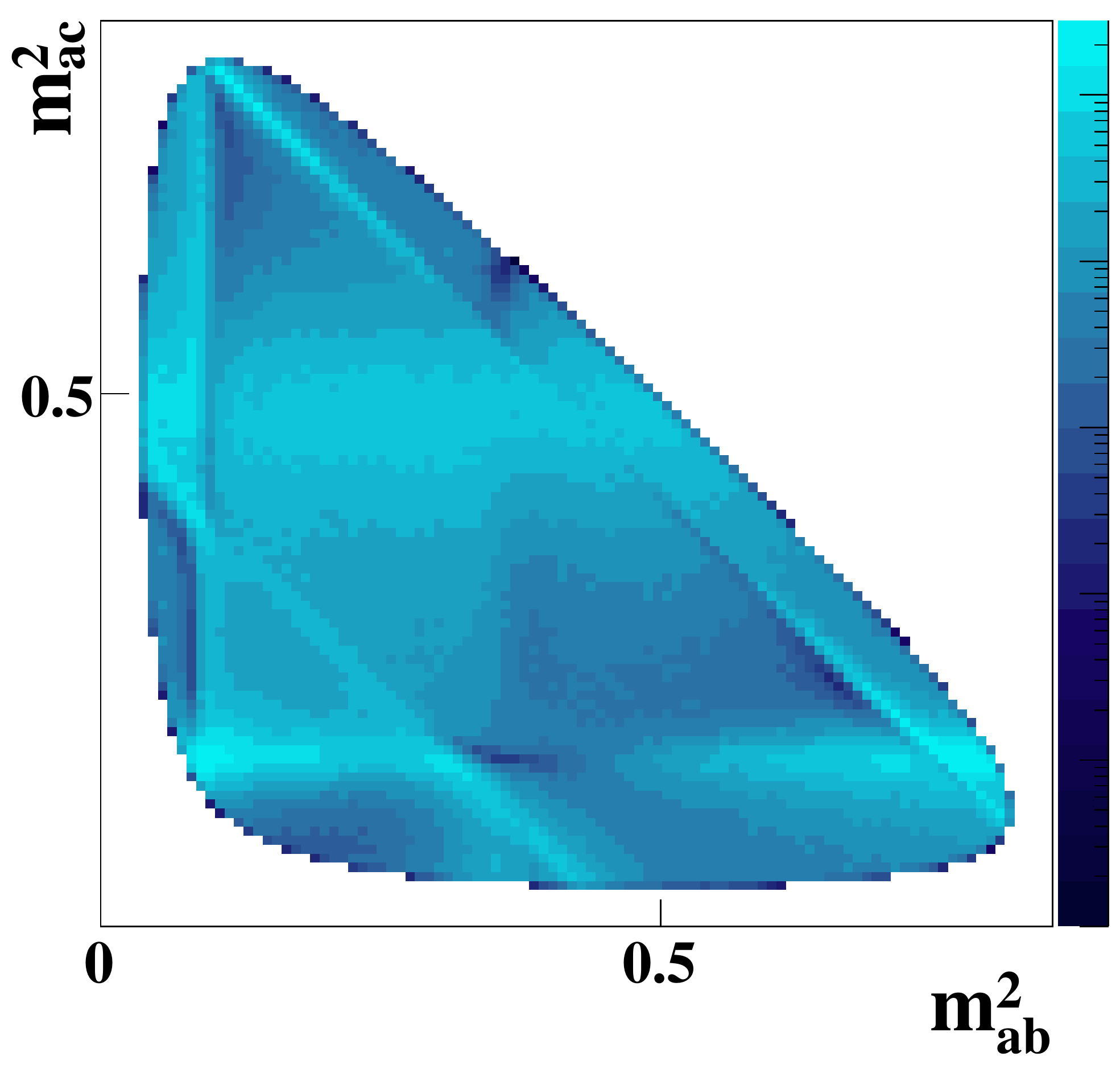}
  \caption[]{\label{fig:dp-ex}
    (Color Online) The Dalitz-plot p.d.f.\ used to generate the data in my toy-model analysis. Note the log scale on the z (color) axis.
  }
\end{figure}

The p.d.f.\ is easily obtained from the total amplitude as $f(\vec{x}) = |{\cal M}(\vec{x})|^2/\int |{\cal M}(\vec{x})|^2 d\vec{x}$, where the normalization to unity is explicit.  Fig.~\ref{fig:dp-ex} shows the Dalitz-plot distribution obtained from this p.d.f.  The details concerning the resonances are not important to this paper; however, it is worth noting that this distribution possesses the complex, rapidly varying structures that are present in many Dalitz-plot (and other high energy physics) analyses.  The presence of such features facillitate testing the robustness of the g.o.f.\ methods discussed below.

\begin{figure*}
  \centering
    \includegraphics[width=0.32\textwidth]{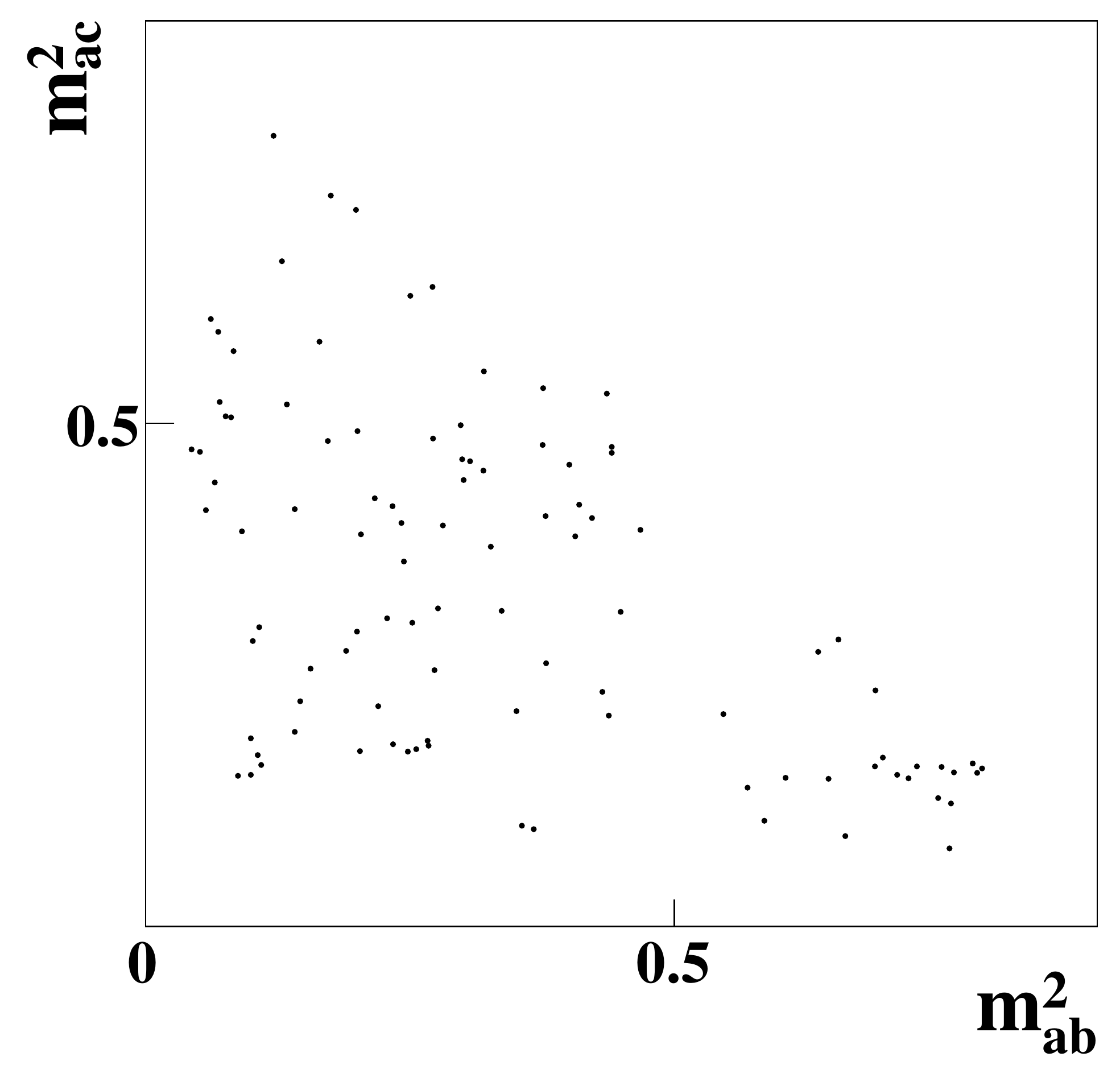}
    \includegraphics[width=0.32\textwidth]{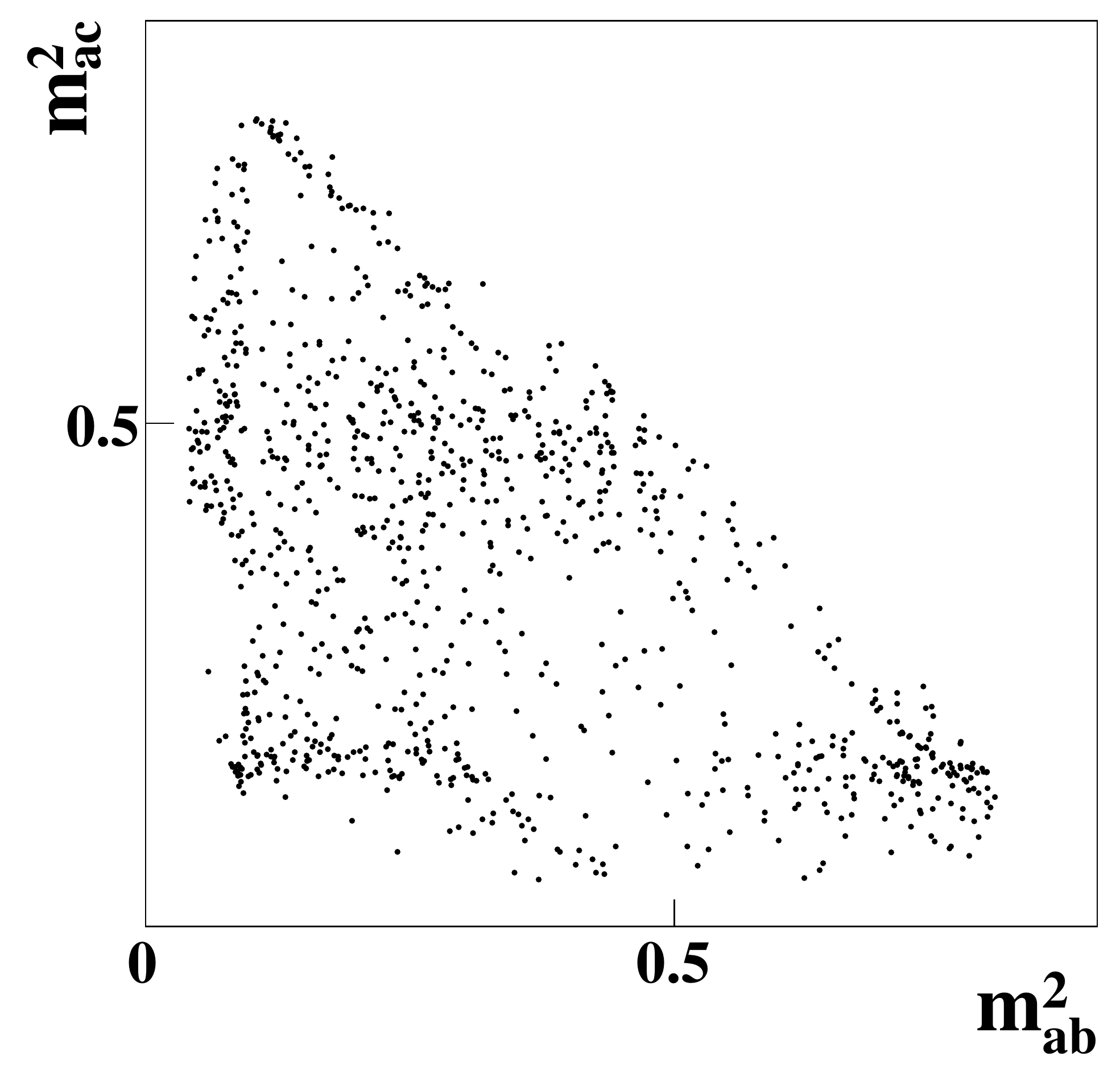}
    \includegraphics[width=0.32\textwidth]{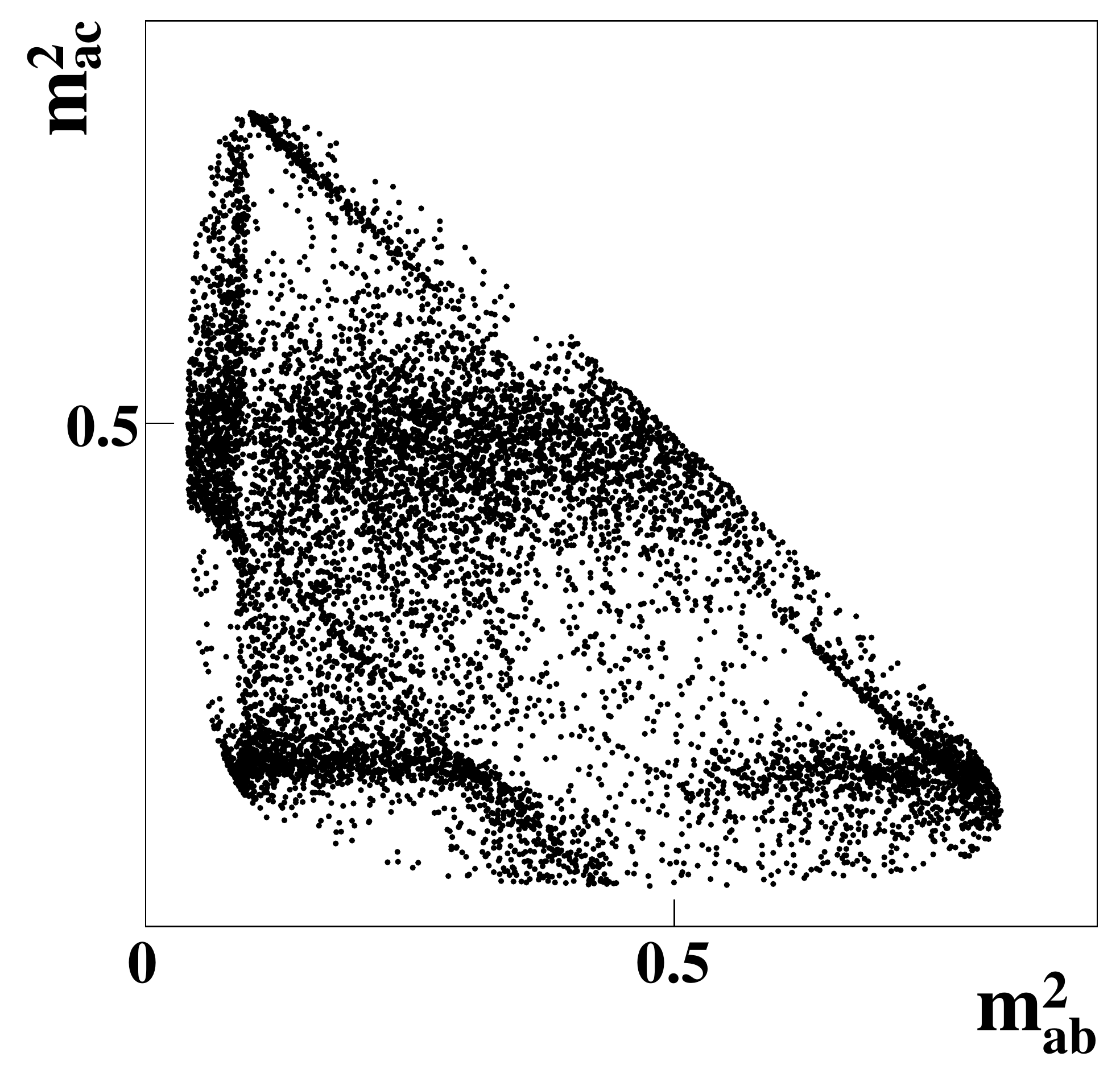}
  \caption[]{\label{fig:dp-lmh}
    Example low (left), medium (middle) and high (right) statistics toy-model data sets.  The number of events generated is 100, 1000 and 10000, respectively.
  }
\end{figure*}

I will consider three different population sizes in this study: low ($n_d = 100$); medium ($n_d=1000$); and high ($n_d=10000$).  Example Dalitz-plot data sets with these three sample sizes are shown in Fig.~\ref{fig:dp-lmh}. Analysis of a Dalitz-plot data set with less than 100 events is difficult due to the sparseness of the data.  Determining g.o.f.\ in a Dalitz-plot analysis with ${n_d \gg 10000}$ events is typically possible even using binned methods. Thus, studying data sets of these three sample sizes should suffice to ascertain the applicability of any unbinned multivariate g.o.f.\ method to a Dalitz-plot analysis. 

An ensemble of 100 data sets of each of the three sample sizes listed above will be produced and analyzed in this study.  For each data set, the g.o.f.\ of the following p.d.f.'s will be examined:
\begin{description}
  \item[Model P.D.F.] \hfill \\
The same p.d.f.\ as used to generate all of the toy-model data sets. {\em I.e.}, it is the parent distribution of every data set examined in this study.  The $p$-value distribution obtained for each ensemble of toy-model data sets must be flat (modulo statistical fluctuations) for any g.o.f.\ method when the test p.d.f.\ is the parent p.d.f.\ (see Section~3).  In a real-world analysis, one does not have access to this p.d.f. It is examined here as an important systematic check of each g.o.f.\ method.
  \item[Fit I  P.D.F.] \hfill \\
The p.d.f.\ obtained for each data set by fitting the toy-model p.d.f., with all resonance parameters free (a total of 13 free parameters), to the data.  Each toy-model data set has its own Fit~I p.d.f. In the absence of test bias, the $p$-value distributions obtained for Fit~I should also be flat; however, because of the fact that each Fit~I p.d.f.\ is obtained from a fit to the data set being analyzed, some test bias is expected.  The consistency of each g.o.f.\ method will be judged by the size of the observed test bias.
  \item[Fit II P.D.F.] \hfill \\
The p.d.f.\ obtained in the same way as that in Fit~I but with the $J^P = 1^-$ resonance in the $bc$ system - that has a 10\% fit fraction - removed.  These p.d.f.'s have a large discrepancy relative to the Model p.d.f.\ but in a small region of phase space. The power of each g.o.f.\ method will, in part, be judged by how well it is able to reject Fit~II.  
  \item[Fit III P.D.F.] \hfill \\
The p.d.f.\ obtained in the same way as that in Fit~I but with the non-resonant term - that has a 1\% fit fraction - removed.  These p.d.f.'s have a small discrepancy relative to the Model p.d.f.\ but in a large region of phase space (all of it).  This p.d.f.\ is very similar to what one would obtain using a slightly deficient background estimation.  The power of each g.o.f.\ method will also be judged by how well it is able to reject Fit~III.  
\end{description}

%% file: methods.tex
\section{Goodness-of-Fit Methods}

The goal of the Dalitz-plot analysis carried out in this paper is to test the g.o.f.\ of each of the p.d.f.'s defined in Section~2. The notation used here, and throughout this paper, is as follows: $f$ denotes the parent p.d.f.\ of the data; $f_0$ denotes the test p.d.f.; $\vec{x}$ denotes the $D$-dimensional vector of variables; and $n_d$ denotes the number of events in a data sample. For each g.o.f.\ method, a test statistic, $T$, is defined that quantifies (in some way) the agreement between the data and the test p.d.f.  For all of the methods presented in this paper, larger values of $T$ correspond to a worse level of agreement ({\em n.b.} this is not a universal property of all g.o.f.\ methods).

The p.d.f.\ of the test statistic, $g(T)$, may depend on the test p.d.f., {\em i.e.}, $g$ may not be distribution free (as it is, {\em e.g.}, for the $\chi^2$ test for a fixed number of degrees of freedom).  The significance of any discrepancy between the data and the test p.d.f.\ is quantified by the $p$-value, which is defined as follows for the case where larger $T$-values correspond to worse levels of agreement:
\begin{equation}
  p = \int_{T}^{\infty} g_{f_0}(T^{\prime}) dT^{\prime}.
\end{equation}
Thus, the $p$-value is the probability of finding a $T$-value corresponding to lesser agreement than the observed $T$-value.  It is important to note that the $p$-value is not the probability that $f = f_0$.  If $f_0$ is, in fact, the parent distribution of the data, {\em i.e.}, if $f = f_0$, then the $p$-value distribution is uniform on the interval between zero and one.  For this case, the $p$-value is the same as the confidence level.  One can reject the hypothesis $f = f_0$ at confidence level $\alpha$ if $p < 1 - \alpha$; {\em e.g.}, the test hypothesis is rejected at 95\% confidence level if $p < 0.05$. 

The statistical literature on g.o.f.\ is vast.  It is not possible to test every available g.o.f.\ method.  Many of the available methods use similar concepts in constructing their g.o.f.\ tests.  I have divided up the methods I have found into five categories: mixed-sample methods; point-to-point dissimilarity methods; distance to nearest-neighbor methods; local-density methods; and kernel-based methods.  I have chosen to implement and study one method from each category to determine its applicability to the Dalitz-plot analysis described in Section~2.  
I note here that I have ignored methods specifically designed to find highly localized discrepancies ({\em e.g.}, unexpected peaks) in the data.  Such methods can be useful, {\em e.g.}, for signal discovery; however, they are not well suited to the analysis performed in this paper.  It is also worth noting here that none of the methods presented in this paper (all of which are distance-based methods) should be used in an analysis that includes both continuous and discrete variables (a rare occurrence in high energy physics).
Finally, the notation used in the original publications is (in many cases) different than that used in this paper.  I have opted for using, as much as possible, a consistent set of notation for all of the methods described in this paper.

\subsection{The Binned $\chi^2$ Method}

Prior to introducing any unbinned methods, I will first examine the performance of the binned $\chi^2$ test in this analysis; this test will be used as the benchmark for all of the other methods studied in this paper.  The $\chi^2$ test typically used in high energy physics was first introduced by Pearson in 1900~\cite{ref:pearson}.  The test statistic is defined as
\begin{equation}
  \label{eq:chi2-def}
  \chi^2 = \sum \limits_{c=1}^{n_c} \frac{(o_c - e_c)^2}{e_c},
\end{equation}
where $n_c$ is the number of cells and $o_c$ ($e_c$) is the number of observed (expected) events in the $c^{th}$ cell. If the number of degrees of freedom, $n_{\rm dof}$, is known, then the $\chi^2$ value can be used to obtain a $p$-value (since the $\chi^2$ distributions for each $n_{\rm dof}$ are known); it is the $p$-value that determines the g.o.f., not the value of $\chi^2/n_{\rm dof}$ (which is often given in high energy physics publications).

If the model to be tested has no free parameters, then $n_{\rm dof} = n_c -1$ (if the model is normalized to the number of observed data events).  If, however, the model has $n_p$ free (independent) parameters that are determined by minimizing the $\chi^2$ statistic defined in Eq.~\ref{eq:chi2-def}, then $n_{\rm dof} = n_c - n_p - 1$ ({\em n.b.}, one should be careful not to double count the normalization here).  In the toy-model analysis performed in this paper, estimators for the free parameters in the p.d.f.'s are obtained from unbinned maximum likelihood fits to the data (not by minimizing $\chi^2$). Unfortunately, in this case the test statistic does not follow a limiting $\chi^2$ distribution.  All that is known about it is the following: $\chi^2(n_{\rm dof}=n_c-n_p-1) \leq \chi^2 \leq \chi^2(n_{\rm dof}=n_c-1)$~\cite{ref:chernoff}; {\em i.e.}, the $\chi^2$ value obtained using the model parameters that maximize the likelihood is generally larger than the minimum $\chi^2$ value.  How much larger depends on the p.d.f.;  the effect this has on the $p$-values depends on $n_c$ and $n_p$.

For the Model p.d.f.\ there are no free parameters which makes determining the $p$-values straightforward.  The rejection power at 95\% confidence level is shown in Table~\ref{table:chi2-power}; the results are as expected.  The free parameters in the Fit~I, Fit~II and Fit~III p.d.f.'s are obtained from unbinned maximum likelihood fits; thus, we can only (analytically) set limits on the rejection power.  These limits are shown in Table~\ref{table:chi2-power}.  For the larger data sets, a larger number of cells can be used which results in a smaller difference between the upper and lower limits.  For $n_d = 100$, the number of free parameters is equal to the number of cells making the upper limit undefined.  While the rejection power cannot be calculated analytically, it can be estimated using any one of the many data-driven techniques found in the statistical literature.  I will postpone giving a detailed discussion on this topic until Section~3.3 (where a full example is provided).  The estimates for the $\chi^2$ rejection power obtained using one of these techniques are given in Table~\ref{table:chi2-power}.  For Fit~II the rejection power of the $\chi^2$ test is excellent for $n_d = 10000$, good for $n_d = 1000$ and poor for $n_d = 100$.  For Fit~III the rejection power is fair for $n_d = 10000$ and poor for $n_d \leq 1000$.  The unbinned multivariate g.o.f.\ techniques presented below will be judged relative to these results.

\begin{table}
  \begin{center}
    \begin{tabular}{|c|c|ccc|}
      \hline
      $n_d$ & Model & Fit I & Fit II & Fit III \\
      \hline
      10000 & 5\% & [3\%-11\%](5\%) & [100\%](100\%) & [39\%-58\%](44\%) \\
      1000 & 4\% & [1\%-28\%](6\%) & [56\%-95\%](67\%) & [4\%-35\%](11\%) \\
      100 & 5\% & [$\geq$ 1\%](5\%) & [$\geq$ 2\%](3\%) & [$\geq$ 2\%](5\%) \\
      \hline
    \end{tabular}
    \\
    \vspace{0.01\textheight}
    \caption[]{\label{table:chi2-power}  
      Rejection power at 95\% confidence level using Pearson's $\chi^2$ method~\cite{ref:pearson}.  The values in square brackets represent the analytical upper and lower limits on the rejection power. The values in parentheses give the rejection power estimates obtained using a data-driven technique. See section 3.1 for details.
    }
  \end{center}   
\end{table}

\subsection{Mixed-Sample Methods}

\begin{figure*}
  \centering
    \includegraphics[width=0.4\textwidth]{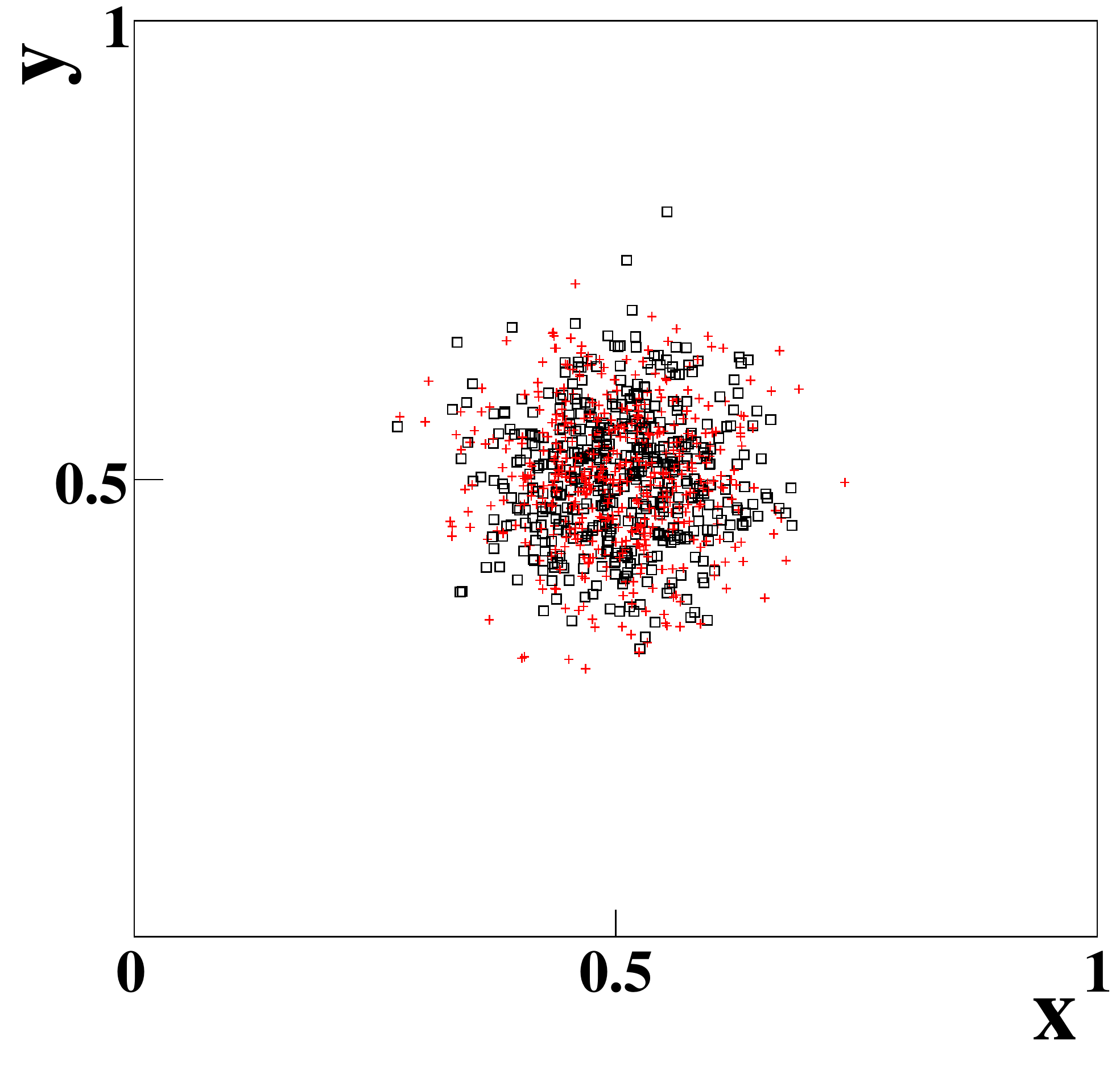}
    \includegraphics[width=0.4\textwidth]{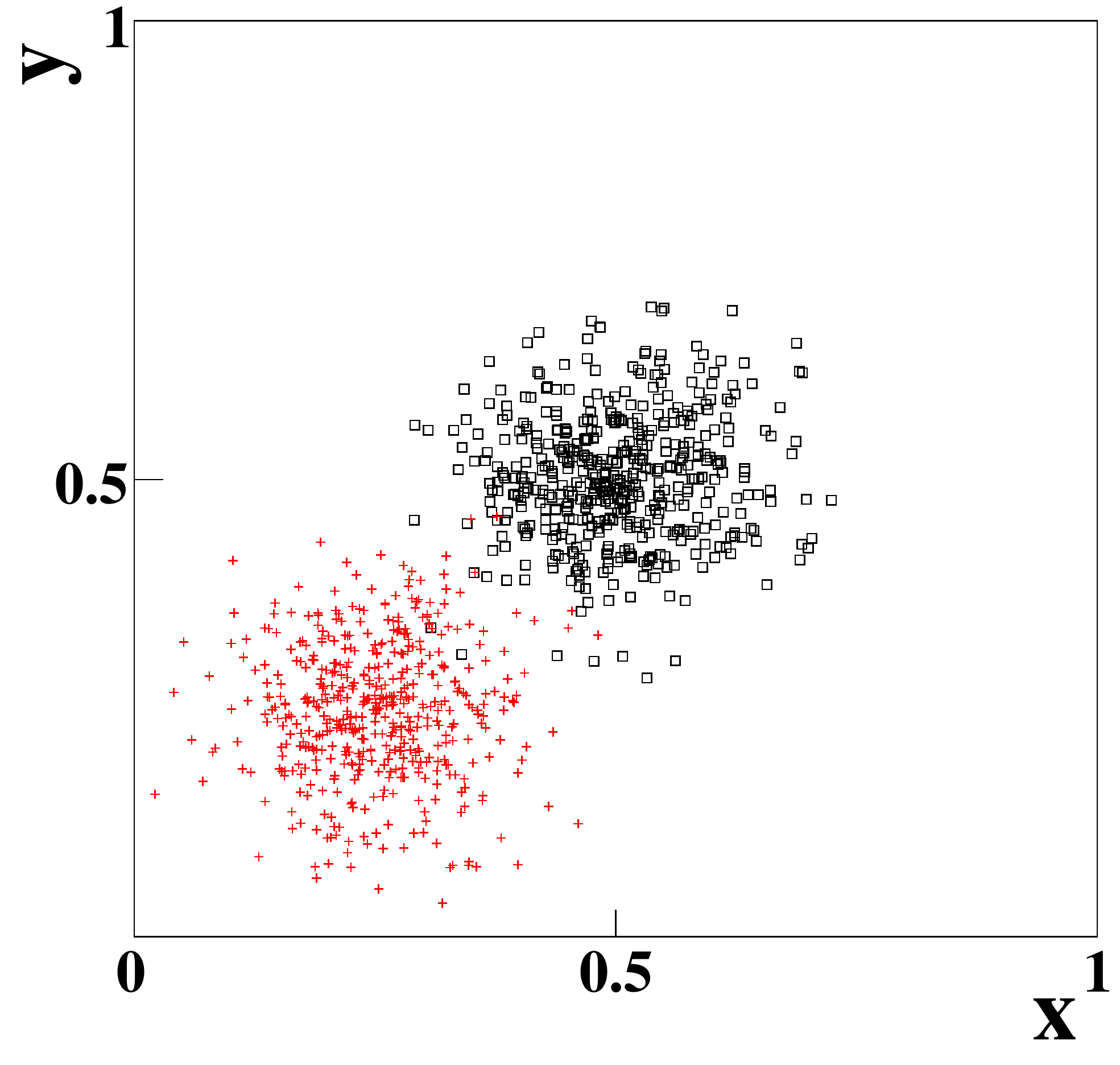}
  \caption[]{\label{fig:schil-ex}
    Example distributions of data randomly sampled from the p.d.f.'s $f_a(\vec{x})$ (black open squares) and $f_b(\vec{x})$ (red crosses) for the cases: (left) $f_a(\vec{x}) = f_b(\vec{x})$; (right) $f_a(\vec{x}) \ne f_b(\vec{x})$.  The two samples are optimally mixed if $f_a(\vec{x}) = f_b(\vec{x})$ but not so if $f_a(\vec{x}) \ne f_b(\vec{x})$.  This fact is exploited by g.o.f.\ tests in the mixed-sample category.
  }
\end{figure*}

If two data sets are combined to form a pooled sample, the mixing of the two samples is only optimal if they share the same parent distribution (see Fig.~\ref{fig:schil-ex}).  This fact can be used to determine g.o.f.~\cite{ref:schil,ref:henze}. The method described below does not require any knowledge concerning the p.d.f.'s of either of the samples; thus, it could be used, {\em e.g.}, to determine the quality of a detector Monte Carlo simulation without the need for a parametric expression of the efficiency.  It could also be used to study the stability of data taken by an experiment by comparing data samples taken during different time periods.

Prior to presenting this category of methods, the concept of {\em nearest-neighbor} events must be introduced.  To determine which events are the nearest neighbors to any given event in a data sample, one first needs to define the distance between events in the multivariate space.  One option is to use the {\em normalized Euclidean distance} which is defined as
\begin{equation}
  |\vec{x}_i - \vec{x}_j|^2 = \sum\limits_{v=1}^D \left(\frac{x_i^v - x_j^v}{w_v}\right)^2,
\end{equation}
where the $w_v$ values are used to weight each of the variates.  Because of the fact that the two invariant mass ranges in the Dalitz-plot analysis considered in this paper are the same, I chose to use $w_v = 1$ for each $v$ (the {\em Euclidean distance}).  Another choice (that is more desirable when the allowed values of the variates are not equivalent) is to set each $w_v$ value to be the root mean square of the data for the $v^{th}$ variate. One could also simply chose to set each $w_v = x^{\rm max}_v - x^{\rm min}_v$. The conclusions drawn from the g.o.f.\ test should not depend on the choice of distance function used, provided a {\em reasonable} choice is made (analogous to the choice of binning scheme when performing the $\chi^2$ test).  Once the distance between events is determined, the $i^{th}$ event's $n_k$ nearest neighbors are simply the events with the $n_k$ smallest distances from the $i^{th}$ event.

Following Ref.~\cite{ref:schil}, let $\{\vec{x}^a_1 \ldots \vec{x}^a_{n_a}\}$ and $\{\vec{x}^b_1 \ldots \vec{x}^b_{n_b}\}$ be two independent random $D$-dimensional samples from the distributions corresponding to the p.d.f.'s $f_a(\vec{x})$ and $f_b(\vec{x})$, respectively. For my toy-model Dalitz-plot analysis, the two data sets will be the data and a Monte Carlo data set sampled from one of the fit p.d.f.'s.  For now I will keep the notation generic as this method is applicable to any situation where one wants to determine whether two data sets share the same parent distribution. 

The statistic that will be used to test the hypothesis $f_a = f_b$ is defined as follows:
  \begin{equation}
    \label{eq:t-schil-def}
    T = \frac{1}{n_k(n_a + n_b)} \sum\limits_{i=1}^{n_a+n_b} \sum\limits_{k=1}^{n_k} I(i,k), \nonumber
  \end{equation}
  where $I(i,k) = 1$ if the $i^{th}$ event and its $k^{th}$ nearest neighbor belong to the same sample and $I(i,k) = 0$ otherwise, and $n_k$ is the number of nearest-neighbor events being considered. The quantity $T$ is then simply the mean fraction of like-sample nearest-neighbor events in the pooled sample of the two data sets.  The expectation value of $T$ is larger for the case $f_a \ne f_b$ due to the lack of complete mixing of the two samples that occurs if their parent distributions are not the same. For the extreme example shown in Fig.~\ref{fig:schil-ex}, one can see that the left panel has $T \approx 1/2$ ($n_a = n_b$) while the right panel has $T \approx 1$.

For the case where $f_a = f_b$, the quantity $(T - \mu_T)/\sigma_T$ has a limiting standard normal distribution; {\em i.e.}, it has a mean of zero and a width of one, where the mean is easily found to be
\begin{equation}
  \label{eq:schil-mean}
  \mu_T = \frac{n_a(n_a-1)+n_b(n_b-1)}{n(n-1)}
\end{equation}
using $n = n_a+n_b$.  For the special case $n_a = n_b$, $\mu_T \approx 1/2$.  The variance is much more difficult to calculate since it depends on the p.d.f.  The limiting value is given by
\begin{equation}
  \label{eq:schil-var-limit}
  \lim_{n, n_k, D \to \infty} \sigma_T^2 = \frac{1}{n n_k} \left(\frac{n_a n_b}{n^2} + 4\frac{n_a^2 n_b^2}{n^4}  \right),
\end{equation}
see Appendix~B for a detailed discussion on this quantity. The convergence to this limit is so fast that Eq.~\ref{eq:schil-var-limit} can be used to obtain a good approximation of $\sigma_T$ even for $D=2$ for certain values of $n_a, n_b$ and $n_k$; this is discussed in detail below in the context of the Dalitz-plot analysis.

As stated above, for the Dalitz-plot analysis considered in this paper the two data sets are the data (whose parent distribution is $f$) and a Monte Carlo sample obtained from the p.d.f.\ to be tested (whose parent distribution is denoted by $f_0$).  The hypothesis to be tested is that $f = f_0$.  Eq.~\ref{eq:t-schil-def} can be rewritten for the Dalitz-plot analysis as
 \begin{equation}
    \label{eq:t-schil-dp}
    T = \frac{1}{n_k(n_d + n_{mc})} \sum\limits_{i=1}^{n_d+n_{mc}} \sum\limits_{k=1}^{n_k} I(i,k), \nonumber
  \end{equation}
  where $I(i,k) = 1$ if the $i^{th}$ event and its $k^{th}$ nearest neighbor are either both data or both Monte Carlo events and $I(i,k) = 0$ otherwise. It is worth noting that $T$ is easy to calculate (it is simply a bookkeeping exercise). 

The expectation value of $T$ is also easy to obtain from Eq.~\ref{eq:schil-mean}. Thus, the only information required to obtain the g.o.f.\ are the values of $n_{mc}$ and $n_k$ that should be used.  Generating more Monte Carlo data reduces the statistical uncertainty on $f_0$.  Similarly, collecting a larger number of nearest-neighbor events reduces the statistical uncertainty on the local population density around each event; however, the power of the method will obviously be reduced if the region of phase space required to collect each event's $n_k$ nearest neighbors becomes too large (analogous to using very wide bins in a $\chi^2$ test).  The constraints on $n_{mc}$ and $n_k$ required to insure the validity of Eq.~\ref{eq:schil-var-limit} are discussed in detail in Appendix~B.  I have found that the values $n_{mc} = 10~n_d$ and $n_k = 10$ satisfy all of the relevant concerns and constraints on these quantities.

\begin{figure*}
  \centering
  \subfigure[]{
    \includegraphics[width=0.4\textwidth]{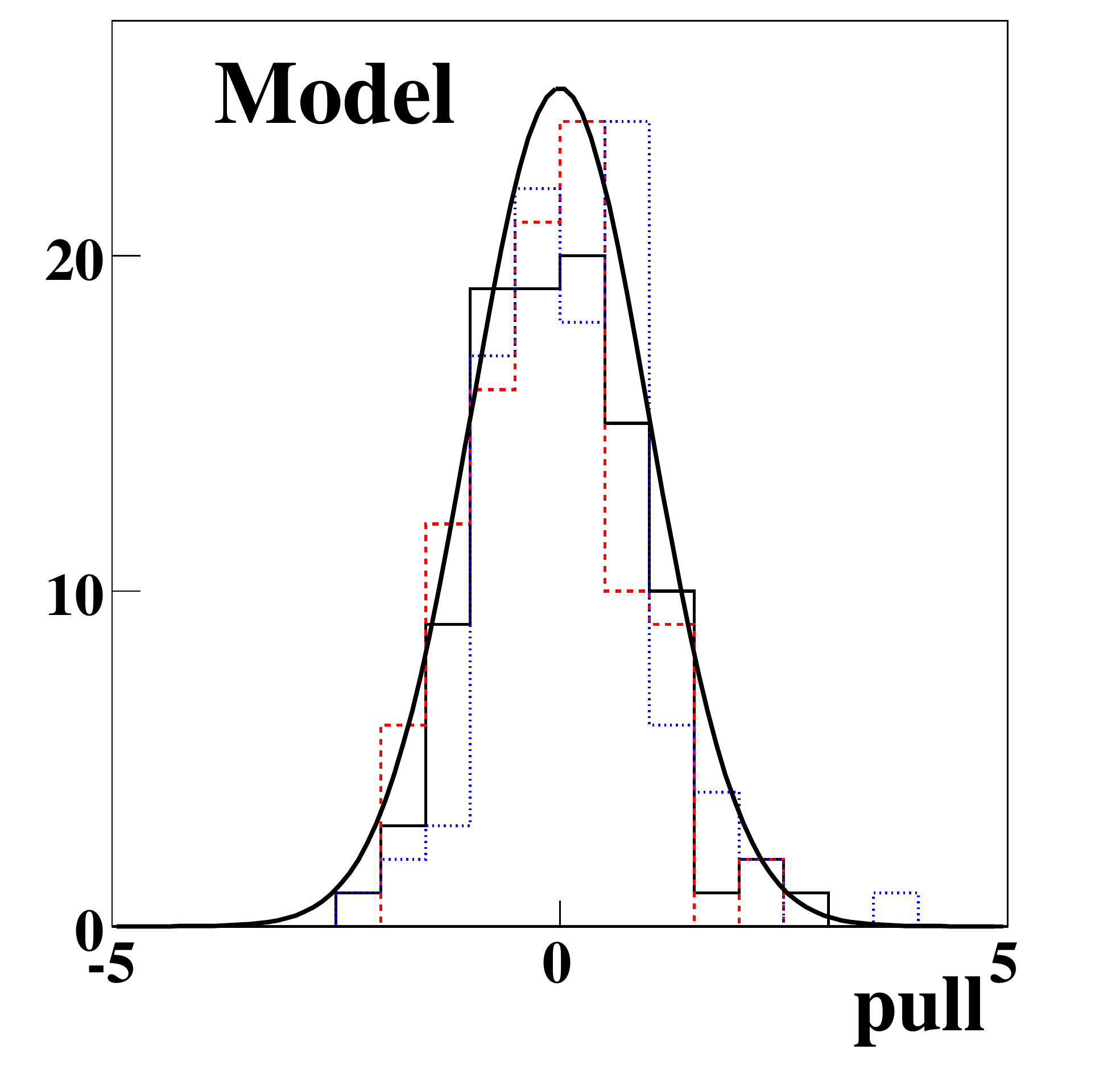}
  }
  \subfigure[]{
    \includegraphics[width=0.4\textwidth]{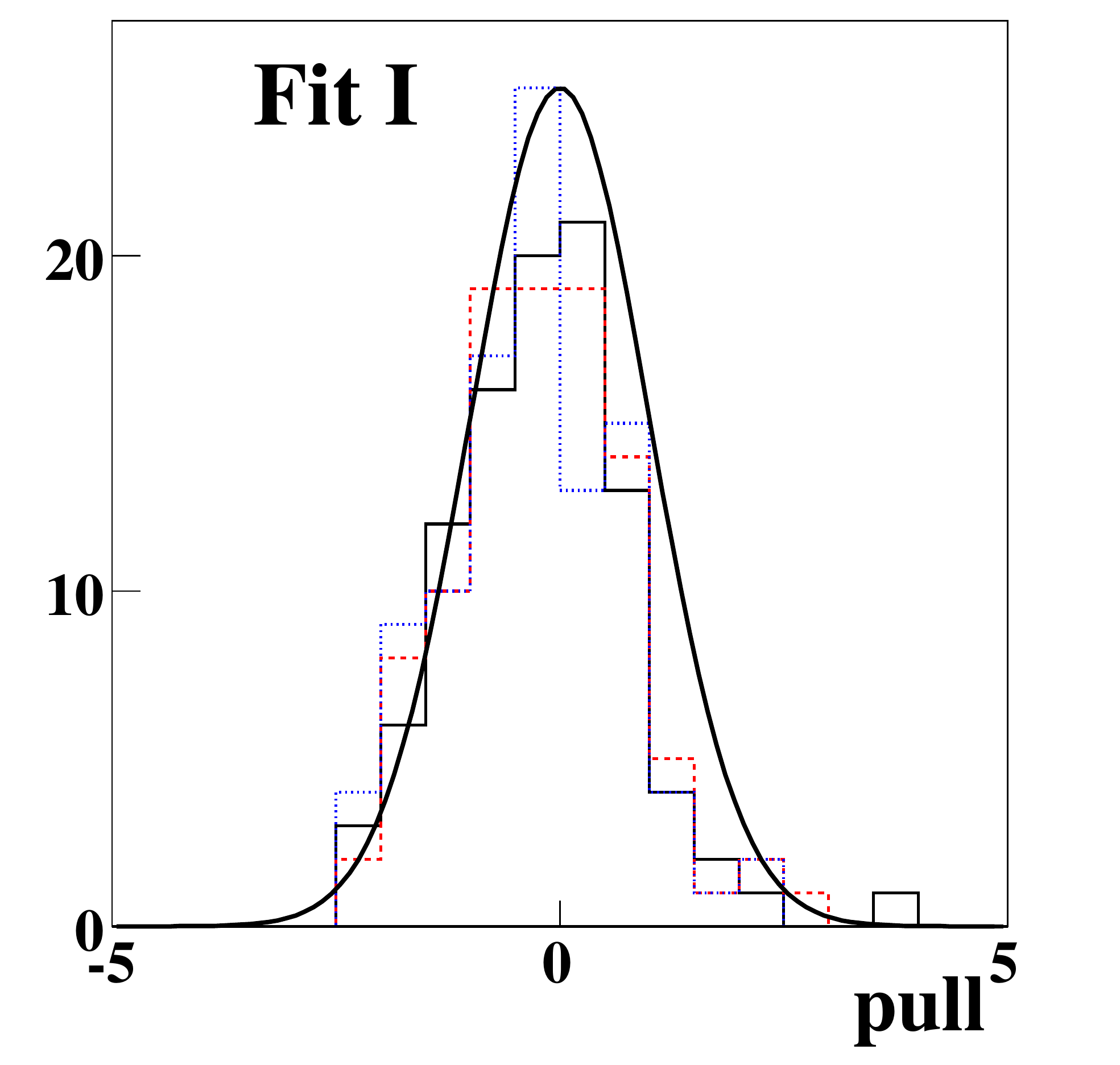} 
  }
  \\
  \subfigure[]{
    \includegraphics[width=0.4\textwidth]{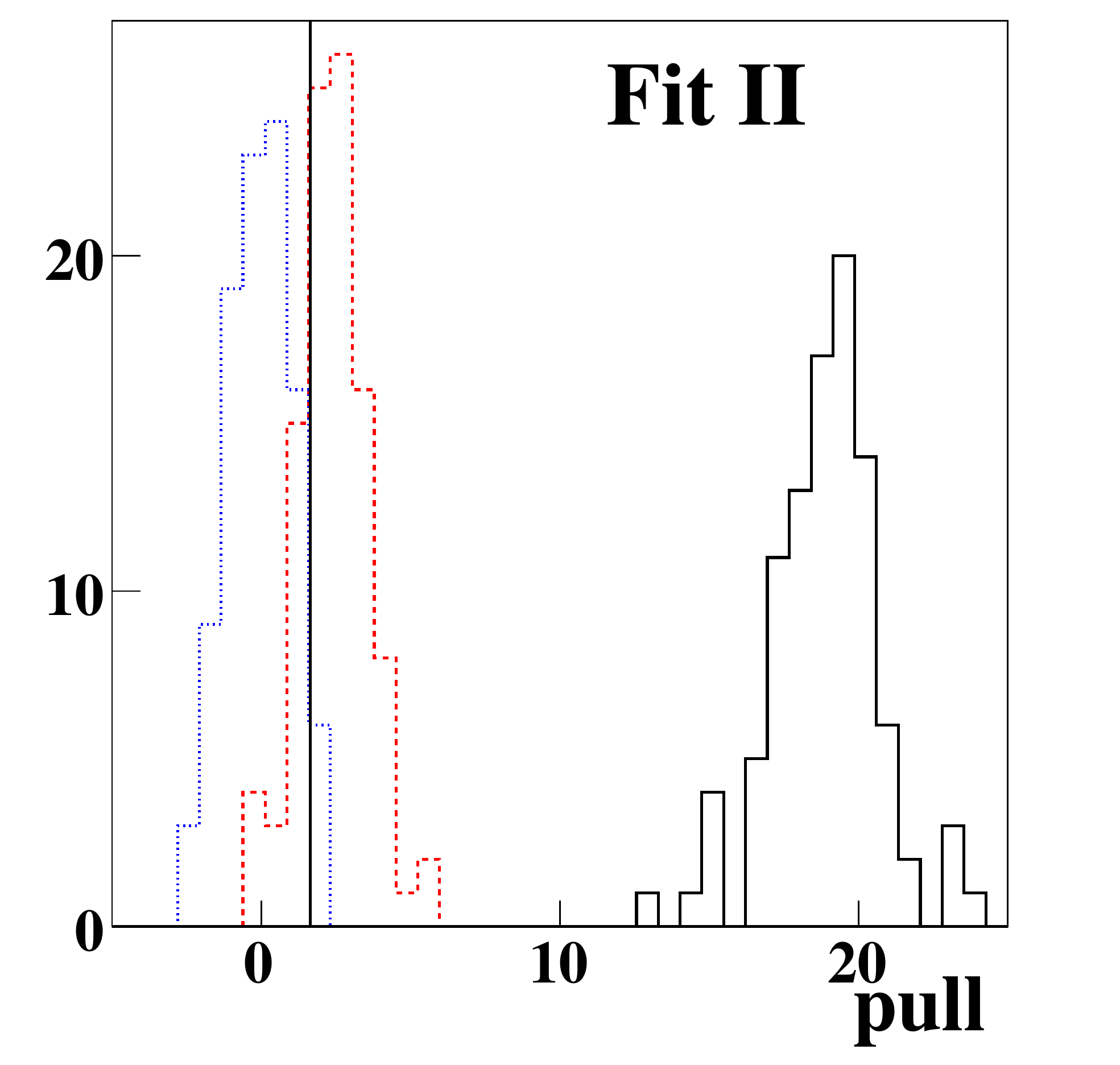} 
  }
  \subfigure[]{
    \includegraphics[width=0.4\textwidth]{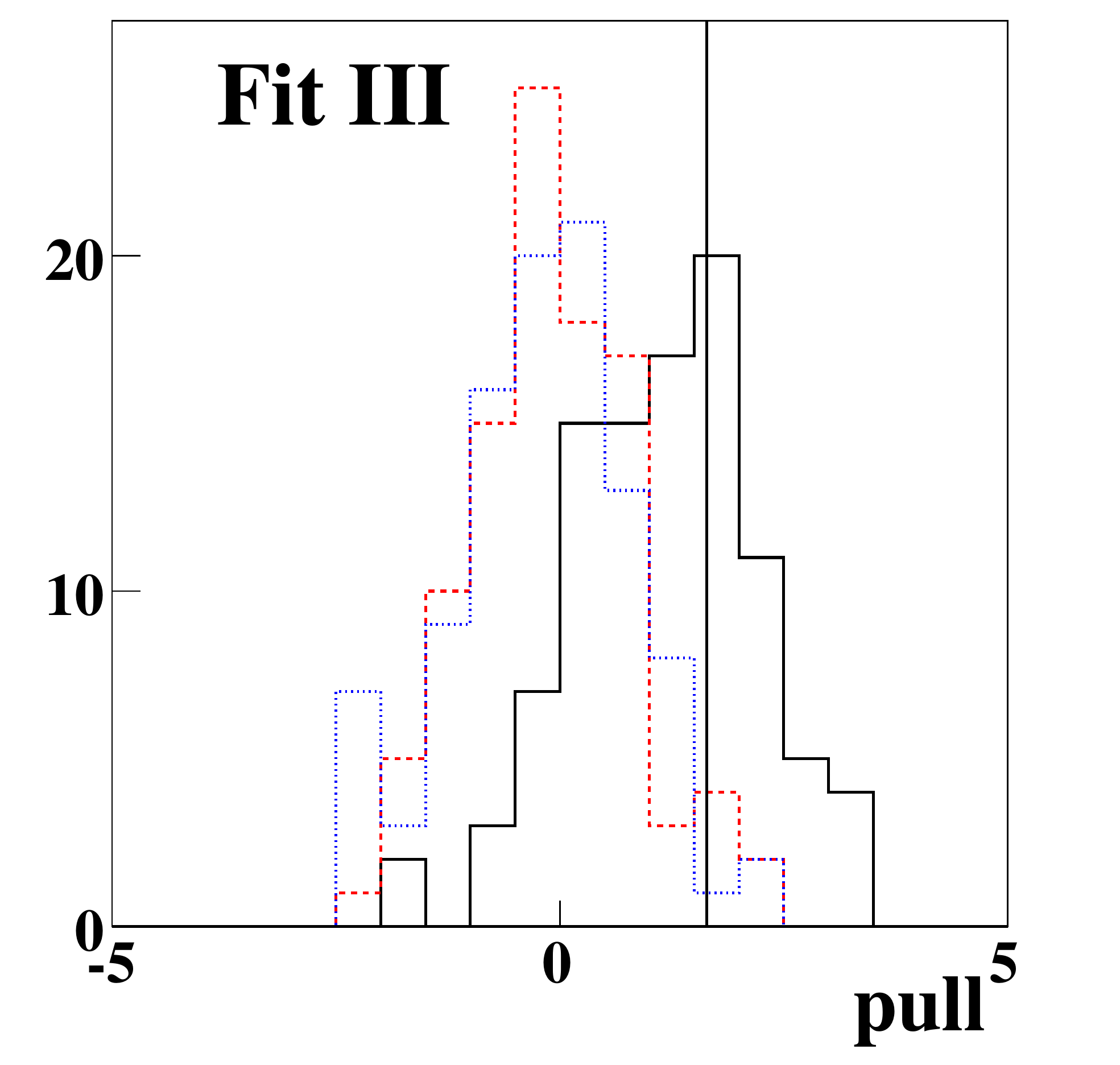} 
  }
  \caption[]{\label{fig:schil-results}
    (Color Online) Pull distributions obtained by mixing low (blue dotted), medium (red dashed) and high (solid black histograms) statistics data sets with Monte Carlo data obtained from the following p.d.f.'s using the mixed-sample g.o.f.\ method of Ref.~\cite{ref:schil}: (a) Model; (b) Fit~I; (c) Fit~II; (d) Fit~III.  The (solid black) curve shown in panels (a) and (b) represents the expected (standard normal) pull distribution.  The (solid black) vertical line shown in panels (c) and (d) represents the 95\% confidence-level cut value; data sets with $(T - \mu_T)/\sigma_T > 1.64$ are rejected at this level.  See Section~3.2 for further discussion on these results.
  }
\end{figure*}

In all of the results that follow the values of $n_{mc} = 10~n_d$ and $n_k = 10$ are used.  For these values, the mean, $\mu_T$, and variance, $\sigma_T^2$, of $T$ are easily found from Eqs.~\ref{eq:schil-mean} and \ref{eq:schil-var-limit}, respectively.  The pull is then given by $(T - \mu_T)/\sigma_T$.  The pull distributions for the low, medium and high statistics ensembles ($n_d = 100, 1000$ and 10000, respectively) obtained by mixing each data set with a Monte Carlo data set ($n_{mc} = 10~n_d$) sampled from the Model p.d.f.\ are shown in Fig.~\ref{fig:schil-results}(a).  The agreement of the results obtained with the expected (standard normal) distribution is excellent.  This is confirmation that the approximation for $\sigma_T$ given in Eq.~\ref{eq:schil-var-limit} is valid for all three sample sizes considered here.  

Fig.~\ref{fig:schil-results}(b) shows the pull distributions (obtained in exactly the same way as for the Model p.d.f.) for the Fit~I p.d.f.'s.  The agreement with the predicted distribution is very good; however, there is a small test bias: $\mu_{\rm pull} \approx -0.3$ for each value of $n_d$.  Recall from Section~2 that such a bias is expected because each Fit~I p.d.f.\ is obtained from a fit to the data.  This means that the agreement between the Fit~I Monte Carlo and the data is slightly better (on average) than that of two data sets randomly sampled from the same parent distribution.  

Because larger values of $T$ are expected if $f \ne f_0$, rejecting the hypothesis $f = f_0$ at level $\alpha$ is a one-sided cut on the pull.  {\em E.g.}, the cuts $(T - \mu_T)/\sigma_T > 1.28$ and 1.64 correspond to rejecting at 90\% and 95\% confidence level, respectively.  The rejection powers at 95\% confidence level for the Model and Fit~I p.d.f.'s are shown in Table~\ref{table:schil-power}. Because of the relatively small number of data sets used in each ensemble, there is a small uncertainty (a few percent) on each value.  The deviation from the expected rejection rate of 5\% is within a few percent for both the Model and Fit~I p.d.f.'s.  This is further confirmation that the approximation for $\sigma_T$ given in Eq.~\ref{eq:schil-var-limit} is valid for all three sample sizes considered for the values of $n_{mc}$ and $n_k$ used in this study.  It also demonstrates that the effect of the small test bias on the rejection performance at 95\% confidence level is only a few percent and can safely be ignored.

Figs.~\ref{fig:schil-results}(c) and (d) show the pull distributions obtained for the Fit~II and Fit~III p.d.f.'s, respectively.  The rejection powers at 95\% confidence level for the Fit~II and Fit~III p.d.f.'s are shown in Table~\ref{table:schil-power}. The rejection power for Fit~II is excellent for $n_d=10000$, good for $n_d=1000$ and poor for $n_d=100$.  For Fit~III the rejection power is fair for $n_d=10000$ and poor for $n_d \leq 1000$.  Thus, this method appears to be better at rejecting a large localized discrepancy than a small omnipresent one.  Overall, its power is comparable to that of the $\chi^2$ test; however, this method does not require any knowledge about the functional form of the p.d.f.  The method presented in Ref.~\cite{ref:schil} is easy to use and understand and has decent rejection power; it would make a useful addition to the high energy physics g.o.f.\ toolkit.

\begin{table}
  \begin{center}
    \begin{tabular}{|c|c|ccc|}
      \hline
      $n_d$ & Model & Fit I & Fit II & Fit III \\
      \hline
      10000 & 3\% & 3\% & 100\% & 35\% \\
      1000 & 2\% & 4\% & 73\% & 5\% \\
      100 & 6\% & 3\% & 5\% & 3\% \\
      \hline
    \end{tabular}
    \\
    \vspace{0.01\textheight}
    \caption[]{\label{table:schil-power}  
      Rejection power at 95\% confidence level using the mixed-sample method of Ref.~\cite{ref:schil}.
    }
  \end{center}   
\end{table}

\subsection{Point-to-Point Dissimilarity Methods}

If the parent p.d.f.\ of the data is known, then the statistic formed from the integral of the quadratic difference between $f$ and $f_0$,
\begin{equation}
  \label{eq:quad-diff}
  T = \frac{1}{2}\int \left( f(\vec{x}) - f_0(\vec{x}) \right)^2 d\vec{x},
\end{equation}
can be used as a measure of g.o.f. Since $f$ is not known, $T$ cannot be calculated.  Of course, if $f$ were known there would also be no reason to perform a fit.  

A more general form of Eq.~\ref{eq:quad-diff} involves correlating the difference between the two p.d.f.'s at different points in the multivariate space using a weighting function, denoted by $\psi(|\vec{x} - \vec{x}^{\prime}|)$, as follows~\cite{ref:cuadras,ref:baringhaus,ref:aslan1,ref:aslan2}:
\begin{equation}
  \label{eq:pt2pt-gen}
  T = \frac{1}{2}\int\int \left(f(\vec{x}) - f_0(\vec{x}) \right)  \left(f(\vec{x}^{\prime}) - f_0(\vec{x}^{\prime}) \right) \psi(|\vec{x} - \vec{x}^{\prime}|) d\vec{x} d\vec{x}^{\prime}.
\end{equation}
Notice that Eq.~\ref{eq:quad-diff} is simply Eq.~\ref{eq:pt2pt-gen} for the case $\psi(|\vec{x} - \vec{x}^{\prime}|) = \delta(|\vec{x} - \vec{x}^{\prime}|)$.  Expanding the term in the integrand yields
\begin{equation}
  \label{eq:pt2pt-gen2}
  T = \frac{1}{2}\int\int \left[f(\vec{x})f(\vec{x}^{\prime}) + f_0(\vec{x})f_0(\vec{x}^{\prime}) -2f(\vec{x})f_0(\vec{x}^{\prime})\right]\psi(|\vec{x}-\vec{y}|) d\vec{x}d\vec{x}^{\prime},
\end{equation}
which can be calculated using only the data and a Monte Carlo data set sampled from $f_0$ as follows: 
\begin{eqnarray}
  \label{eq:pt2pt-gen3}
  T = \frac{1}{n_d(n_d-1)}\sum\limits_{i,j>i}^{n_d} \psi(|\vec{x}^d_i - \vec{x}^d_j|) \hspace{3.0in}\nonumber \\
 \hspace{1.0in}  + \frac{1}{n_{mc}(n_{mc}-1)}\sum\limits_{i,j>i}^{n_{mc}} \psi(|\vec{x}^{mc}_i - \vec{x}^{mc}_j|) 
  - \frac{1}{n_dn_{mc}}\sum\limits_{i,j}^{n_d,n_{mc}} \psi(|\vec{x}^d_i - \vec{x}^{mc}_j|).
\end{eqnarray}
Thus, $T$ is very easy to calculate.  I also note here that the expectation value of $T$ is larger for the case $f \ne f_0$.  

The following choices for the functional form of $\psi(x)$ are used in the statistical literature: Ref.~\cite{ref:cuadras} uses $\psi(x) = x^2$;  Ref.~\cite{ref:baringhaus} uses $\psi(x) = x$; Refs.~\cite{ref:aslan1,ref:aslan2} use $\psi(x) = \frac{1}{x}$, $\psi(x) = -\log{x}$ and $\psi(x) = e^{-x^2/2\sigma^2}$.  Ref.~\cite{ref:aslan1}, which was written by physicists, observes that for the case $\psi(x) = \frac{1}{x}$ Eq.~\ref{eq:pt2pt-gen} is the electrostatic energy of two charge distributions of opposite sign. Ref.~\cite{ref:aslan1} also notes that the electrostatic energy is minimized if the charges neutralize each other, {\em i.e.}, if $f = f_0$.  This was the motivating factor behind the derivation of their method. 

The optimal choice for the weighting function depends on the p.d.f.\ to be tested.  Since Dalitz-plot p.d.f.'s vary rapidly, I chose to use $\psi(x) = e^{-x^2/2\sigma^2}$; thus, from this point forward I will follow Ref.~\cite{ref:aslan1} which alters Eq.~\ref{eq:pt2pt-gen3} slightly by writing
\begin{equation}
  T = \frac{1}{n_d^2}\sum\limits_{i,j>i}^{n_d} \psi(|\vec{x}^d_i - \vec{x}^d_j|) - \frac{1}{n_d n_{mc}}\sum\limits_{i,j}^{n_d,n_{mc}} \psi(|\vec{x}^d_i - \vec{x}^{mc}_j|).
\end{equation}
The replacement of $1/n_d(n_d-1)$ with $1/n_d^2$ is made due to the better small number properties of the latter expression.  The term in Eq.~\ref{eq:pt2pt-gen3} that depends only on the Monte Carlo is dropped because its statistical fluctuations should be negligible (assuming $n_{mc} \gg n_d$).  Perhaps, from a theoretical perspective, it would be better to keep this term; however, in practice I found that including it greatly increased the processing time but had no effect on the performance of the method.

Ref.~\cite{ref:aslan1} also suggests that, rather than using a constant value for $\sigma$ in $\psi(x)$, the choice $\sigma(\vec{x}) \propto 1/f_0(\vec{x})$ improves the power of the test. Thus, I have chosen to use the following weighting function:
\begin{equation}
  \label{eq:psi-def}
  \psi(|\vec{x}_i - \vec{x}_j|) = e^{-|\vec{x}_i - \vec{x}_j|^2/2\sigma(\vec{x}_i)\sigma(\vec{x}_j)},
\end{equation}
where $\sigma(\vec{x}) = \bar{\sigma}/ (f(\vec{x}) \int d\vec{x}^{\prime})$. I have included the factor of $\int d\vec{x}^{\prime}$, which is simply the area of the Dalitz plot in this analysis, because the mean value of $f(\vec{x}) \int d\vec{x}^{\prime}$ is one.  This makes the interpretation of $\bar{\sigma}$ much easier. I note here that in my tests of this method the variable $\sigma(\vec{x})$ did perform significantly better than using a constant $\sigma$.

Given this choice for $\psi(x)$, $T$ can now be calculated from the data and a Monte Carlo data set sampled from the test p.d.f.  The number of Monte Carlo events generated should be much larger than the number of data ($n_{mc} \gg n_d$) to ensure that statistical fluctuations in the Monte Carlo are negligible.  Unlike for the mixed-sample method, there is no inherent limit on $n_{mc}$ here.  The only limiting factor is the amount of processing time. I will postpone the discussion on the lone nuisance parameter, $\bar{\sigma}$, until later but note here that its optimal value can be estimated by examining $f_0$; {\em i.e.}, it can be obtained from the physics or interest.

Once the Monte Carlo is generated and a value for $\bar{\sigma}$ is chosen, $T$ can be calculated; however, the distribution of $T$ for the case $f = f_0$ is not known which means that the $p$-value cannot be calculated.  Although it cannot be calculated, the $p$-value can be estimated using a re-sampling method known as the {\em permutation test}.  This approach involves combining the data and Monte Carlo data into a pooled sample of size $n_d + n_{mc}$.  A sample of size $n_d$ is then randomly drawn from the pooled sample and temporarily labeled ``data'' while the remaining $n_{mc}$ events are labeled ``Monte Carlo.''  The test statistic, denoted $T_{\rm perm}$, is then calculated with these designations for each event.  This process is then repeated $n_{\rm perm}$ times to obtain $\{T^1_{\rm perm}\ldots T^{n_{\rm perm}}_{\rm perm}\}$.  The $p$-value is then simply the fraction of times where $T < T_{\rm perm}$.  For a more detailed discussion on this technique, see Appendix~C.  I note here that, if Monte Carlo generation is not too expensive, one could instead generate an ensemble of Monte Carlo data sets to obtain an approximation of the $T$ distribution and, in turn, the $p$-value.  

\begin{figure*}
  \centering
  \subfigure[]{
    \includegraphics[width=0.4\textwidth]{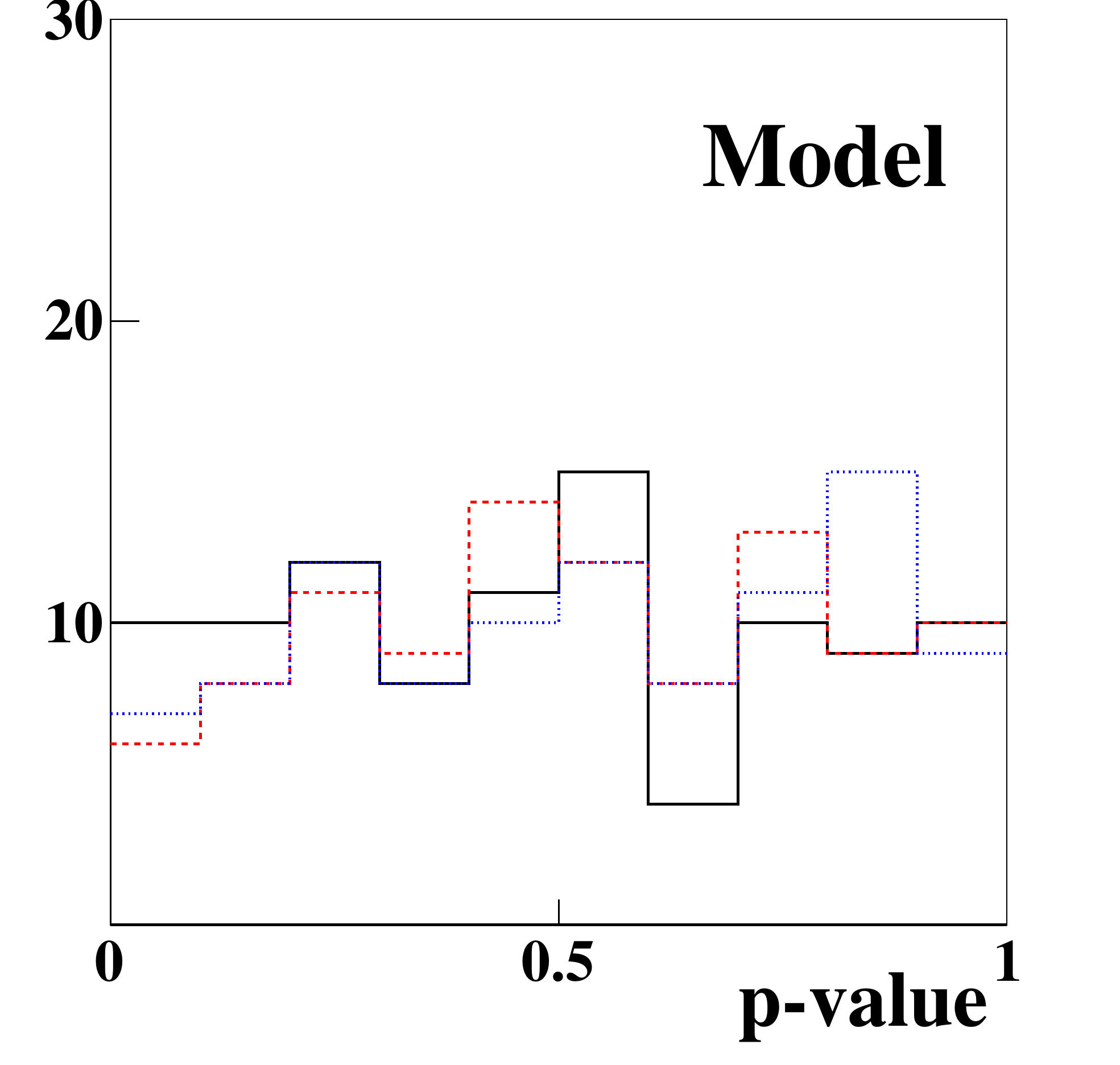}
  }
  \subfigure[]{
    \includegraphics[width=0.4\textwidth]{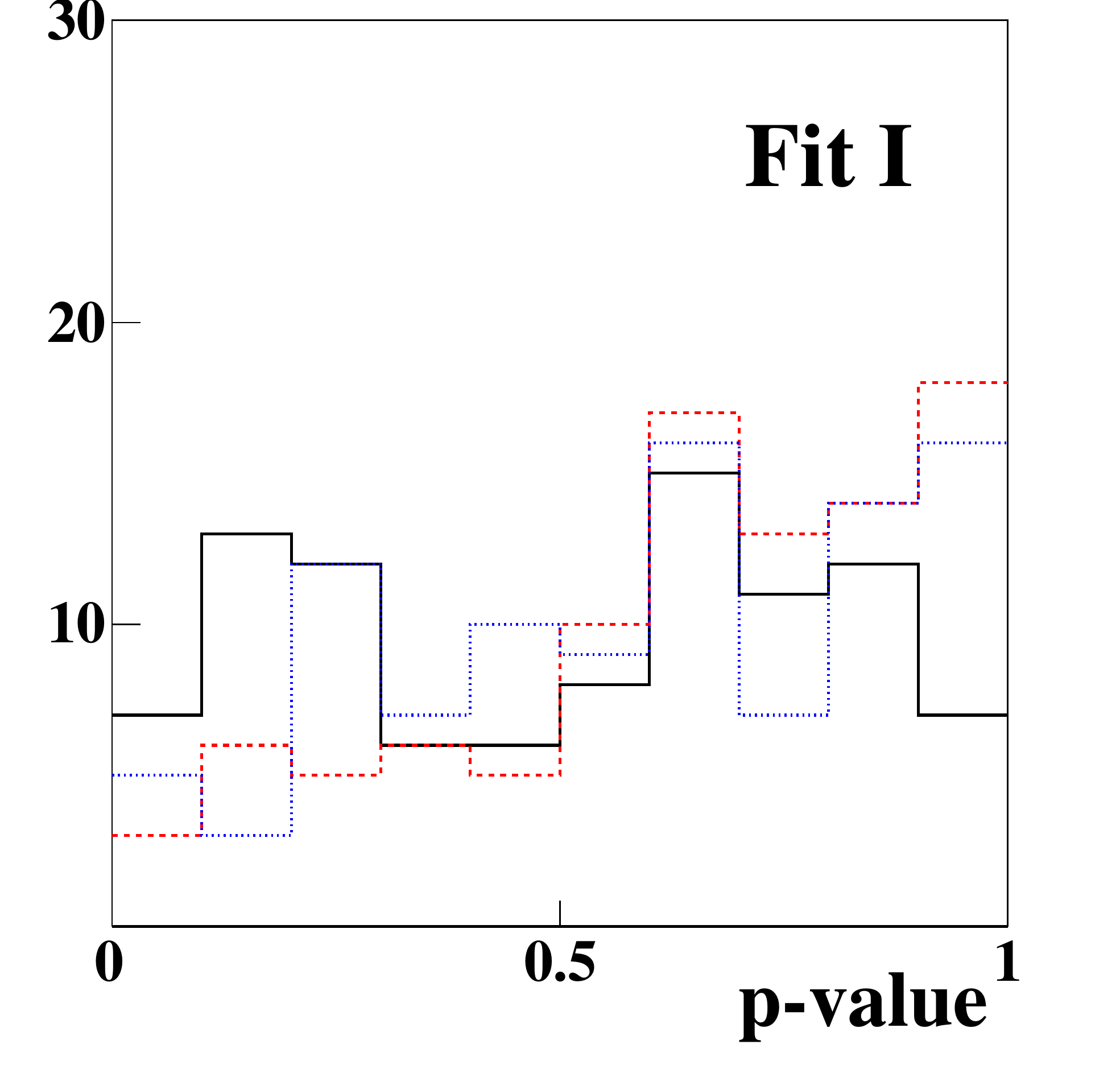} 
  }
  \\
  \subfigure[]{
    \includegraphics[width=0.4\textwidth]{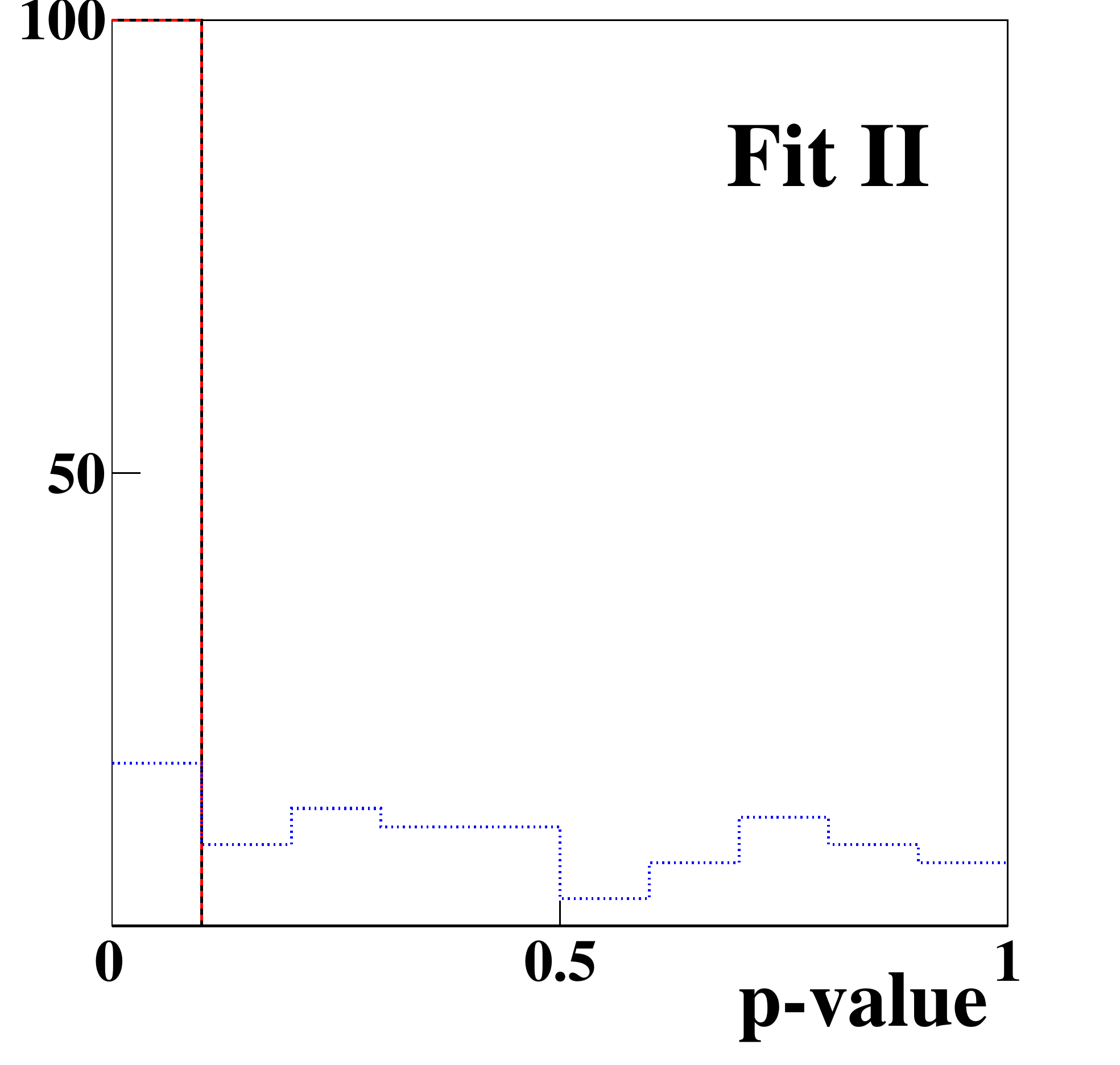} 
  }
  \subfigure[]{
    \includegraphics[width=0.4\textwidth]{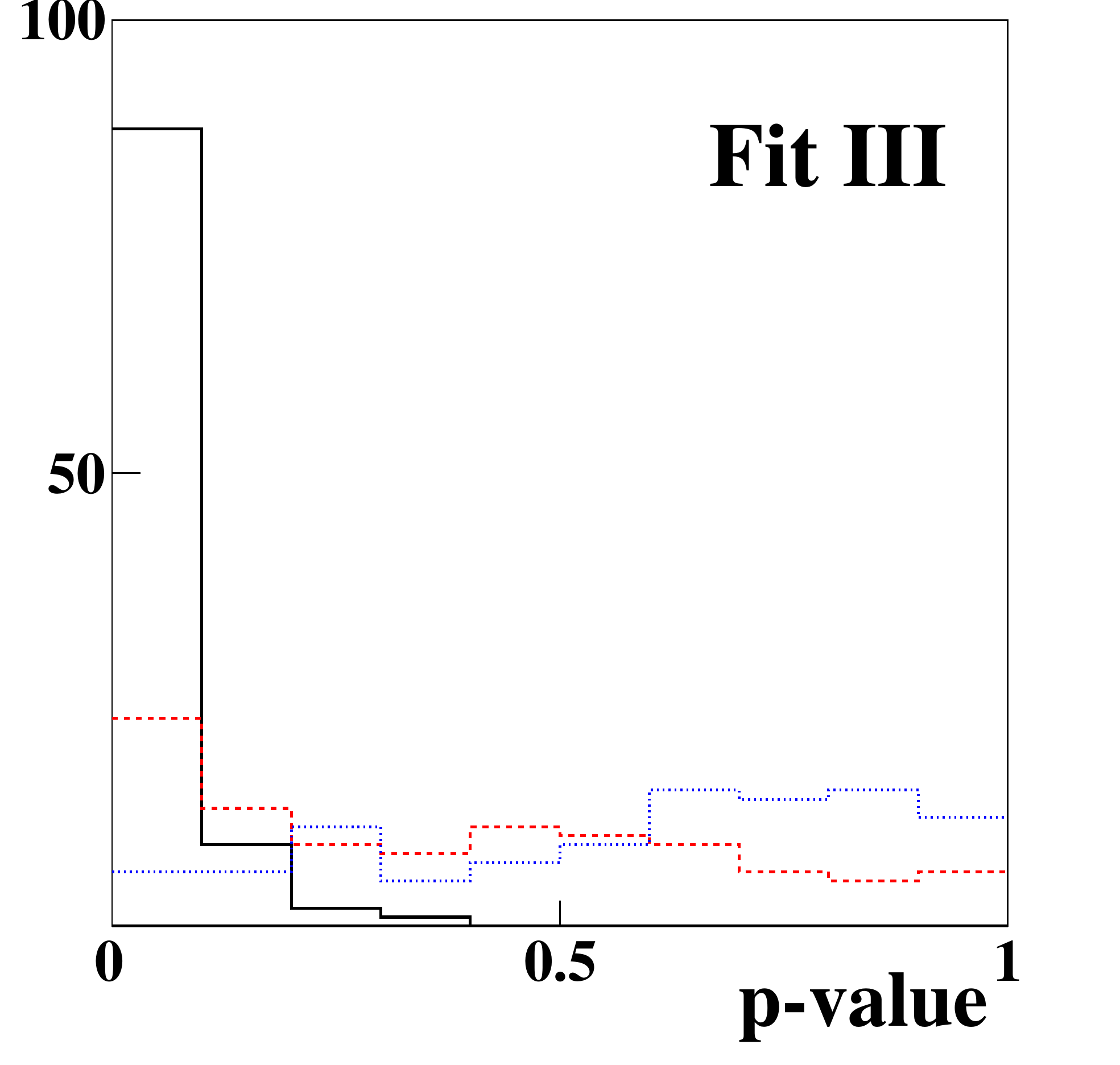} 
  }
  \caption[]{\label{fig:aslan-results}
     (Color Online) $p$-value distributions obtained from low (blue dotted), medium (red dashed) and high (solid black histograms) statistics data sets from the following p.d.f.'s using the point-to-point dissimilarity g.o.f.\ method of Ref.~\cite{ref:aslan1}: (a) Model; (b) Fit~I; (c) Fit~II; (d) Fit~III.  Data sets whose $p$-values are less than 0.05 are rejected at 95\% confidence level (by definition).  See Section~3.3 for further discussion on these results.
  }
\end{figure*}

In all of the results that follow the value $\bar{\sigma} = 0.01$ is used (this quantity is discussed in detail below). The $p$-value distributions for the low, medium and high statistics ensembles ($n_d = 100, 1000$ and 10000, respectively) obtained using each data set and a Monte Carlo data set sampled from the Model p.d.f.\ are shown in Fig.~\ref{fig:aslan-results}(a).  The agreement of the results obtained with the predicted (flat) distribution is excellent.  This is confirmation that the permutation technique does produce valid $p$-values for all three sample sizes considered.  Fig.~\ref{fig:aslan-results}(b) shows the $p$-value distributions for the Fit~I p.d.f.'s (obtained in exactly the same way as for the Model p.d.f.).  The agreement with the predicted distribution is very good; however, there is a small test bias (a small positive slope) for $n_d \le 1000$.  Again, such a bias is expected because each Fit~I p.d.f.\ is obtained from a fit to the data. 

Rejection of the hypothesis $f = f_0$ at level $\alpha$ is simply done by requiring the $p$-value be less than $1-\alpha$ ({\em e.g.}, $p < 0.05$ at 95\% confidence level).  The rejection powers at 95\% confidence level for the Model and Fit~I p.d.f.'s are given in Table~\ref{table:aslan-power}.  The deviation from the expected rejection rate of 5\% is within a few percent for both the Model and Fit~I p.d.f.'s for $\bar{\sigma} = 0.01$.  This is further confirmation that the permutation technique is valid for all three sample sizes considered in this study.  It also demonstrates that the effect of the small test bias on the rejection performance at 95\% confidence level is only a few percent.
Figs.~\ref{fig:aslan-results}(c) and (d) show the $p$-value distributions obtained for the Fit~II and Fit~III p.d.f.'s.  The rejection powers at 95\% confidence level for these p.d.f.'s are given in Table~\ref{table:aslan-power}. The rejection power for Fit~II is excellent for $n_d \geq 1000$ and fairly poor for $n_d=100$.  For Fit~III the rejection power is good for $n_d=10000$, fair for $n_d = 1000$ and poor for $n_d = 100$.  These are impressive results; the rejection power is far greater than that of the binned $\chi^2$ test.

\begin{table}[h!]
  \begin{center}
    \begin{tabular}{|c|c|ccc|}
      \hline
      $n_d$ & Model & Fit I & Fit II & Fit III \\
      \hline
      10000  & [0:5:4:6]\% & [8:2:4:1]\% & [100:100:100:100]\% & [41:78:81:77]\% \\
      1000  & [7:6:3:9]\% & [5:4:2:3]\% & [93:100:100:71]\% & [9:12:15:29]\% \\
      100  & [5:6:3:2]\% & [5:5:2:1]\% & [11:14:10:1]\% & [5:4:3:1]\% \\
      \hline
    \end{tabular}
    \\
    \vspace{0.01\textheight}
    \caption[]{\label{table:aslan-power}  
      Rejection power at 95\% confidence level for $\bar{\sigma}$ = [0.001:0.005:0.01:0.05] using the point-to-point dissimilarity method of Ref.~\cite{ref:aslan1}. See Section~3.3 for discussion on the value of  $\bar{\sigma}$.
    }
  \end{center}   
\end{table}

The results obtained using $\bar{\sigma} = 0.01$ are impressive, but how does one know what value to choose for $\bar{\sigma}$?   Table~\ref{table:aslan-power} shows the rejection power at 95\% confidence level for the $\bar{\sigma}$ values 0.001, 0.005, 0.01 and 0.05.  The units of $\bar{\sigma}$ are those of mass squared (see Eq.~\ref{eq:psi-def}); thus, the quantity $\sqrt{\bar{\sigma}}$, which has approximate values 0.03, 0.07, 0.1 and 0.22, has units of mass.  The method presented in this section performs best for $0.07 \lesssim \sqrt{\bar{\sigma}} \lesssim 0.1$.  The typical resonance width in the Dalitz-plot model used in this analysis is $\bar{\Gamma} = \sum {\rm ff}_r \Gamma_r/\sum{\rm ff}_r \approx 0.06$, where ${\rm ff}_r$ and $\Gamma_r$ are the fit fractions and widths of the resonances, respectively.  The preferred range for $\bar{\sigma}$ can be rewritten using this quantity as $\bar{\Gamma} \lesssim \sqrt{\bar{\sigma}} \lesssim 2\bar{\Gamma}$.  

This result is not surprising.  The widths of the resonances serve as a measure of how rapidly the p.d.f.\ varies.  If $\sqrt{\bar{\sigma}} < \bar{\Gamma}$ then the p.d.f.\ is approximately constant in the Gaussian region around each event.  If $\sqrt{\bar{\sigma}} > 2\bar{\Gamma}$ then the finer structure in the p.d.f.\ is lost in the comparison. From this one can conclude that the physics of interest can be used to estimate the optimal value of $\bar{\sigma}$.  In practice, it would be advisable to obtain $p$-values for several $\bar{\sigma}$ values in the expected optimal region.  The conclusions drawn about the quality of the fit should not depend on this quantity (provided a {\em reasonable} choice is made).  If a strong dependence is observed, then further study using Monte Carlo may be necessary.  I note here that for other types of high energy physics analyses a different choice for $\psi(|\vec{x}_i - \vec{x}_j|)$ may perform better. Additional Monte Carlo studies may be necessary in these cases.

This method has excellent rejection power for both large localized discrepancies and small omnipresent ones, even for fairly low-statistics data sets.  Conceptually, it is not as easy to understand as some other methods, {\em e.g.}, the mixed-sample method described in Section~3.2.  It also requires a rather large amount of processing time ($\mathcal{O}(1~{\rm hr})$ for $n_d = 10000$) due to the fact that the use of the permutation technique is required.  These downsides are not enough to out-way its excellent performance.  For a Dalitz-plot (or similar) analysis, this method is a very powerful g.o.f.\ tool.

\subsection{Distance to Nearest Neighbor Methods}

The distance from any event to its nearest neighbor is inversely proportional to the magnitude of the parent p.d.f in the region around the event. {\em I.e.}, in a region where the parent p.d.f.\ is larger (smaller) the density of events will also be larger (smaller); thus, the events will be closer together (farther apart) on average.  This fact can be used to construct a g.o.f.\ test.  

Ref.~\cite{ref:bickel-1983} defines the following statistic for the $i^{th}$ event in a data set:
\begin{equation}
  \label{eq:bickel-u}
  U_i = {\rm exp}\left(-n_d \int_{|\vec{x} - \vec{x}_i| < R^{nn}_i} f_0(\vec{x}) d\vec{x}\right) \simeq {\rm exp}\left(-n_d f_0(\vec{x}_i) V_D(R^{nn}_i)\right) \nonumber,
\end{equation}
where $R^{nn}_i$ is the distance from the $i^{th}$ event to its nearest neighbor and $V_D(R) \propto R^D$ is the $D$-dimensional hyper-spherical volume of radius $R$.  The approximation 
\begin{equation}
\int_{|\vec{x} - \vec{x}_i| < R^{nn}_i} f_0(\vec{x}) d\vec{x} \simeq f_0(\vec{x}_i) V_D(R^{nn}_i), 
\end{equation}
which is valid if the hypersphere centered at $\vec{x}_i$ with radius $R^{nn}_i$ is sufficiently small such that $f_0(\vec{x})$ is approximately constant inside of it, is made to avoid having to do the integral.  Given the power of modern computers, it is possible to omit this substitution and do the integral numerically; however, I found that this had no effect on the results.  For the case $f = f_0$, the distribution of $U$ values is approximately uniform (see Appendix~D for a detailed discussion).

Fig.~\ref{fig:bickel-example} shows the $U$ distributions obtained for a single high statistics ($n_d = 10000$) data set.  For the Model and Fit~I p.d.f.'s the distributions are in good agreement with the expected (uniform) one.  The Fit~II $U$ distribution has a significant deviation from this, while the Fit~III distribution does not.  Based on these plots one would expect (for $n_d = 10000$) this method to have good rejection power for Fit~II and poor rejection power for Fit~III.

\begin{figure*}
  \centering
  \subfigure[]{
    \includegraphics[width=0.4\textwidth]{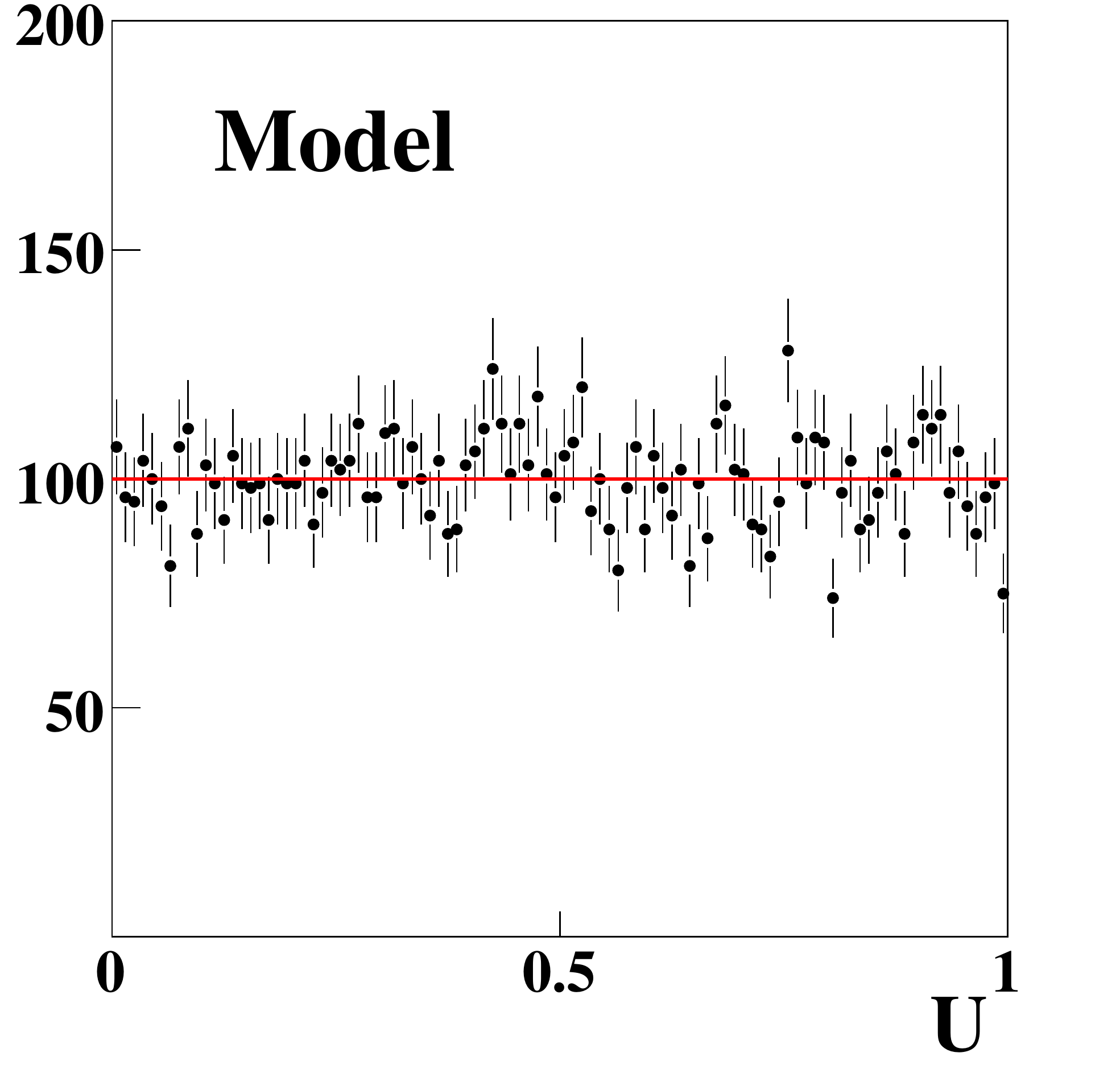}
  }
  \subfigure[]{
    \includegraphics[width=0.4\textwidth]{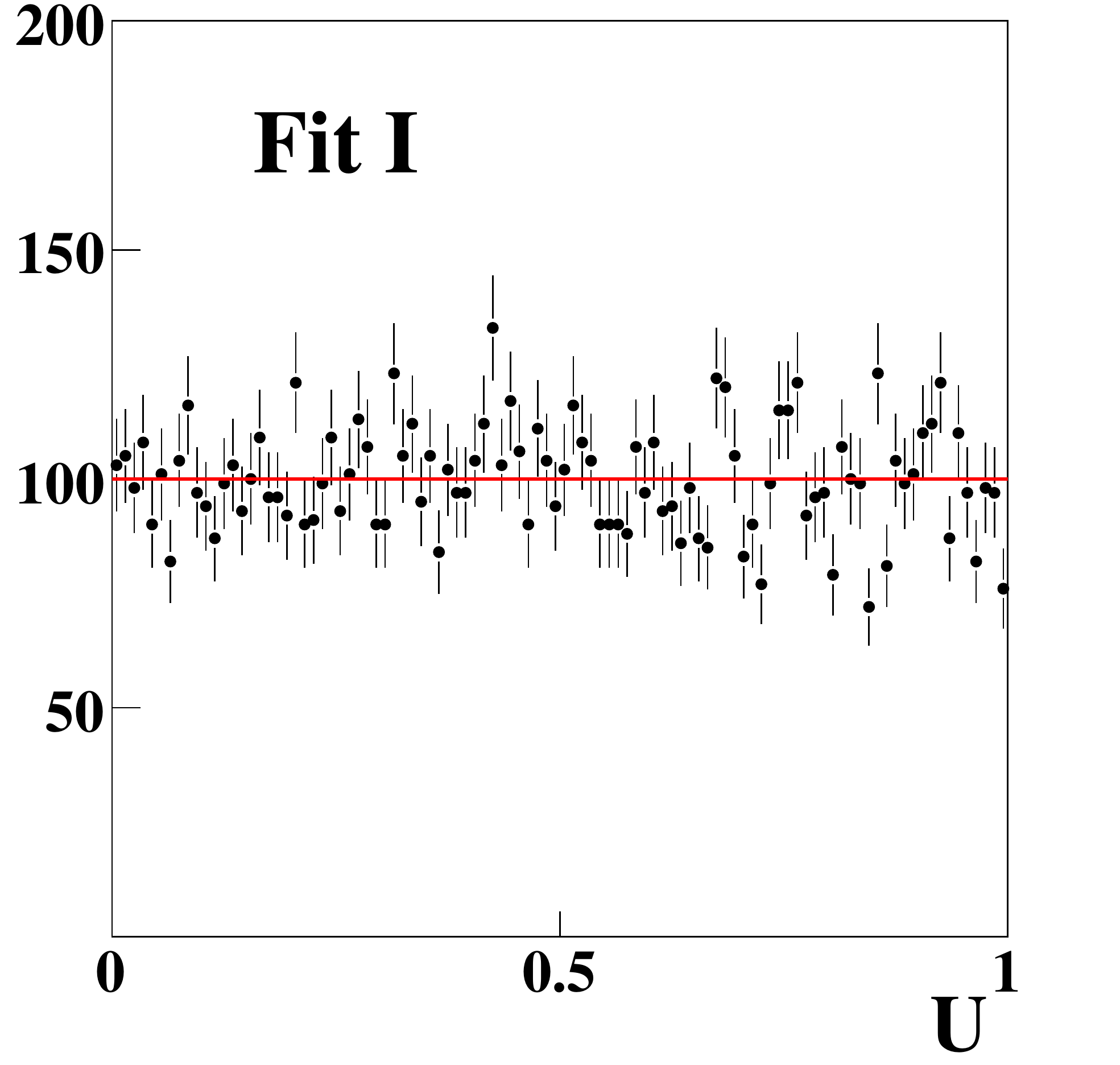} 
  }
  \\
  \subfigure[]{
    \includegraphics[width=0.4\textwidth]{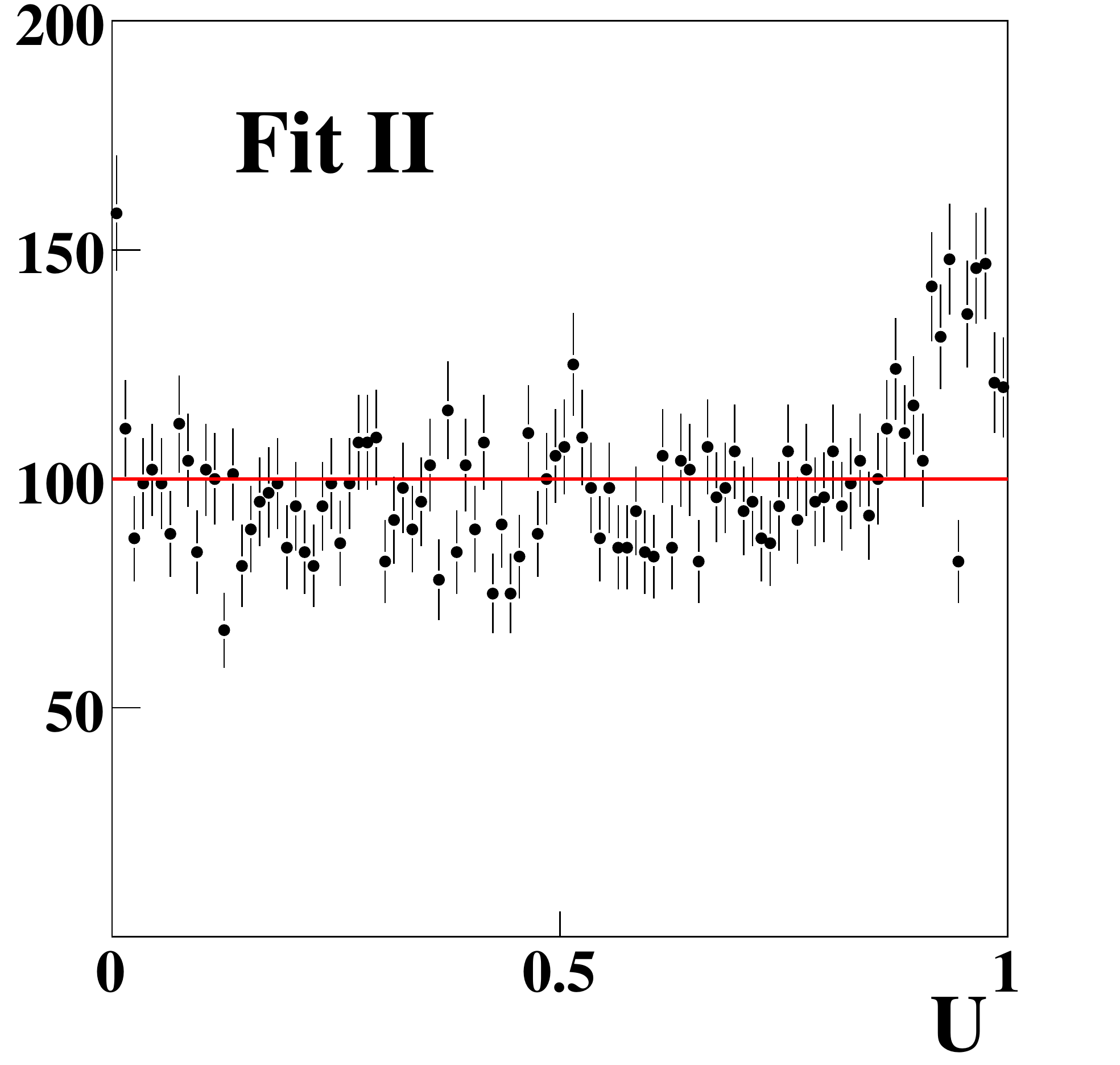} 
  }
  \subfigure[]{
    \includegraphics[width=0.4\textwidth]{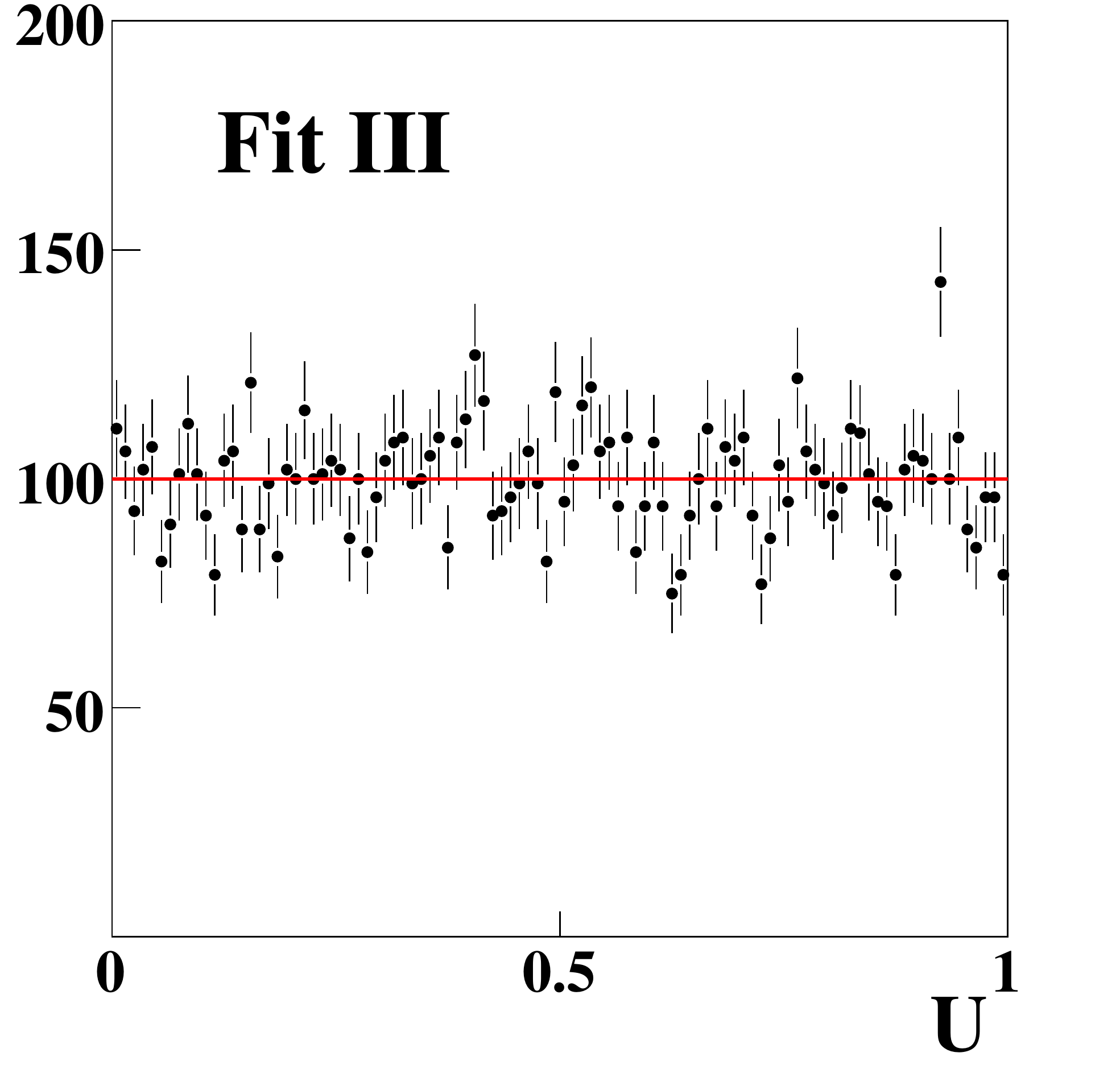} 
  }
  \caption[]{\label{fig:bickel-example}
    Distributions for the $U$ statistic of Ref.~\cite{ref:bickel-1983} defined in Eq.~\ref{eq:bickel-u} for a single high statistics ($n_d = 10000$) data set for the following p.d.f.'s: (a) Model; (b) Fit~I; (c) Fit~II; (d) Fit~III.  The (solid red) line shows the expected (uniform) distribution.  
  }
\end{figure*}

\begin{figure*}
  \centering
  \subfigure[]{
    \includegraphics[width=0.4\textwidth]{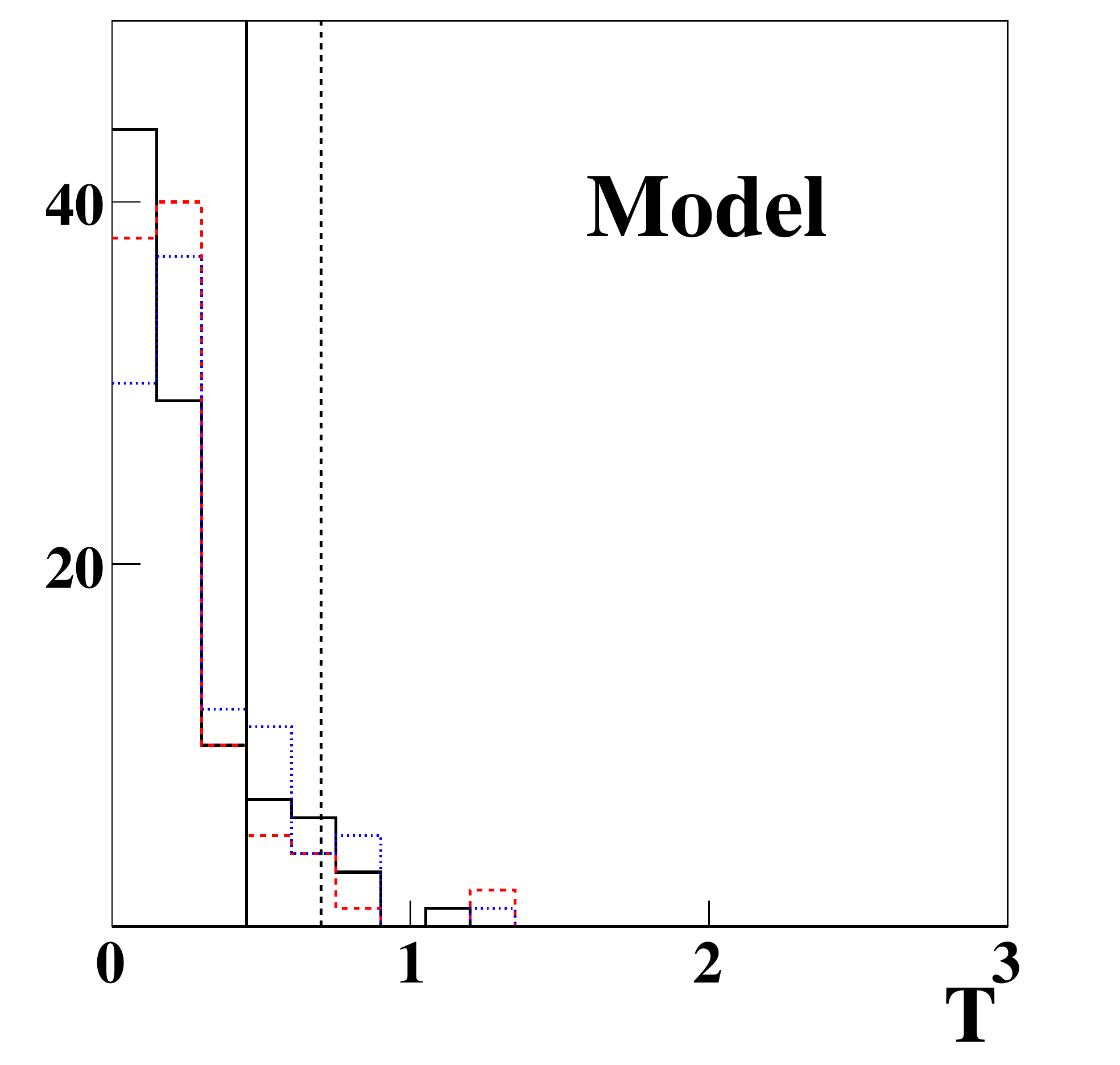}
  }
  \subfigure[]{
    \includegraphics[width=0.4\textwidth]{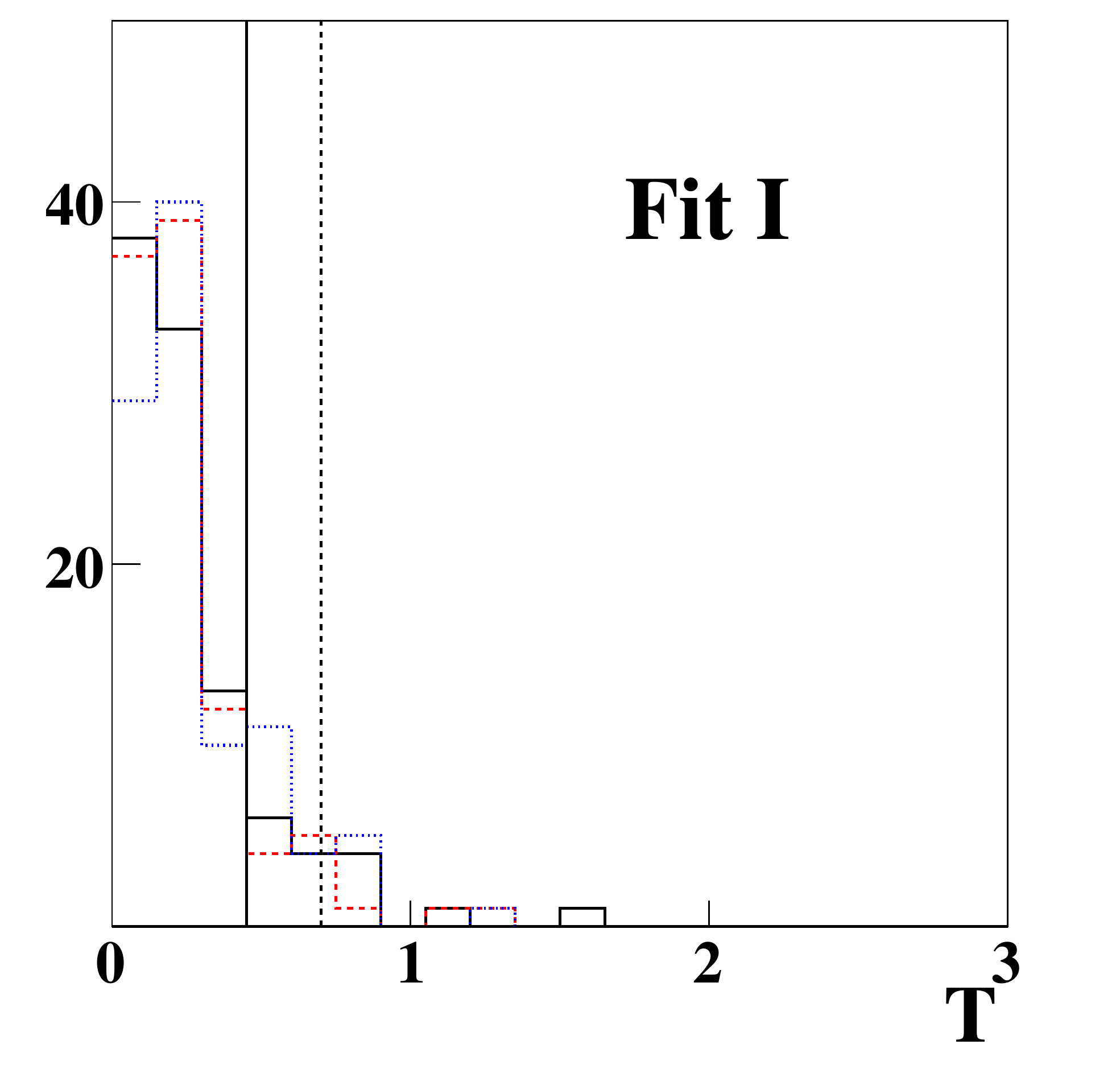} 
  }
  \\
  \subfigure[]{
    \includegraphics[width=0.4\textwidth]{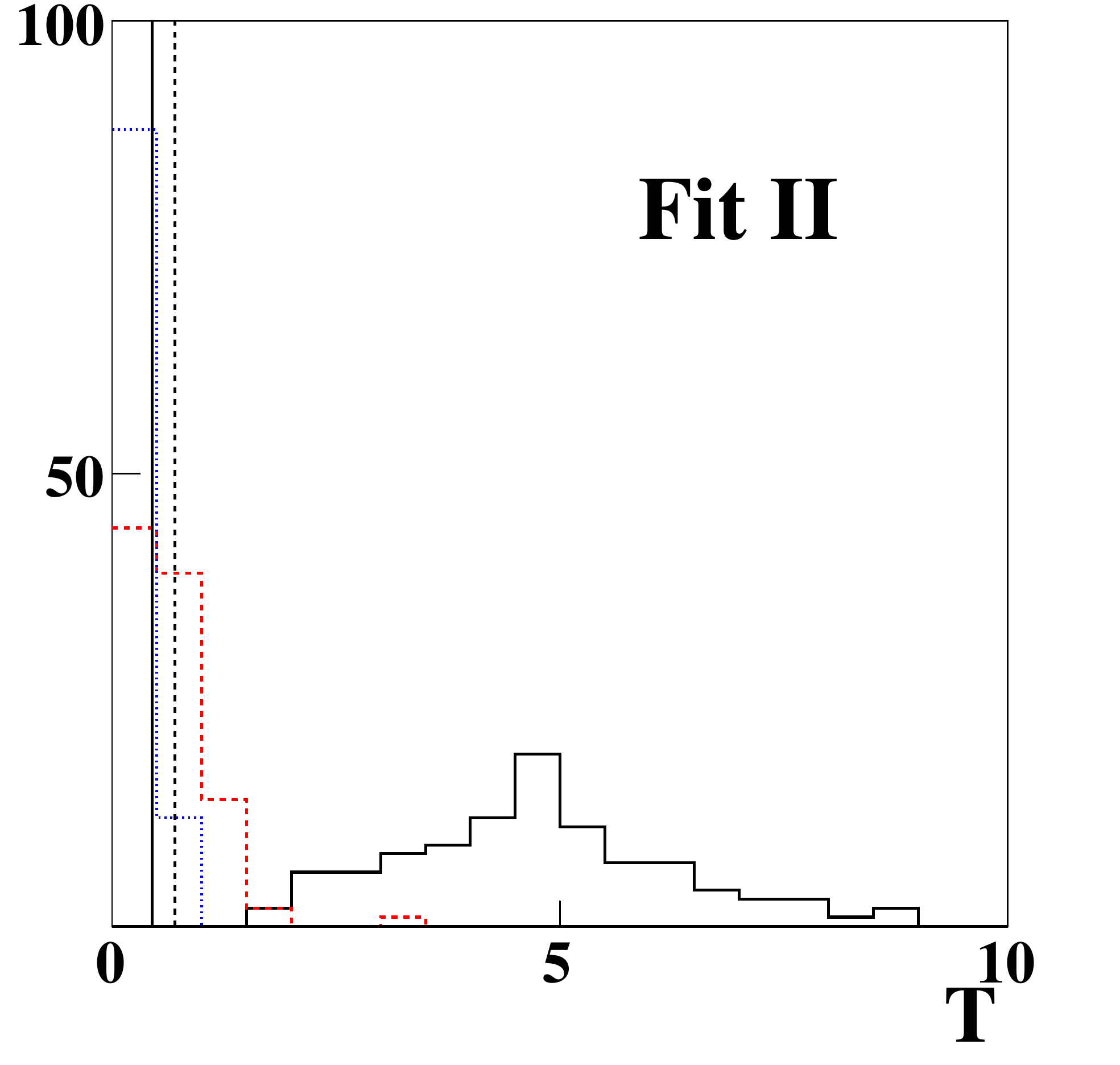} 
  }
  \subfigure[]{
    \includegraphics[width=0.4\textwidth]{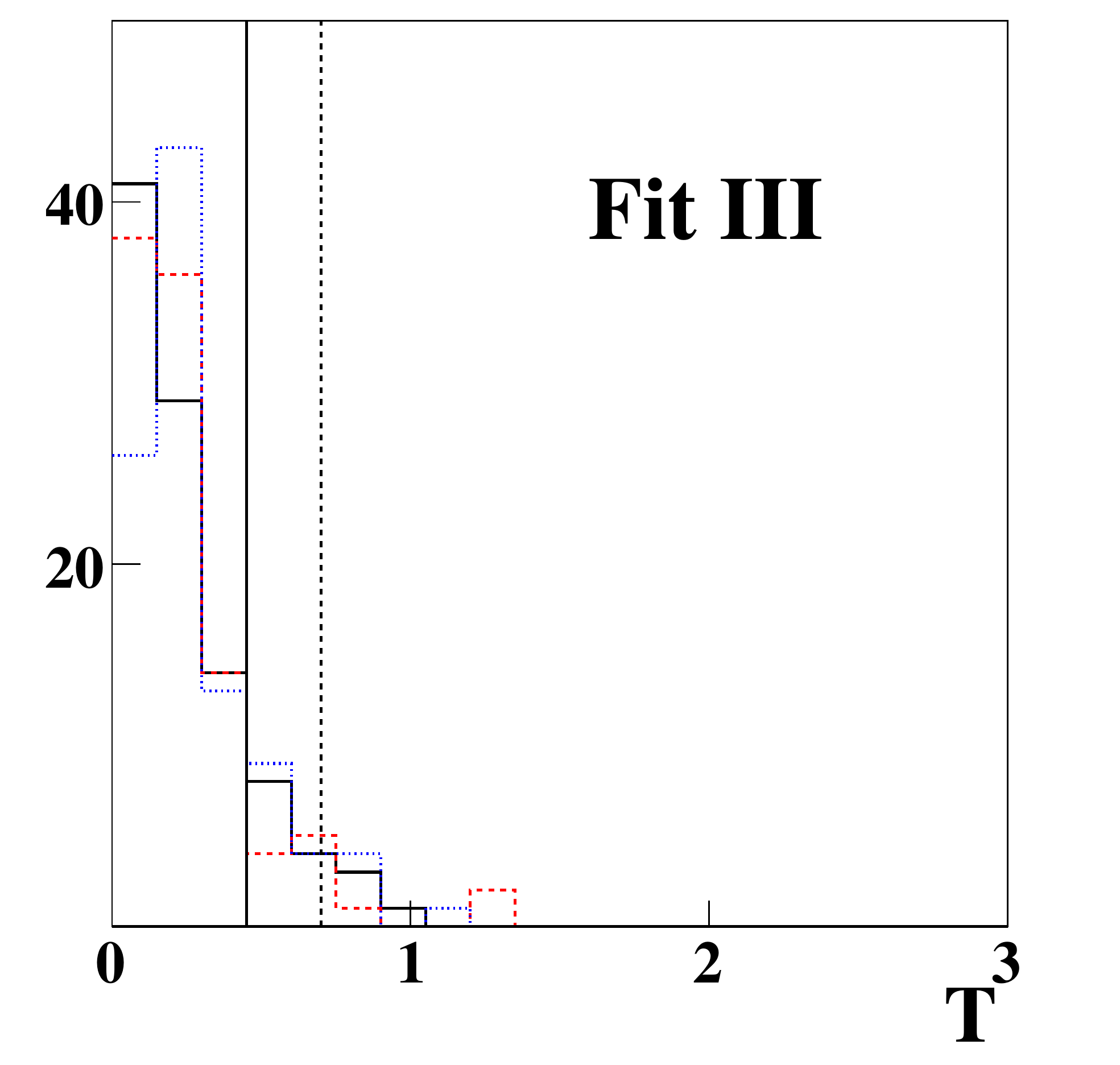} 
  }
  \caption[]{\label{fig:bickel-results}
    (Color Online) $T$ distributions obtained from low (blue dotted), medium (red dashed) and high (solid black histograms) statistics data sets from the following p.d.f.'s using the distance to nearest-neighbor g.o.f.\ method of Ref.~\cite{ref:bickel-1983}: (a) Model; (b) Fit~I; (c) Fit~II; (d) Fit~III.  The expected and observed locations of a 95\% confidence-level cut are shown by the solid and dashed lines, respectively.  See Section~3.4 for further discussion on these results.
  }
\end{figure*}

Obtaining a g.o.f.\ value simply involves testing the uniformity of the one-dimensional $U$ distributions.  This can be done in a number of ways ({\em e.g.}, using a $\chi^2$ test); the method suggested in Ref.~\cite{ref:bickel-1983} is to use the statistic
\begin{equation}
  T = \sum_i^{n_d} (U^{\prime}_i - i/n_d)^2, 
\end{equation}
where $\{U^{\prime}_i\}$ is the set of ordered $U$ values. Fig.~\ref{fig:bickel-results}(a) shows the $T$ distributions obtained using this method for the Model p.d.f.  Because of the fact that the uniformity of the $U$ distributions is only approximate, the observed location of the 95\% confidence-level cut is not at the value expected if the p.d.f.\ of the $U$ distribution was truly uniform.  This is discussed in more detail below.  Fig.~\ref{fig:bickel-results} also shows the $T$ distributions for the Fit~I, Fit~II and Fit~III p.d.f.'s.  The rejection powers at 95\% confidence level for all four p.d.f.'s are given in Table~\ref{table:bickel-power}.  For Fit~II, the rejection power is excellent for $n_d = 10000$, fair for $n_d = 1000$ and poor for $n_d = 100$.  The rejection power is poor for Fit~III for all data set sizes considered in this study. 

\begin{table}[]
  \begin{center}
    \begin{tabular}{|c|c|ccc|}
      \hline
      $n_d$ & Model & Fit I & Fit II & Fit III \\
      \hline
      10000 & 5(17)\% & 7(16)\% & 100(100)\% & 6(16)\% \\
      1000 & 3(12)\% & 4(12)\% & 38(61)\% & 5(12)\% \\
      100 & 7(21)\% & 7(21)\% & 6(14)\% & 7(18)\% \\
      \hline
    \end{tabular}
    \\
    \vspace{0.01\textheight}
    \caption[]{\label{table:bickel-power}  
      Rejection power for $T > 0.7(0.45)$ corresponding to the observed (expected) 95\% confidence level limit of the distance to nearest neighbor method of Ref.~\cite{ref:bickel-1983}.
    }
  \end{center}   
\end{table}

This method is very easy to use, has no nuisance parameters and requires very little processing time. Unfortunately, it is not very powerful.  The theoretical 95\% confidence-level cut rejects too many data sets due to the fact that the $U$ distributions are only approximately uniform for the case $f = f_0$.  Because of this, I do not think the $p$-values are worth calculating (especially given the power of the previous two methods); however, that does not mean this method is useless.  Producing the $U$ distribution is fast and easy and can reveal any large discrepancies between the fit p.d.f.\ and the data.  For this reason, this method does (at least) have a place as a diagnostic tool.  Furthermore,  it could be useful to publish the $U$ distribution for a very high-dimensional analysis to provide an easy to interpret demonstration of the qualitative agreement between the data and the fit p.d.f.

\subsection{Local-Density Methods}

The local density of events in a region around each event in the data set can be compared to the density expected from a test p.d.f.\ to determine the g.o.f.  This idea was introduced in Ref~\cite{ref:ripley} as a way of testing a two-dimensional distribution for complete spatial randomness, {\em i.e.}, testing whether a distribution is consistent with a uniform Poisson process.  For this (2-$D$ homogeneous) case the expected number of events contained inside a circle of radius $r$ around the $i^{th}$ event in the data set is given by
\begin{equation}
  \label{eq:ripley-expected}
  \left< \sum\limits_{j=1,j \ne i}^{n_d} I(|\vec{x}_i - \vec{x}_j|<r) \right> = (n_d-1)\frac{\pi r^2}{A},
\end{equation}
where $I({\rm true}) = 1$, $I({\rm false}) = 0$ and $A$ is the total area that the events are allowed to occupy.  Eq.~\ref{eq:ripley-expected} simply states that the expected number of events inside of the circle is the total number of events multiplied by the fraction of the total allowed area occupied by the circle used to collect the events.  If the circle centered at $\vec{x}_i$ with radius $r$ intersects the boundary of the allowed data region, then an edge correction factor is also required (this is discussed in detail below).

Ref.~\cite{ref:ripley} uses the sum of the Eq.~\ref{eq:ripley-expected} values for each event to define the $K$ function as follows:
\begin{equation}
  \label{eq:ripley-k}
  K(r) = \frac{A}{n_d^2} \sum\limits_{i=1}^{n_d}\sum\limits_{j\ne i} I(|\vec{x}_i - \vec{x}_j|<r)/a(i,j),
\end{equation}
where $a(i,j)$ is the edge correction factor for the circle centered at $\vec{x}_i$ with radius $|\vec{x}_i - \vec{x}_j|$. There is a lot of discussion in the literature concerning different ways of calculating $a(i,j)$.  Ref.~\cite{ref:ripley} suggests using the fraction of the circumference of the circle that lies inside the allowed data region.  I found that randomly sampling points within the circle and counting the fraction that fall in the allowed data region works best. This method is not discussed in the references I have read; however, this is most likely due to the limited computing power available at the time these references were written.  With the power of modern computers, this approach is quite feasible (although, it is still worth while to first check whether any part of the circle exits the allowed data region prior to doing the calculation).

The expectation value of Eq.~\ref{eq:ripley-k} is easily calculated to be $\left<K(r)\right> = \pi r^2 (n_d-1)/n_d \approx \pi r^2$.  Typically in two dimensions the quantity $L(r) = \sqrt{K(r)/\pi}$, introduced in Ref.~\cite{ref:besag}, is used instead; this quantity has $\left< L(r) \right> \approx r$.  The g.o.f.\ is then determined by examining how well the $K(r)$ or $L(r)$ distribution agrees with the expected one.  For a Poisson process, larger values of $K$ and $L$ are expected if the process is non-uniform. The reasoning is identical to that used above for the mixed-sample methods. If the process is non-uniform, then the events will tend to cluster together.  This results in there being more events (on average) inside the circles drawn around each event which, in turn, leads to larger $K$ and $L$ values.  Using the $K$ and $L$ distributions to determine g.o.f.\ is discussed in more detail below.

An extension for the inhomogeneous case ({\em i.e.}, for non-uniform p.d.f.'s) for $D$ dimensions is provided in Ref.~\cite{ref:baddeley}.  The generalized $K$ function is written as
\begin{equation}
  \label{eq:baddeley-k}
  K(r) = \frac{1}{V_D n_d^2} \sum\limits_{i=1}^{n_d}\sum\limits_{j\ne i} \frac{I(|\vec{x}_i - \vec{x}_j|<r)}{v(i,j) f_0(\vec{x}_i) f_0(\vec{x}_j)},
\end{equation}
where $V_D$ is the total allowed $D$-dimensional hyper-volume and $v(i,j)$ is the $D$-dimensional equivalent of $a(i,j)$ in Eq.~\ref{eq:ripley-k}; {\em i.e.}, it is the allowed hyper-volume fraction of a hypersphere centered at $\vec{x}_i$ with radius $|\vec{x}_i - \vec{x}_j|$.  The factor of $V_D$ appears in the denominator of Eq.~\ref{eq:baddeley-k} while the factor of $A$ appears in the numerator of Eq.~\ref{eq:ripley-k}.  This difference is simply due to the inverse-hyper-volume units of the p.d.f.\ factors included in Eq.~\ref{eq:baddeley-k} (that are not present in Eq.~\ref{eq:ripley-k}).

\begin{figure*}
  \centering
  \subfigure[]{
    \includegraphics[width=0.4\textwidth]{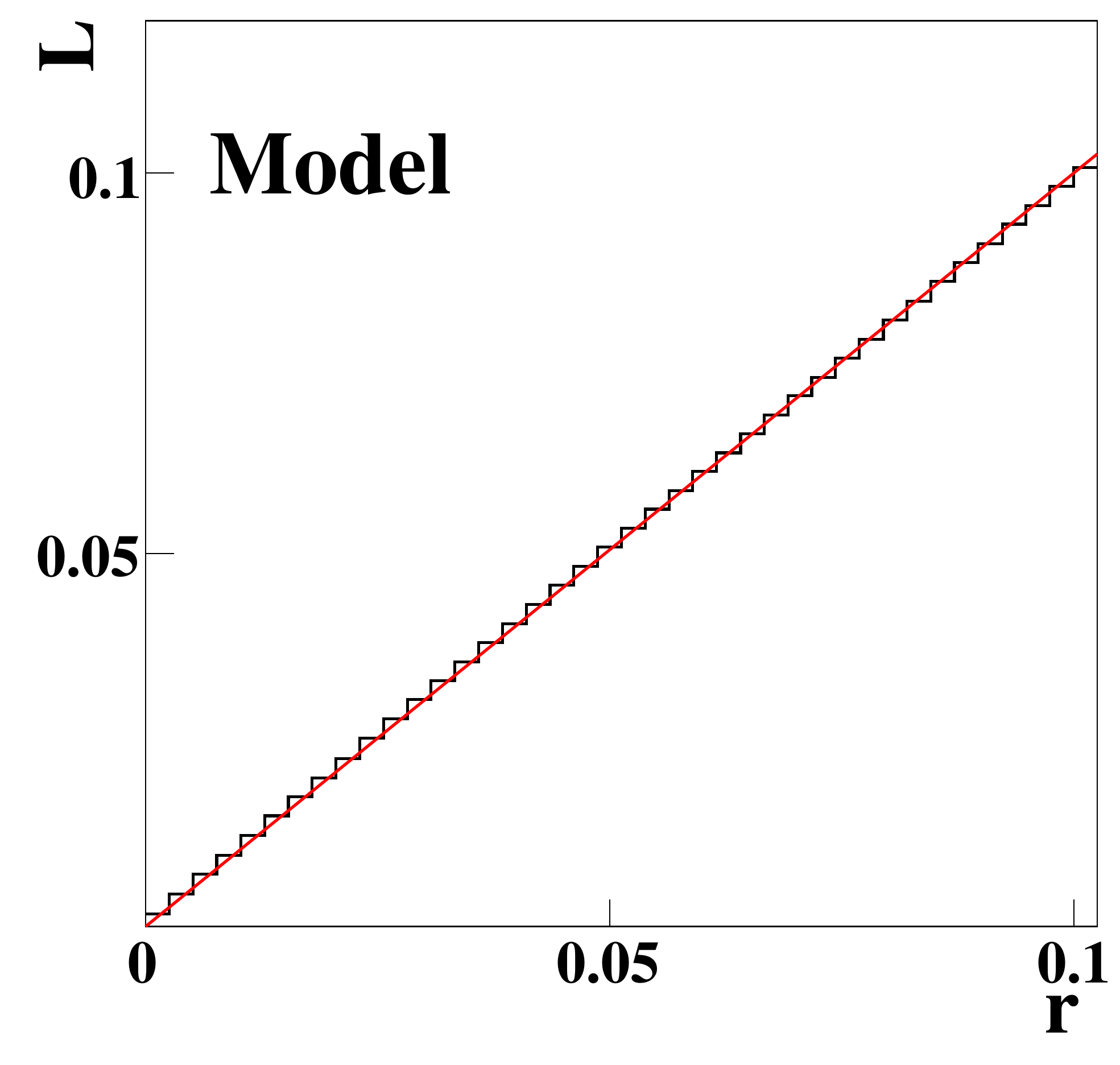}
  }
  \subfigure[]{
    \includegraphics[width=0.4\textwidth]{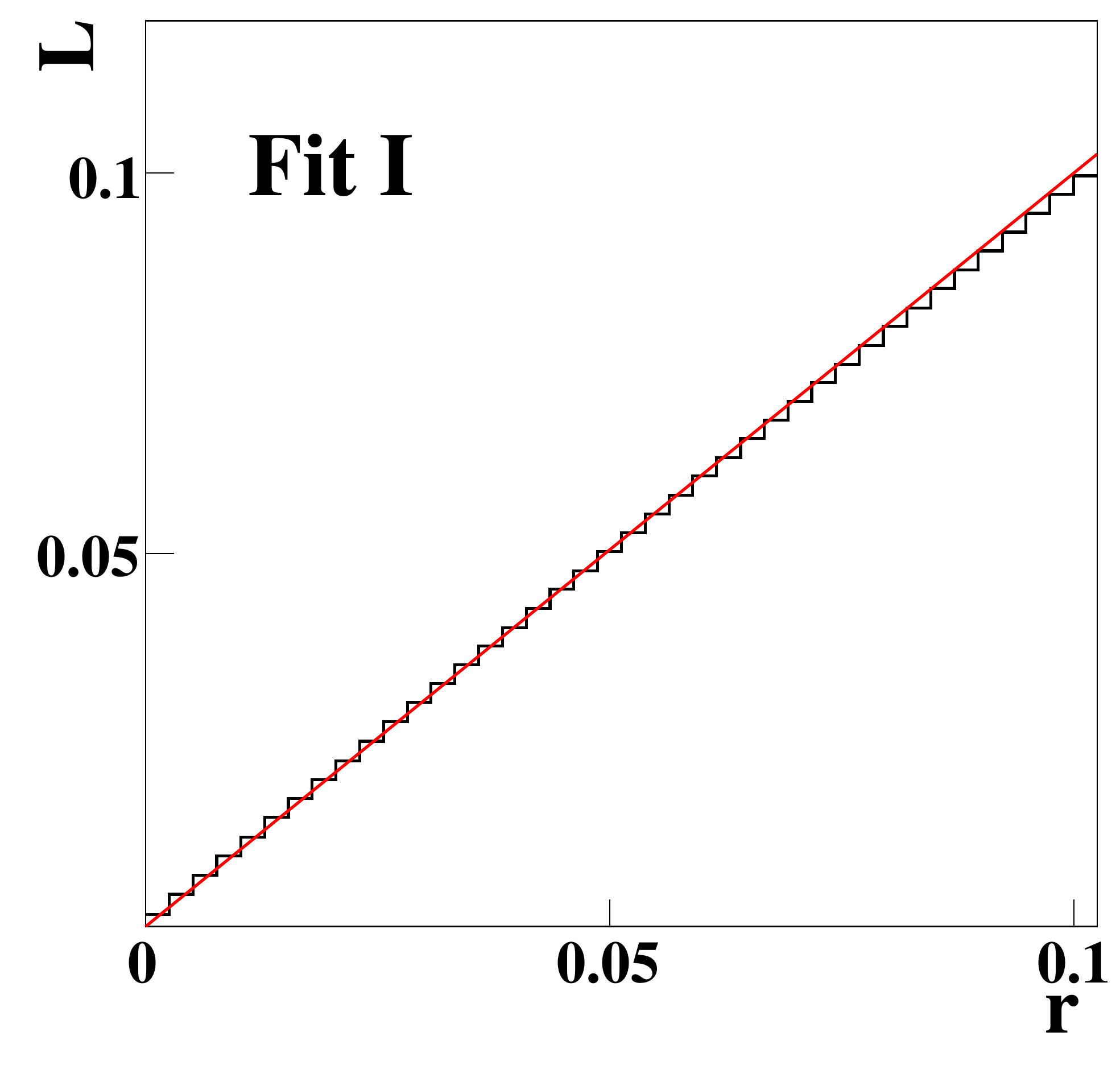} 
  }
  \\
  \subfigure[]{
    \includegraphics[width=0.4\textwidth]{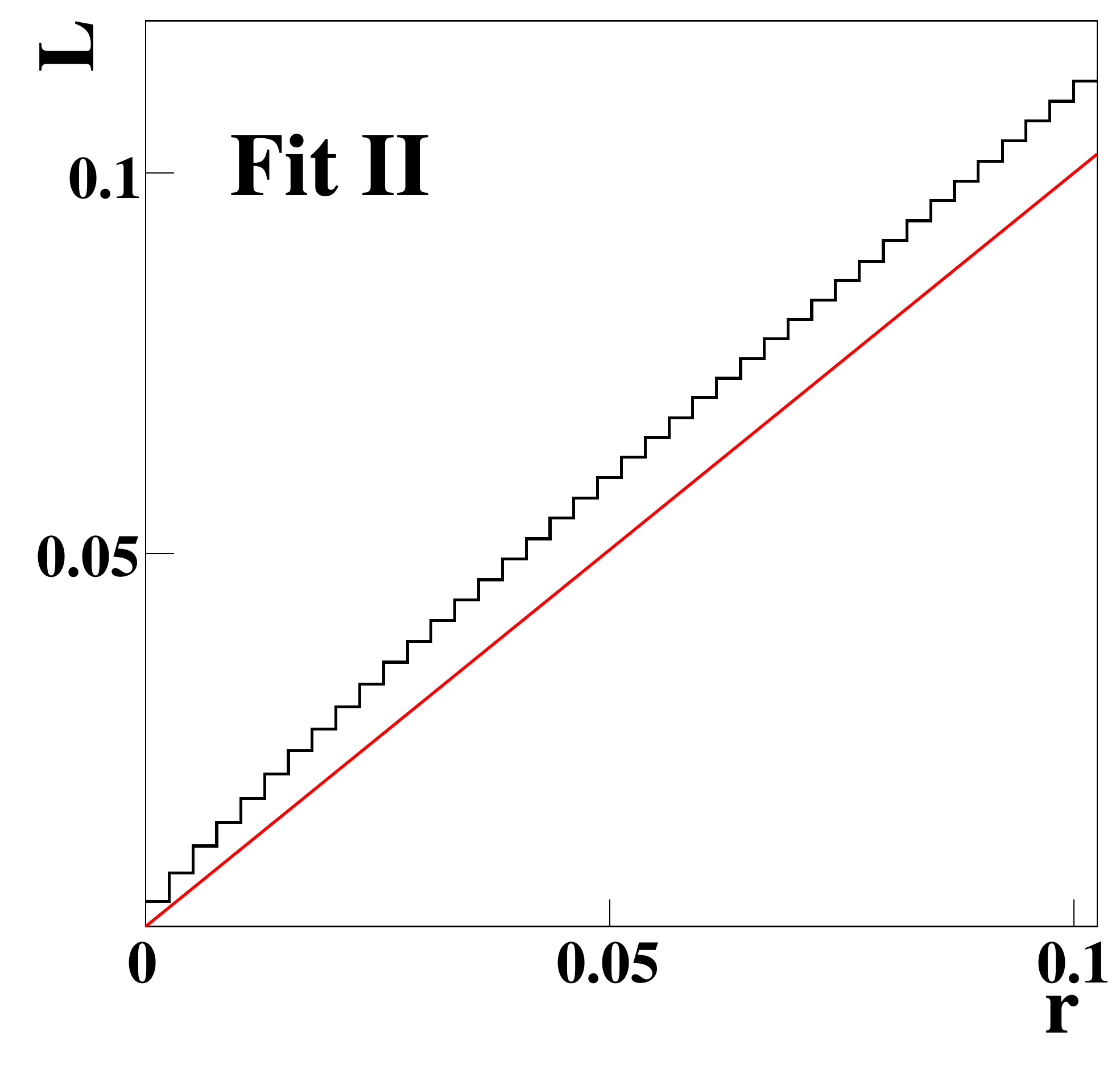} 
  }
  \subfigure[]{
    \includegraphics[width=0.4\textwidth]{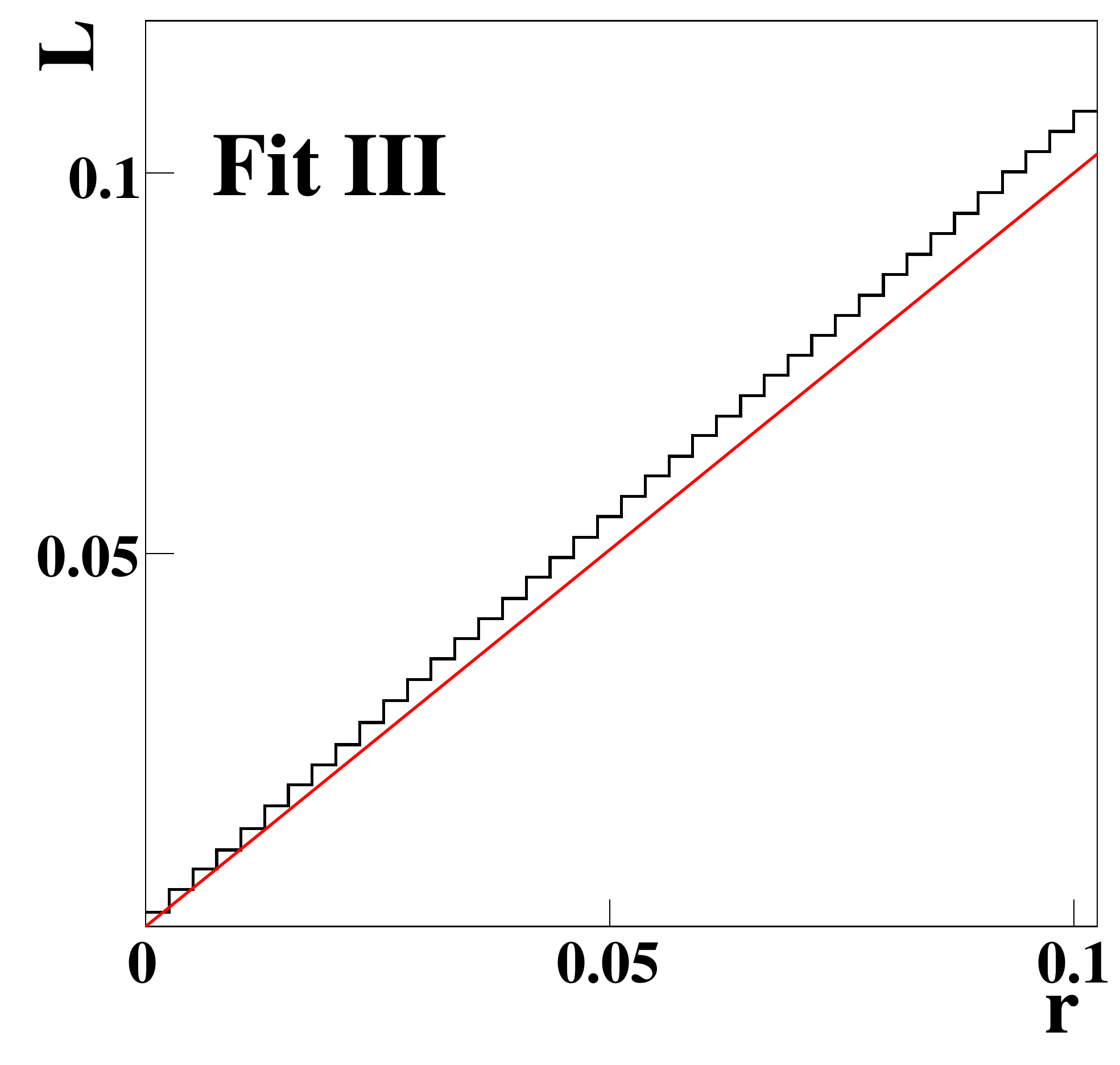} 
  }
  \caption[]{\label{fig:baddeley-example}
    Distributions for the $L(r)$ statistic of Ref.~\cite{ref:baddeley} for a single high statistics ($n_d = 10000$) data set for the following p.d.f.'s: (a) Model; (b) Fit~I; (c) Fit~II; (d) Fit~III.  The (solid red) line shows the expected ($L(r)=r$) distribution.  
  }
\end{figure*}

For a Dalitz-plot analysis $V_D$ is the total area of the Dalitz plot and $v(i,j)$ is the fraction of the circle centered at $\vec{x}_i$ with radius $|\vec{x}_i - \vec{x}_j|$ that is inside the kinematically allowed region of the Dalitz plot.  Because of the fact that a  Dalitz-plot analysis is two-dimensional, the quantity $L(r)$ (defined in the same way as for the inhomogeneous case) can be used. Fig.~\ref{fig:baddeley-example} shows the $L(r)$ distributions for each of the p.d.f.'s examined in this study along with the expected (linear) distribution.  The $L$ functions obtained using the Model and Fit~I p.d.f.'s are in excellent agreement with the expected result.  Notice that for large values of $r$ the Fit~I $L$ function dips below the $L(r) = r$ line.  Recall that larger values of $L$ indicate a discrepancy between the fit and parent p.d.f.'s; thus, this dip is not evidence of a discrepancy in the fit p.d.f.  It is actually evidence of a small test bias.  The fact that the test bias increases with increasing $r$ is expected.  For large values of $r$, large regions of phase space are used to collect each event's neighbors resulting in a much coarser comparison between the fit p.d.f.\ and the data.  This is not a pathology; it simply means that the $K$ and $L$ statistics become less meaningful for large values of $r$ (analogous to a histogram with only a few large bins).  

Figs.~\ref{fig:baddeley-example}(c) and (d) also show the $L(r)$ distributions for the Fit~II and Fit~III p.d.f.'s, respectively. Both have $L(r) > r$ for all $r$ values considered.  To obtain a g.o.f.\ value, the significance of these deviations from the expected distribution needs to be quantified.  It is important to realize that the values $L(r_1)$ and $L(r_2)$ are not independent measurements.  If $r_1 < r_2$, then all of the weighted events used to obtain $L(r_1)$ are also used to obtain $L(r_2)$.  Because of this one cannot use, {\em e.g.}, a $\chi^2$ test to determine the significance of any deviations of $L(r)$ from $L(r)=r$.

Ref.~\cite{ref:ripley} suggests a procedure that requires sampling an ensemble of Monte Carlo data sets from the test p.d.f.\ ($f_0$ in this case) each with $n_{mc} = n_d$.  For the data and each Monte Carlo data set the maximum deviation from the expected distribution,
\begin{equation}
  T = \left(L(r) - r\right)_{\rm max},
\end{equation}
is then calculated.  Recall that for a Poisson process, larger values of $T$ correspond to a lesser level of agreement between the fit and parent p.d.f.'s; thus, a one-sided cut on $T$ is employed.  The fraction of the Monte Carlo data sets whose $T$ value is larger than that of the data is then used as the $p$-value.  There are still two nuisance parameters that need to be determined: the step size in $r$ and the maximum value of $r$.  Neither of these quantities appears to be very important. As discussed above, the test bias (which drives $T$ downwards) increases with increasing $r$; thus, it is unlikely that the value used for $T$ will come from a very large $r$ value.  I chose $r_{\rm max}$ such that a circle with radius $r_{\rm max}$ contained (on average) about 10\% of the events.  The step size also determines the minimum value of $r$ at which events are collected.  This simply needs to be chosen to be large enough such that some events do contribute to $K$ or $L$ for $r_{\rm min}$.  I note here that if Monte Carlo generation is expensive, than some form of data-driven method for determining the significance of $T$ could be used instead.

\begin{figure*}
  \centering
  \subfigure[]{
    \includegraphics[width=0.4\textwidth]{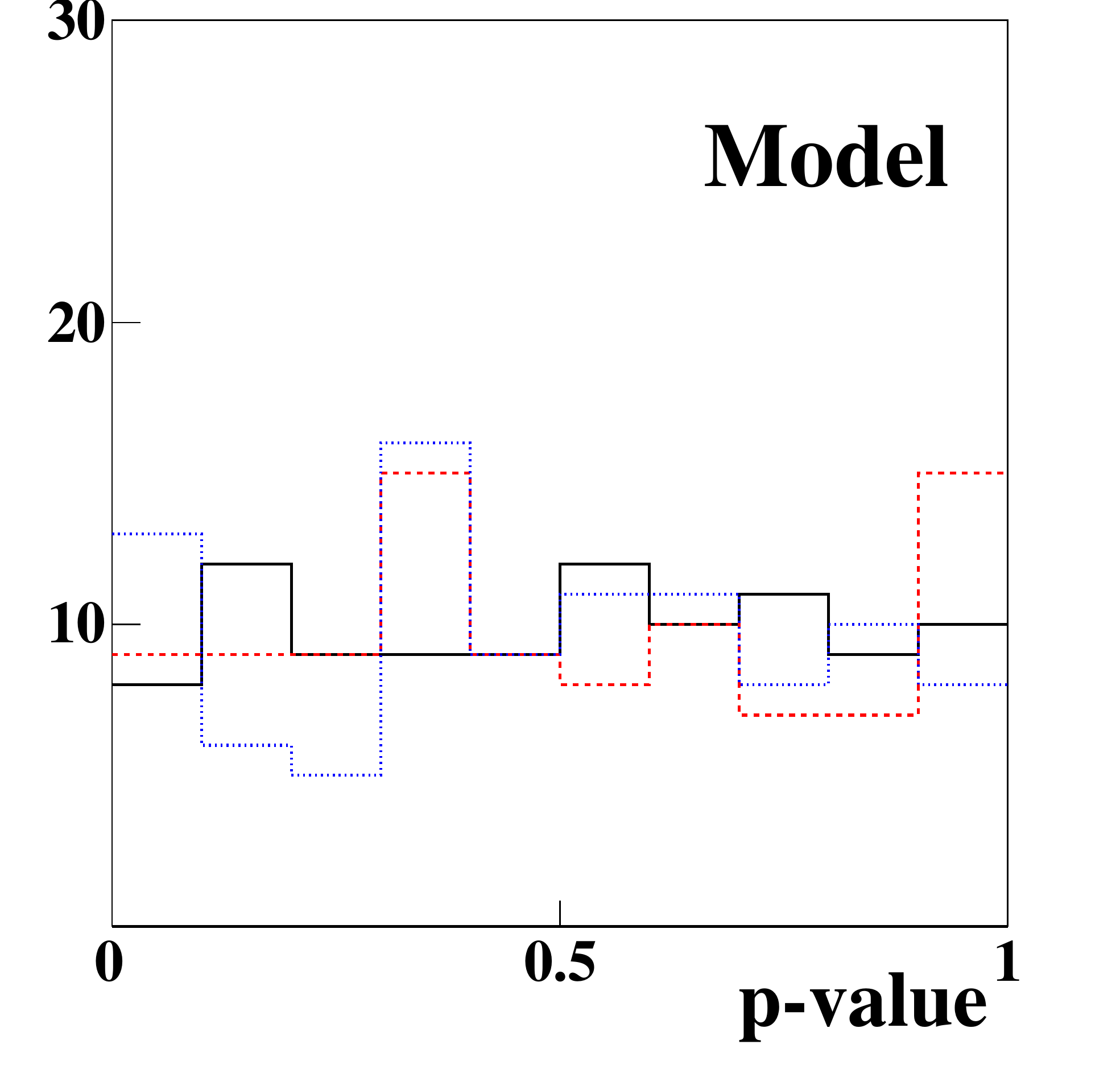}
  }
  \subfigure[]{
    \includegraphics[width=0.4\textwidth]{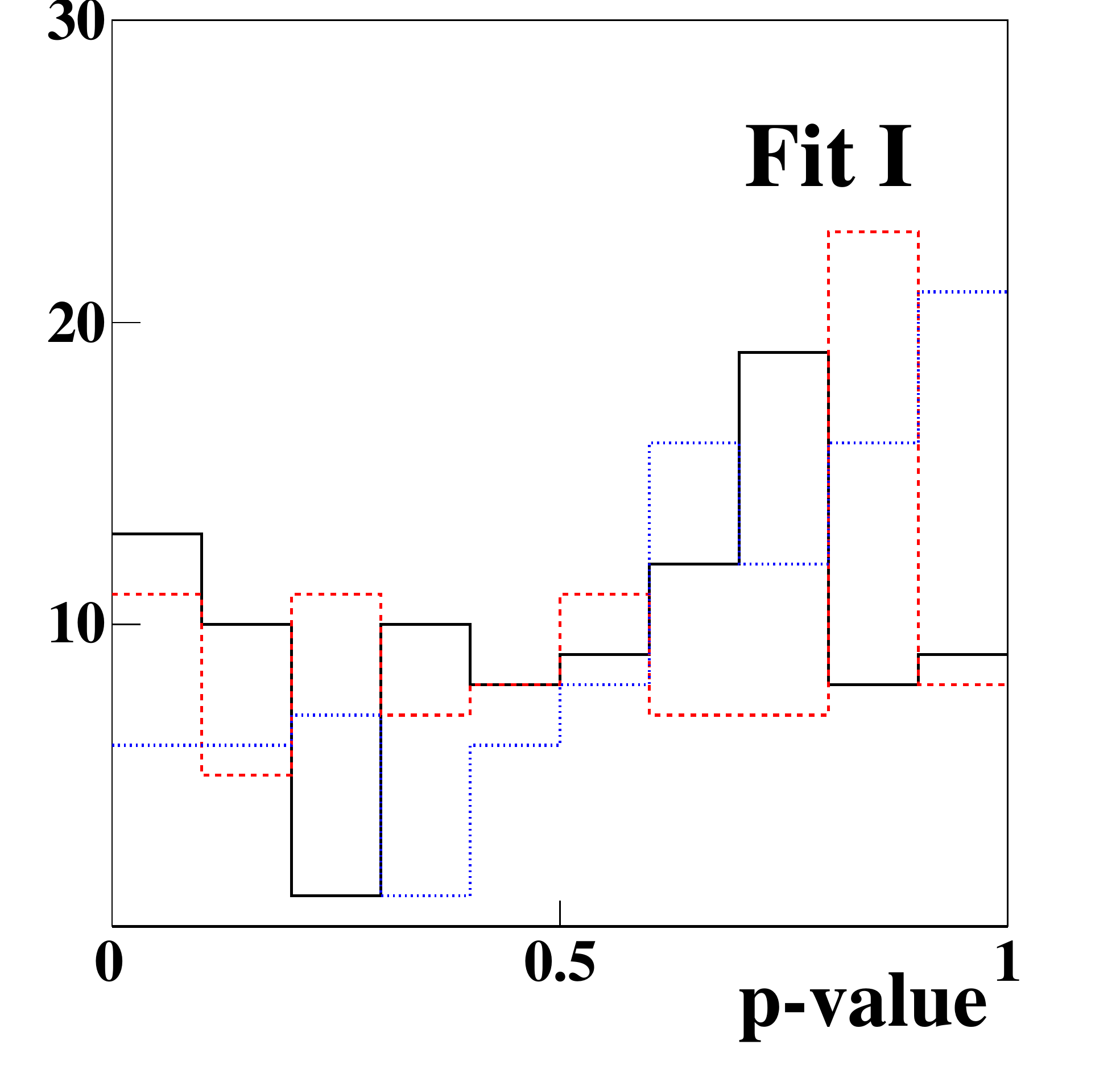} 
  }
  \\
  \subfigure[]{
    \includegraphics[width=0.4\textwidth]{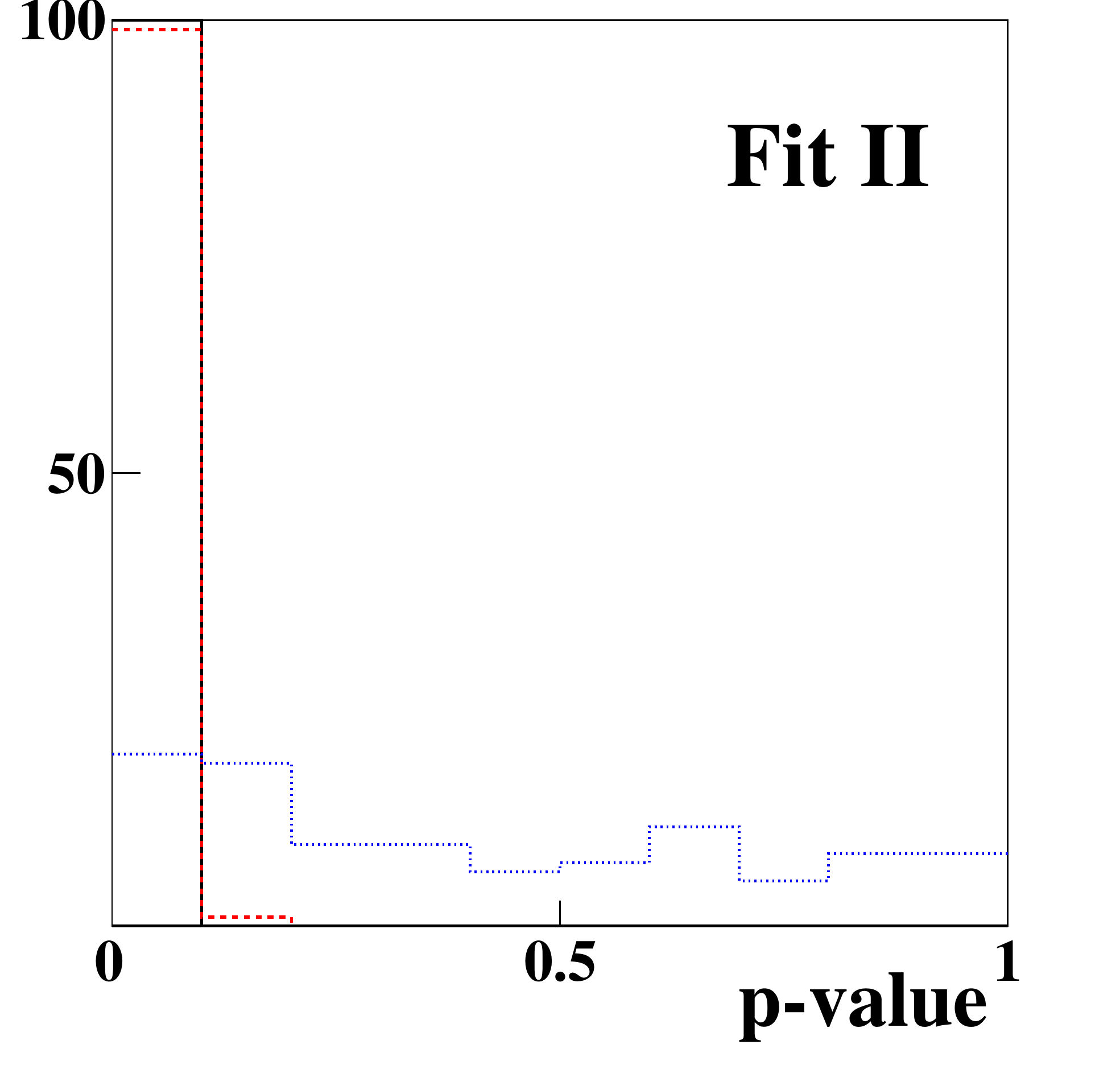} 
  }
  \subfigure[]{
    \includegraphics[width=0.4\textwidth]{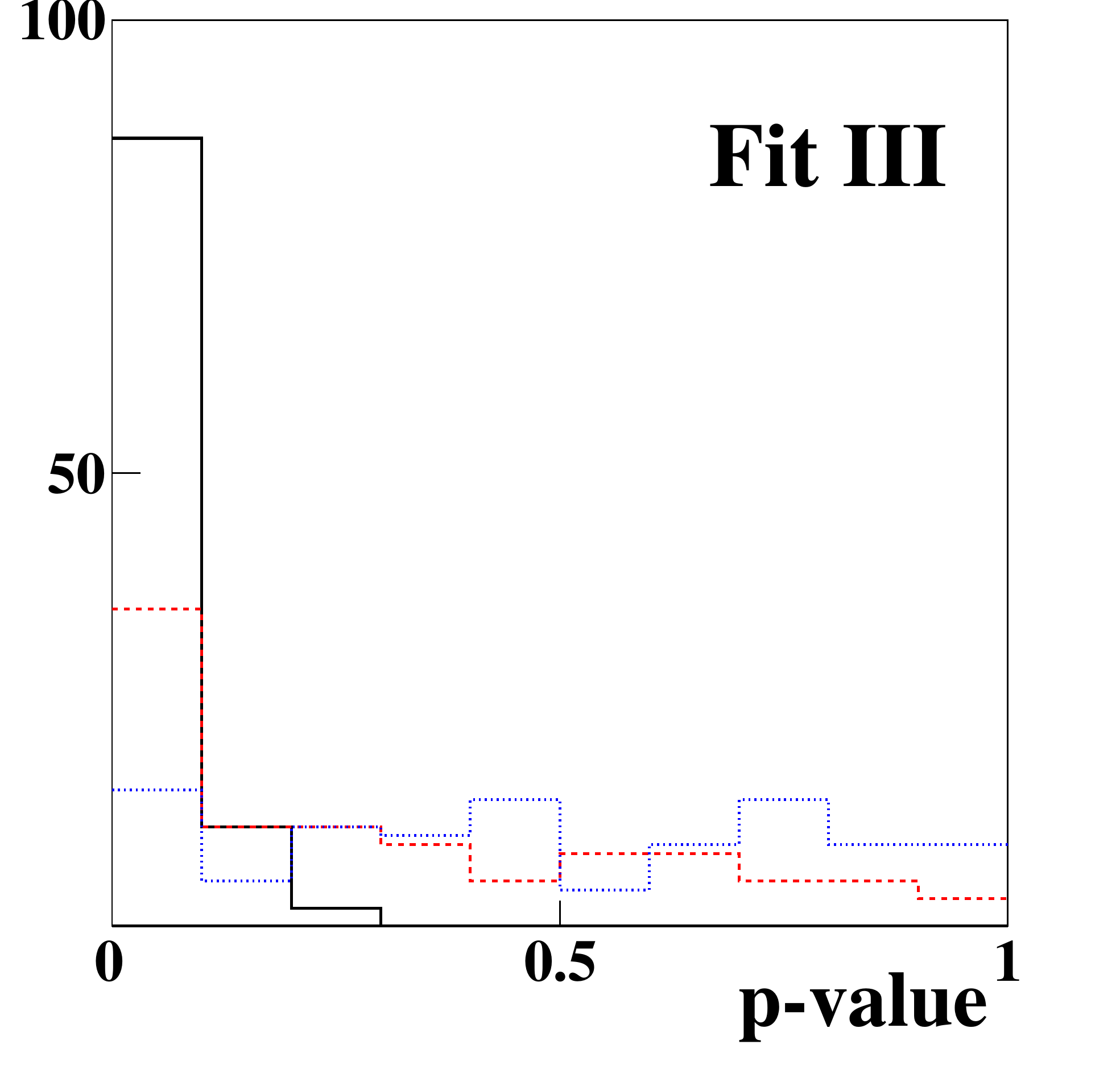} 
  }
  \caption[]{\label{fig:baddeley-results}
    (Color Online) $p$-value distributions obtained from low (blue dotted), medium (red dashed) and high (solid black histograms) statistics data sets from the following p.d.f.'s using the local-density g.o.f.\ method of Ref.~\cite{ref:baddeley}: (a) Model; (b) Fit~I; (c) Fit~II; (d) Fit~III.  Data sets whose $p$-values are less than 0.05 are rejected at 95\% confidence level (by definition).  See Section~3.5 for further discussion on these results.
  }
\end{figure*}

Fig.~\ref{fig:baddeley-results}(a) shows the $p$-value distribution obtained for the Model p.d.f. The distribution is in good agreement with the expected (uniform) one.  Fig.~\ref{fig:baddeley-results}(b) shows the $p$-value distribution obtained for the Fit~I p.d.f.  The agreement with the expected distribution is very good; however, there is a small test bias for $n_d \leq 1000$.  This is, again, expected and is small enough to safely be ignored. Figs.~\ref{fig:baddeley-results}(c) and (d) show the $p$-value distributions obtained for the Fit~II and Fit~III p.d.f.'s, while the rejection power at 95\% condidence level for all four p.d.f.'s is given in Table~\ref{table:baddeley-power}.  The rejection power for Fit~II is excellent for $n_d = 10000$, very good for $n_d = 1000$ and poor for $n_d = 100$.  For Fit~III, the rejection power is good for $n_d = 10000$, fair for $n_d = 1000$ and poor for $n_d = 100$.  Overall, these results are impressive; this method is much more powerful than the binned $\chi^2$ method.  The poor performance at $n_d = 100$ is not surprising since this method relies on using each event's neighbors to obtain an estimate of the local density.  For very low statistics data, this density estimation is difficult due to the small number of neighbor events contained within each hypersphere (or circle for the Dalitz-plot analysis).

\begin{table}[]
  \begin{center}
    \begin{tabular}{|c|c|ccc|}
      \hline
      $n_d$ & Model & Fit I & Fit II & Fit III \\
      \hline
      10000 & 5\% & 2\% & 100\% & 71\% \\
      1000 & 6\% & 3\% & 84\% & 18\% \\
      100 & 7\% & 1\% & 1\% & 3\% \\
      \hline
    \end{tabular}
    \\
    \vspace{0.01\textheight}
    \caption[]{\label{table:baddeley-power}  
      Rejection power at 95\% confidence level of the local-density method of Ref.~\cite{ref:baddeley}.
    }
  \end{center}   
\end{table}

This method has excellent rejection power for large localized discrepancies and good rejection power for small omnipresent ones (excluding low statistics data sets).  It is also fairly easy to understand conceptually.  Determining the $p$-values requires a fair amount of processing time (regardless of whether an ensemble of Monte Carlo data sets or a data-driven method is used); however, even without calculating the $p$-values, the method can still be useful.  Using just the data and no Monte Carlo one can produce the $K$ or $L$ distribution.  About 50\% of my toy-model data sets have $L(r) < r$ $\forall~r$ for Fit~I.  This is expected given that the $L$ values are highly correlated and that the test bias increases with increasing $r$. Thus, if one is fitting data with the true parent p.d.f.\ (with some free parameters), then there is about a 50\% chance that $L(r) < r$ $\forall~r$ at which point one can say that $p \gtrsim 0.5$.  Of course, one should be suspicious if there appears to be a large test bias, {\em i.e.}, a large downwards turn in the $K$ or $L$ distribution.  In this way, one can obtain a quick estimate of the g.o.f.\ using this method.  The $K$ or $L$ distribution plot would be a useful addition to any publication.  One can also include the 95\% confidence-level band on the plot for reference (see Ref.~\cite{ref:ripley} for examples).

\subsection{Kernel-Based Methods}

In Section~3.3 I noted that the integral of the quadratic difference between $f$ and $f_0$,
\begin{equation}
  T = \frac{1}{2}\int \left( f(\vec{x}) - f_0(\vec{x}) \right)^2 d\vec{x},
\end{equation}
could be used as a measure of g.o.f.\ if the parent p.d.f.\ of the data were known.  Since $f$ is not known, $T$ cannot be calculated; however, if $f$ can be approximated, then $T$ can also be approximated.  This is the approach taken by kernel-based g.o.f.\ methods.

A probability density estimate (p.d.e.) can be obtained through the use of a kernel function defined as follows for $D$ dimensions:
\begin{equation}
  \label{eq:kernel-def}
  f_{n_d}(\vec{x}) = \frac{1}{n_d b(n_d)^D}\sum\limits_{i=1}^{n_d} w\left(\frac{|\vec{x} - \vec{x}_i|}{b(n_d)}\right),
\end{equation}
where $b(n_d)$ is the {\em bandwidth} and $w$ is a weighting function.  If the ranges and standard deviations of the variates are not similar, then a different bandwidth for each variate should be used.  For the toy-model Dalitz-plot analysis performed in this paper, the ranges of the variates are equal and the standard deviations are similar; thus, I will use the same bandwidth for $m_{ab}^2$ and $m_{ac}^2$.

\begin{figure*}
  \centering
  \subfigure[]{
    \includegraphics[width=0.3\textwidth]{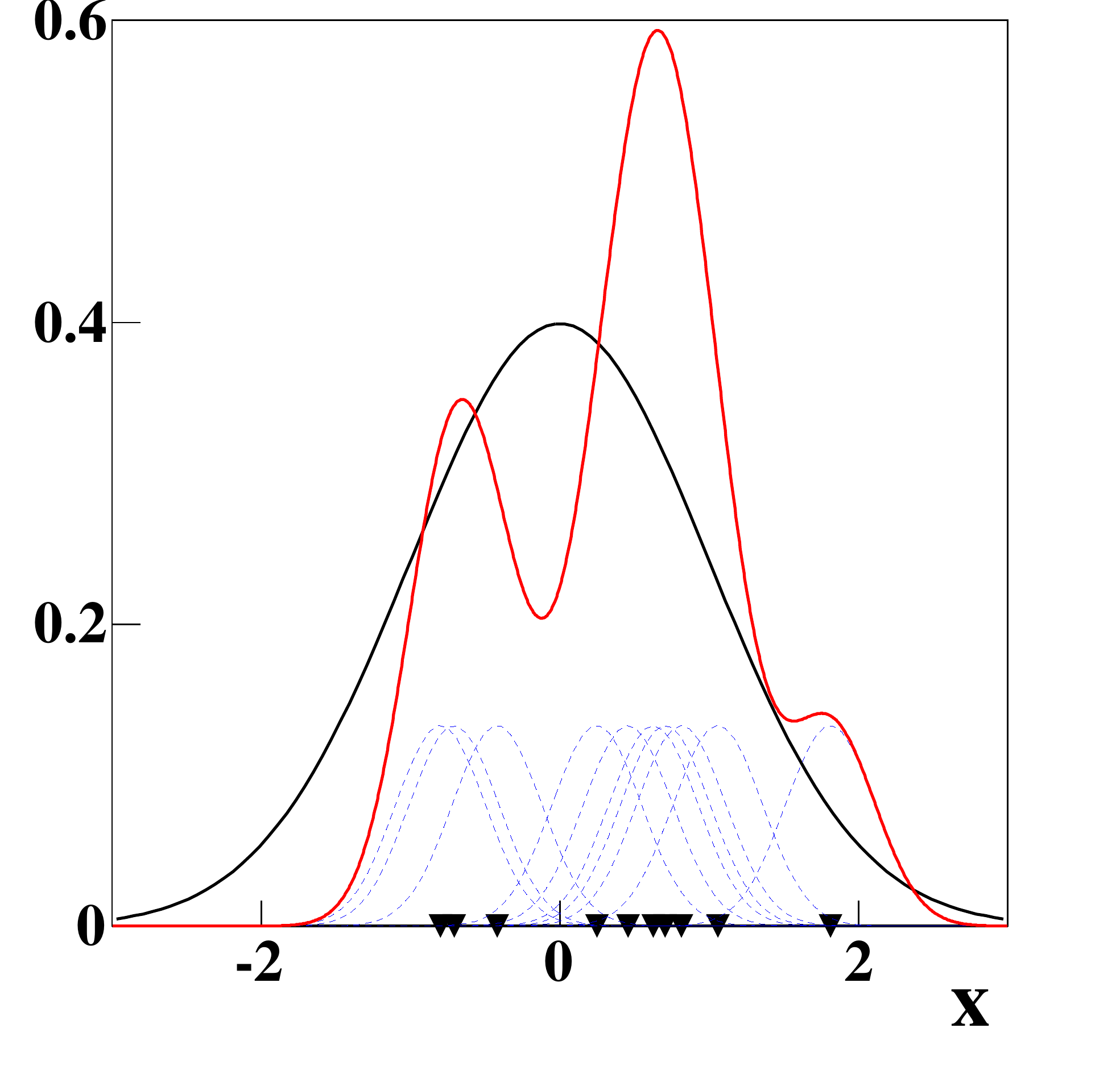}
  }
  \subfigure[]{
    \includegraphics[width=0.3\textwidth]{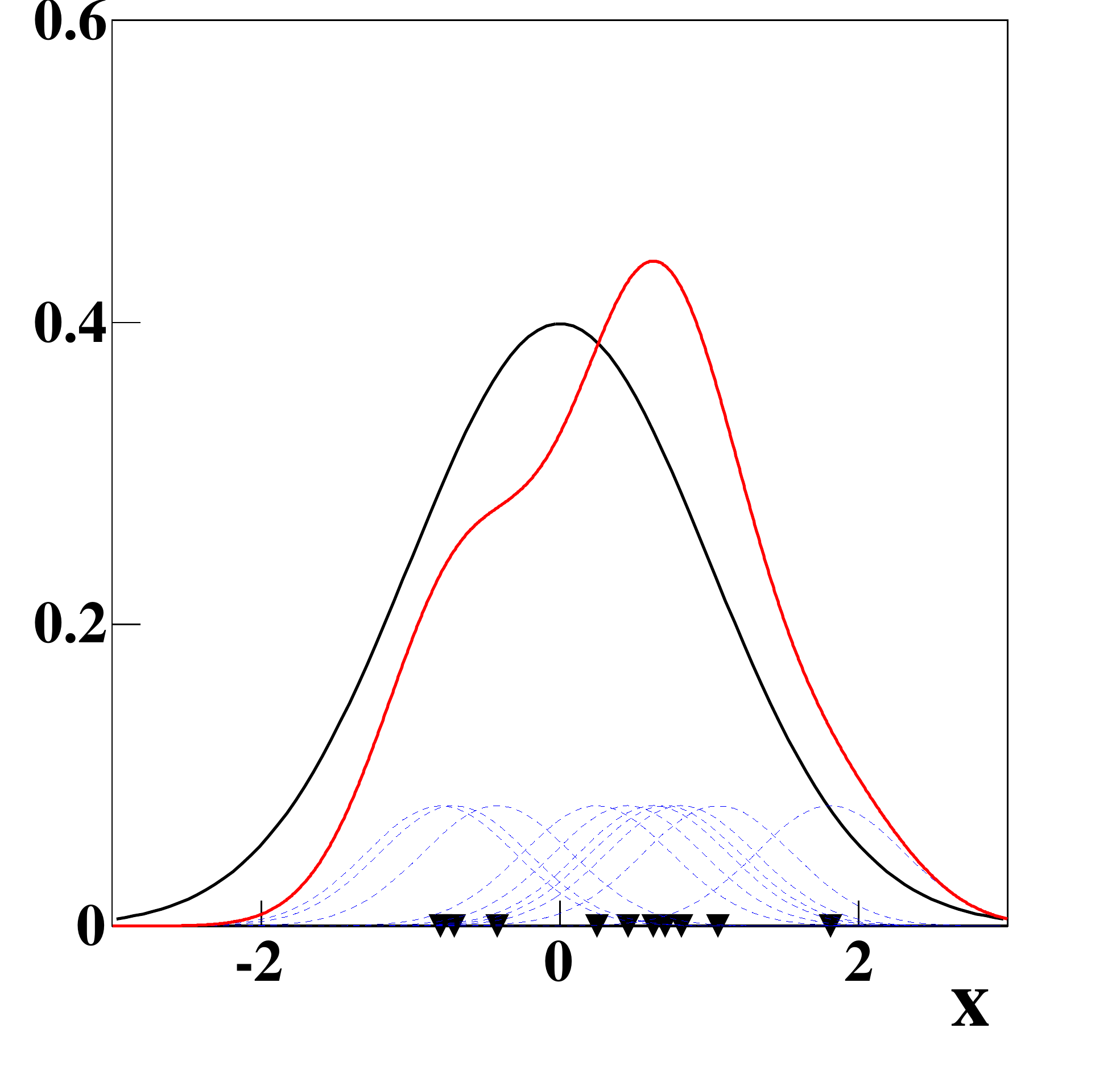} 
  }
  \subfigure[]{
    \includegraphics[width=0.3\textwidth]{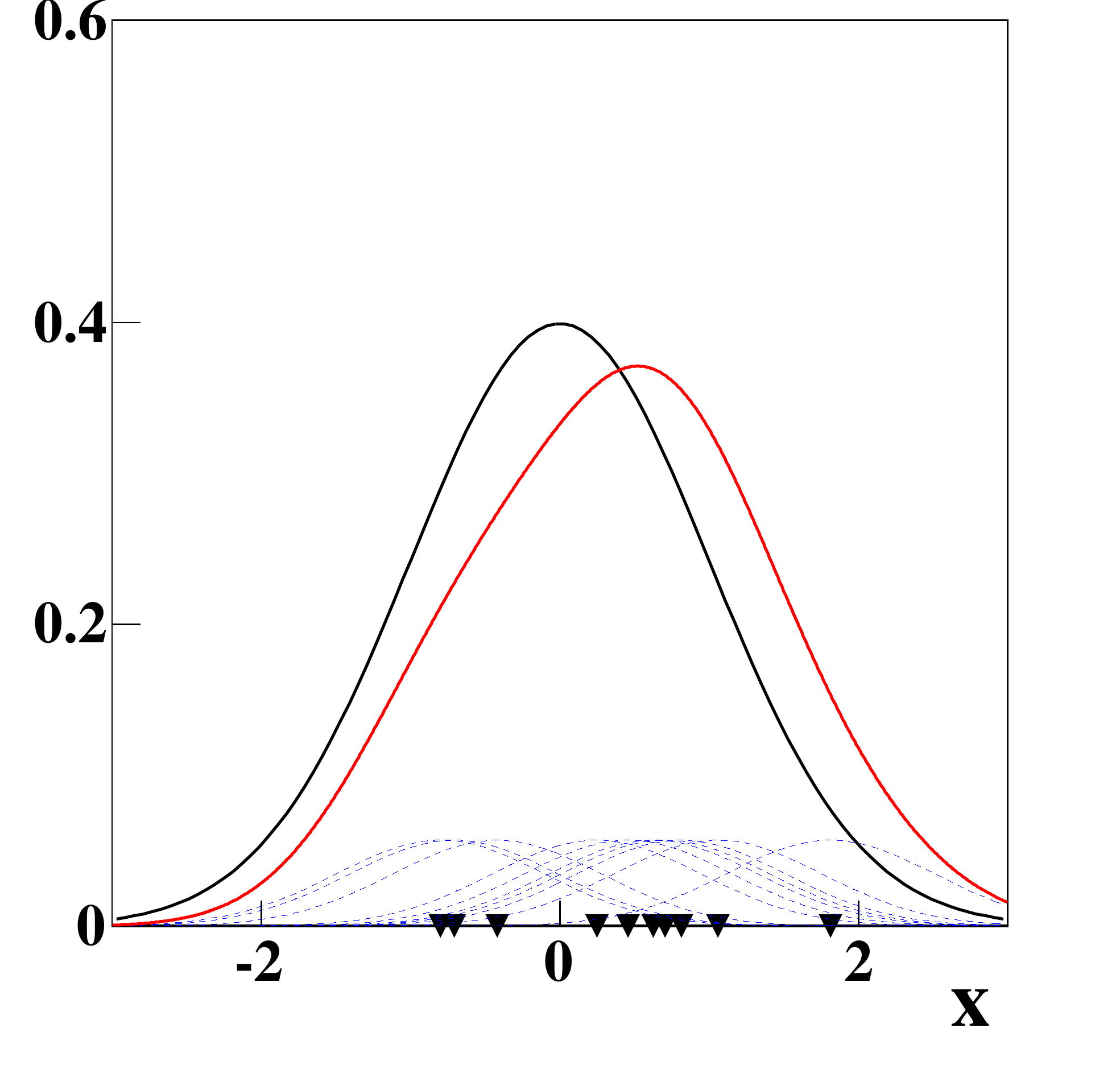} 
  }
  \caption[]{\label{fig:kernel-example}
    (Color Online) Example kernel-based p.d.e.'s for the p.d.f.\ $f(x) \propto e^{-x^2/2}$ (solid black lines) obtained for a very small data set ($n_d = 10$).  The location of the data points (sampled randomly from $f(x)$) is indicated on each plot (black triangles along the $x$-axis). The kernels are obtained using Eq.~\ref{eq:kernel-def} with weighting function $w(x) \propto e^{-x^2/2}$ and using the following bandwidths: (a) $b(n_d) = 0.3$; (b) $b(n_d) = 0.5$; (c) $b(n_d) = 0.7$.  The weighting function at each data point is shown on the plots (blue, dashed lines).  The p.d.e.\ (red, solid lines) is formed by summing the values of the weighting functions at each value of $x$.
  }
\end{figure*}

A simple example p.d.e.\ is shown in Fig.~\ref{fig:kernel-example} for illustrative purposes.  The parent p.d.f.\ to be approximated is $f(x) \propto e^{-x^2/2}$.  A very small data set ($n_d = 10$) is randomly sampled from this p.d.f.\ (the extremely small sample size was chosen so the construction of the p.d.e.\ could be illustrated on the plot).  The first step in kernel-based p.d.e.\ construction is choosing a weighting function.  In principle this could take on just about any functional form (see Ref.~\cite{ref:bickel-1973} for the limited list of restrictions); however, in practice it is typically chosen to be either a Gaussian line shape or uniform with a cutoff window.  Fig.~\ref{fig:kernel-example} shows the p.d.e.'s obtained using a standard normal Gaussian weighting function for three different bandwidths. The quality of the p.d.e.\ is highly dependent on the value chosen for the bandwidth (discussion on how to choose the bandwidth is given below).   

Once the p.d.e.\ $f_{n_d}(\vec{x})$ has been constructed, then the g.o.f.\ can be obtained by examining the statistic~\cite{ref:fan}
\begin{equation}
  \label{eq:bickel-kernel-T}
  T = n_d b(n_d)^{D/2} \int (f_{n_d}(\vec{x}) - f_0(\vec{x}))^2 d\vec{x}.
\end{equation}
Ref.~\cite{ref:gourieroux} suggests replacing  $f_0(\vec{x})$ by $f_{0n_d}(\vec{x})$ (the expectation value of the p.d.e.\ at $\vec{x}$) to remove the bias that arises from using $f_0(\vec{x})$. {\em I.e.}, there is no guarantee that the kernel-based p.d.e.\ is not a biased estimate of the true p.d.f.\ (especially near the edges of the allowed data region); thus, it is better to use the p.d.e.\ of $f_0$ instead of $f_0$ itself.  The test statistic defined in Eq.~\ref{eq:bickel-kernel-T} has an expected mean of $\mu_T = b(n_d)^{-D/2}\int w^2(z)dz$ and an expected variance of
\begin{equation}
  \sigma_T^2 = 2 \int \left( \int w(y+z)w(z)dz\right)^2 dy \int f^2_0(\vec{x})d\vec{x}.
\end{equation}
Unfortunately, the theoretical  mean and variance values given above are often not accurate for finite sample sizes; thus, the $p$-value obtained using them is not reliable~\cite{ref:fan}.  I found this to be true in my analysis.  The quality of the $\mu_T$ and $\sigma_T$ values given above varied drastically as a function of $b(n_d)$. Given that the value of $b(n_d)$ must be chosen (somewhat arbitrarily) by the experimenter, this is a disastrous result.  

Since one cannot trust the $p$-values obtained using the theoretical (limiting) $T$ distribution, some form of data-driven  method must be used to calculate the $p$-value.  I have chosen to use the permutation test used in Section~3.3.  I have not explored whether some other re-sampling method ({\em e.g.}, bootstrapping, jackknifing, {\em etc.}; see Ref.~\cite{ref:efron}) would perform better.  The application of the permutation test for this method is identical to the point-to-point dissimilarity method of Ref.~\cite{ref:aslan1}.  A Monte Carlo data set is sampled from $f_0$ and used, along with the data, to calculate $T$ from Eq.~\ref{eq:bickel-kernel-T}.  A set of random permutations of the labels ``data'' and ``Monte Carlo'' (keeping $n_d$ and $n_{mc}$ fixed) are then used to estimate the distribution of $T$ and, in turn, the $p$-value.  A detailed discussion on this technique is provided in Appendix~C.

The results presented below were obtained using a normal Gaussian weighting function.  I also tried using a uniform weighting function with a cutoff window; this had little effect on the results.  I found that the choice of bandwidth is much more important than the choice of weighting function.  There are a number of data-driven methods found in the statistical literature for determining the optimal bandwidth.  The most common is to use the value of $b(n_d)$ that minimizes the mean integrated squared error (m.i.s.e.)~\cite{ref:mise},
\begin{equation}
  \label{eq:mise}
  \left< \int (f(\vec{x}) - f_{n_d}(\vec{x}))^2 d\vec{x} \right>.
\end{equation}
For the Dalitz-plot p.d.f.\ studied here, the values that minimize m.i.s.e.\ are $b(n_d=10000) = 0.02$, $b(n_d=1000) = 0.025$ and $b(n_d=100) = 0.05$.  

\begin{figure*}
  \centering
  \subfigure[]{
    \includegraphics[width=0.4\textwidth]{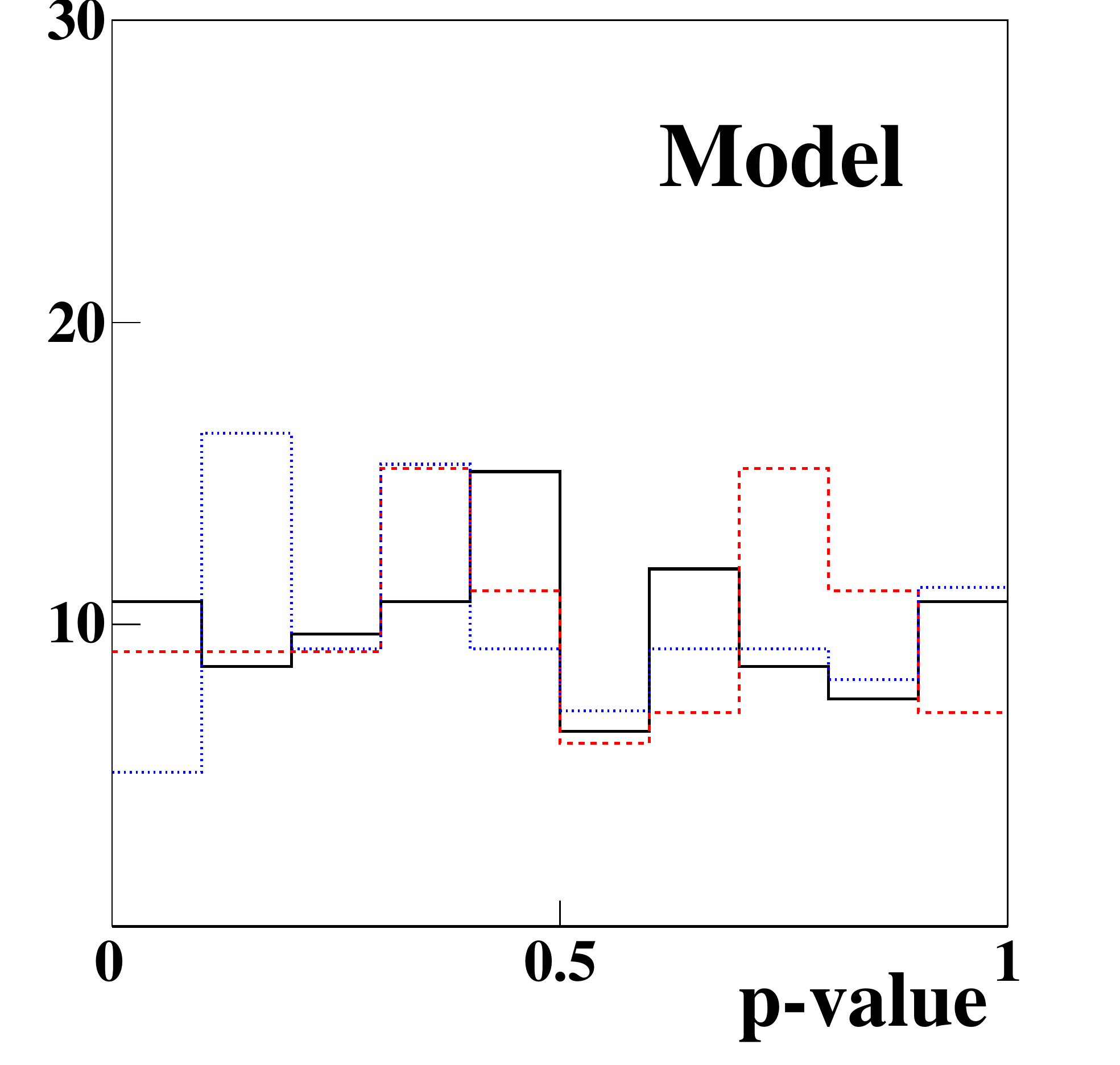}
  }
  \subfigure[]{
    \includegraphics[width=0.4\textwidth]{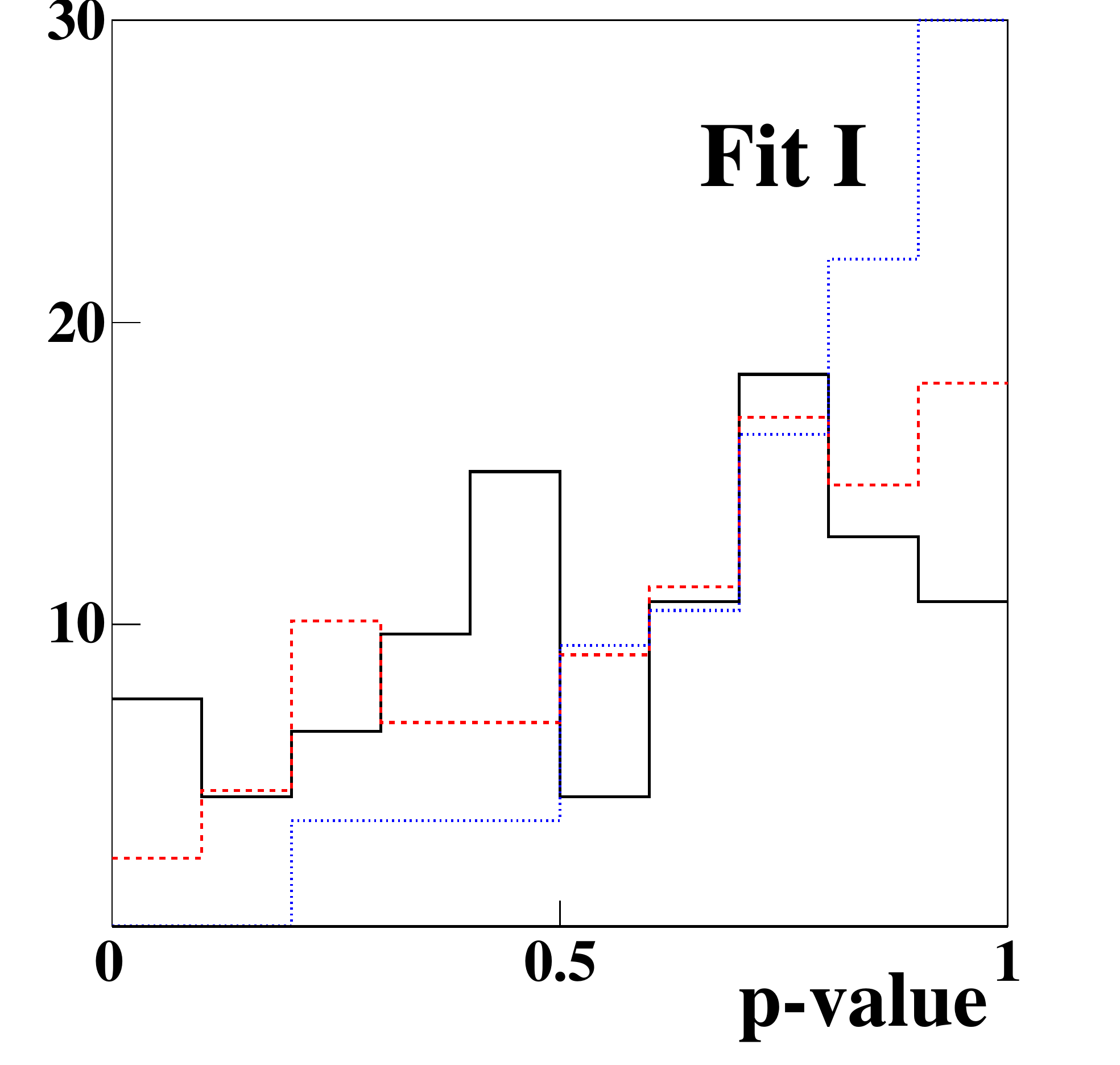} 
  }
  \\
  \subfigure[]{
    \includegraphics[width=0.4\textwidth]{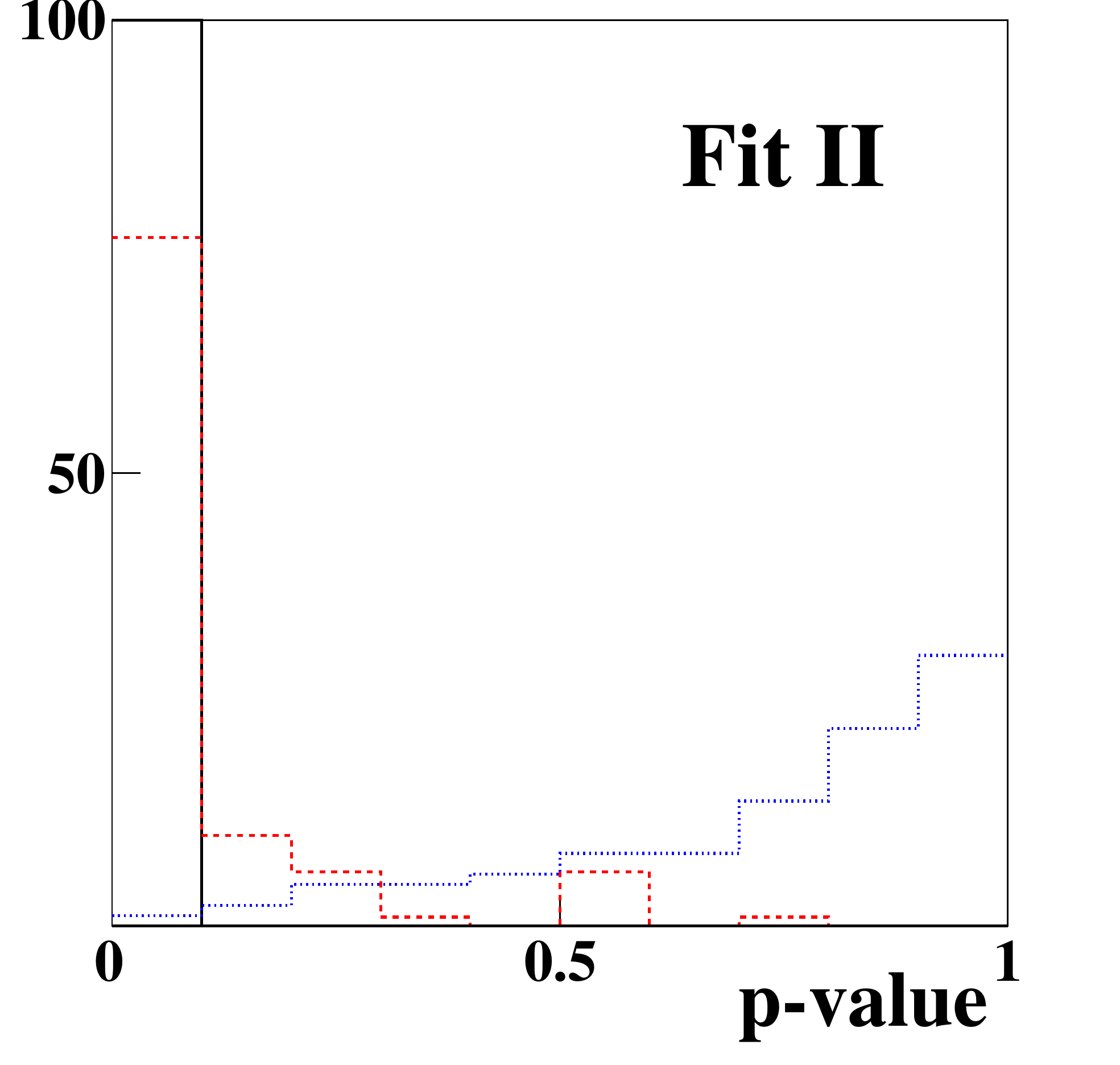} 
  }
  \subfigure[]{
    \includegraphics[width=0.4\textwidth]{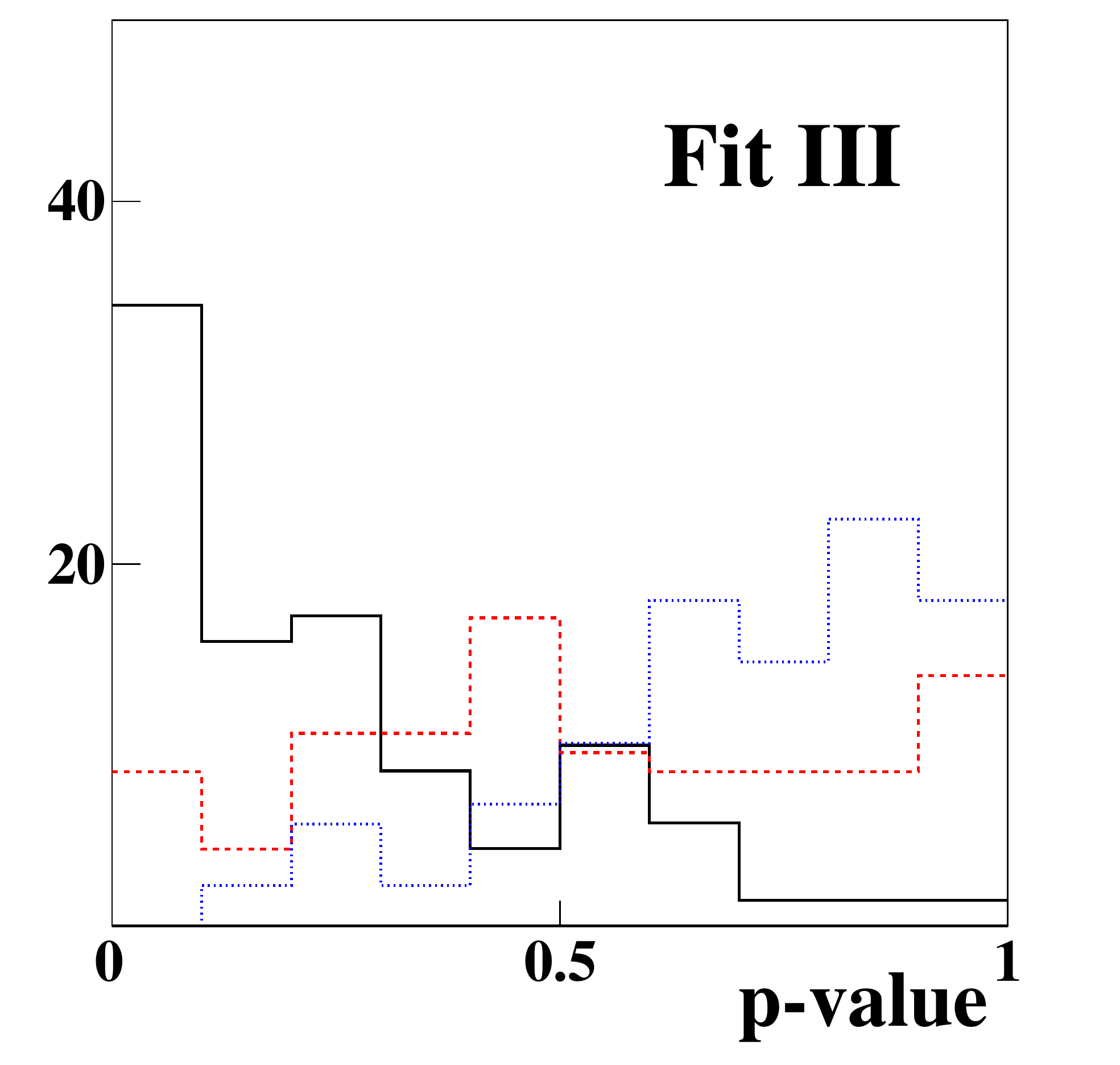} 
  }
  \caption[]{\label{fig:kernel-results}
    (Color Online) $p$-value distributions obtained from low (blue dotted), medium (red dashed) and high (solid black histograms) statistics data sets for the following p.d.f.'s using the kernel-based g.o.f.\ method of Ref.~\cite{ref:fan}: (a) Model; (b) Fit~I; (c) Fit~II; (d) Fit~III.  Data sets whose $p$-values are less than 0.05 are rejected at 95\% confidence level (by definition).  See Section~3.6 for further discussion on these results.
  }
\end{figure*}

Fig.~\ref{fig:kernel-results} shows the $p$-value distributions obtained using the bandwidths that minimize m.i.s.e., while the rejection power at 95\% confidence level for each p.d.f.\ is given in Table~\ref{table:kernel-power}.  
The $p$-value distribution obtained for the Model p.d.f.\ is in good agreement with the expected (uniform) one.  This validates the use of the permutation test for this method.  The test bias obtained for Fit~I is negligible for $n_d \geq 1000$; however, it is sizable for $n_d = 100$ and renders the use of this method (for this bandwidth) invalid for this sample size. The same reasoning used in the previous section to argue that the test bias increases with increasing hyper-spherical radius applies here as well.  As the bandwidth increases, a larger fraction of the data set contributes to the p.d.e.\ at each value of $\vec{x}$.  While this does improve the quality of the p.d.e., it also results in an increased test bias. 
For $n_d \geq 1000$, the rejection powers for Fit~II and Fit~III are comparable to the binned $\chi^2$ test, but they are much lower than the unbinned methods described in Sections~3.3 and 3.5.
The method presented in this section requires all of the overhead of the point-to-point dissimilarity method of Ref.~\cite{ref:aslan1}; however, it is not as powerful or reliable. It is easy to understand conceptually, but this alone is not sufficient to recommend its use in a Dalitz-plot (or similar) analysis.


\begin{table}[h!]
  \begin{center}
    \begin{tabular}{|c|c|ccc|}
      \hline
      $n_d$ & Model & Fit I & Fit II & Fit III \\
      \hline
      10000 & 4\% & 5\% & 100\% & 23\% \\
      1000 & 3\% & 2\% & 58\% & 5\% \\
      100 & 3\% & 0\% & 0\% & 0\% \\
      \hline
    \end{tabular}
    \\
    \vspace{0.01\textheight}
    \caption[]{\label{table:kernel-power}  
      Rejection power at 95\% confidence level of the kernel-based method of Ref.~\cite{ref:fan}. 
    }
  \end{center}   
\end{table}

%% file: conc.tex
\section{Discussion}

In this paper I have studied the performance of a variety of unbinned multivariate g.o.f.\ tests when applied to a real-world high energy physics analysis (a Dalitz-plot analysis).  The vastness of the statistical literature on this topic makes it impossible to study all of the available tests.  Instead, I chose to categorize the tests based on the underlying concept used to determine the g.o.f.  In each of these categories one method was tested and the following results were obtained:

\begin{description}
\item[Mixed-Sample Methods] \hfill \\
The method presented in Ref.~\cite{ref:schil} is easy to use and conceptually it is easy to understand.  It is excellent at rejecting large localized discrepancies but fairly poor at rejecting small omnipresent ones.  The $p$-values can be calculated analytically.  This method would make a nice addition to the high energy physics g.o.f.\ toolkit.
\item[Point-to-Point Dissimilarity Methods] \hfill \\
The method presented in Refs.~\cite{ref:aslan1,ref:aslan2} has excellent rejection power for both large localized discrepancies and small omnipresent ones.  Determining the $p$-value requires re-sampling the data (using the permutation test) which uses a relatively large amount of processing time.  The method is not as easy to understand conceptually as some of the other methods tested in this paper.  These downsides are not enough to out-way its excellent performance; this is a very powerful g.o.f.\ tool.
\item[Distance to Nearest-Neighbor Methods] \hfill \\
The method presented in Ref.~\cite{ref:bickel-1983} is easy to use, requires very little processing time and is conceptually fairly easy to understand; however it is not very powerful.  The $U$ statistic it defines does provide a useful easy-to-visualize diagnostic tool (especially for very high dimensional analyses), but its quantitative usefulness as a g.o.f.\ test is limited.  
\item[Local-Density Methods] \hfill \\
The method presented in Ref.~\cite{ref:baddeley} has excellent rejection power for large localized discrepancies and good rejection power for small omnipresent ones. It is fairly easy to understand conceptually and provides a nice visual element in the $K$ and $L$ distributions.  Determining the $p$-values requires either generating an ensemble of Monte Carlo data sets or re-sampling the data; both require a large amount of processing time.  This is a very useful g.o.f.\ tool.
\item[Kernel-Based Methods] \hfill \\
The method presented in Ref.~\cite{ref:fan} requires all of the overhead of the point-to-point dissimilarity method of Refs.~\cite{ref:aslan1,ref:aslan2} but is nowhere near as powerful or reliable.  It is easy to understand conceptually; however, this is not sufficient justification to make it a useful high energy physics g.o.f.\ tool.

\end{description}

In Section~1 I noted that no g.o.f.\ test is the most powerful in all situations (even in one dimension). Thus, there certainly is not a universal unbinned multivariate g.o.f.\ road map suitable for all high energy physics analyses; however, that does not mean that some general guidance on how to apply the g.o.f.\ methods studied in this paper cannot be provided.  The following is an approximate road map for applying these g.o.f.\ methods to a high energy physics analysis:
\begin{itemize}
\item Start by plotting the $U$ distribution from the distance to nearest-neighbor method of Ref.~\cite{ref:bickel-1983}.  This is easy to do and requires very little processing time and no Monte Carlo data.  Any clear deviations from uniformity indicate that the fit p.d.f.\ is not in good agreement with the data.  This is especially useful for high-dimensional analyses where it can often be difficult to obtain even a qualitative comparison between the data and the fit p.d.f.  
\item Next plot the $K(r)$ or $L(r)$ distribution from the local-density method of Ref.~\cite{ref:baddeley}.  This also requires a small amount of processing time and no Monte Carlo data.  If the values are less than the expected ones ({\em e.g.}, if $L(r) < r$ $\forall~r$) then the $p$-value will be at least approximately 0.5.  Thus, one would accept the fit unless a large fit bias is suspected due to a pronounced downward turn in the $K$ or $L$ distribution.
\item Next, generate a Monte Carlo data set from the fit p.d.f.\ and obtain the $p$-values from the mixed-sample method of Ref.~\cite{ref:schil} and the point-to-point dissimilarity method of Refs.~\cite{ref:aslan1,ref:aslan2}.  This requires a relatively large amount of processing time; however, access to both of these $p$-values should be sufficient to accept or reject the test hypothesis.
\item Finally, generate an ensemble of Monte Carlo data sets and calculate the $p$-value using the local-density method of Ref.~\cite{ref:baddeley}.  At this point the significance bands can also be added to the $K$ or $L$ distribution plots (a nice addition if these are to be published).  If Monte Carlo generation is too expensive, then a re-sampling method can be used instead.
\end{itemize}
All of this information can then be used to either accept or reject the hypothesis that the fit p.d.f.\ is the parent p.d.f.\ of the data.  Exactly how this is done is analysis dependent.  Clearly, the ideal situation is that all of the tests either accept or reject this hypothesis making the conclusion obvious.  If, however, the results are mixed, then one needs to carefully examine how each of the g.o.f.\ methods applies to the specific p.d.f.\ being tested and attempt to resolve (or, at least, understand) the conflict.  The agreement between the g.o.f.\ methods for the Dalitz-plot analysis performed in this paper (following the road map above) was excellent. 

I conclude this section by pointing out that all of the results obtained in this paper are for a 2-D analysis.  As the number of dimensions increases, the power of any distance-based g.o.f.\ method (including the binned $\chi^2$ test) decreases due to the curse of dimensionality.  For very high-dimensional analyses, the unbinned methods recommended in this paper should still be more powerful than the binned $\chi^2$ method; however, it is uncertain how many dimensions they can handle before becoming ineffective.  Ref.~\cite{ref:aslan1} demonstrates that their method performs very well (for some simple p.d.f.'s) in 4-D, but none of the references cited in this paper goes beyond this level of dimensionality. Prior to using any of these methods in an analysis with $D > 4$, it would be advisable to first study their effectiveness in Monte Carlo.    

\section{Conclusions}

In conclusion, the statistical literature on unbinned multivariate g.o.f.\ tests is vast.  Rather than simply applying the $\chi^2$ test to every analysis or attempting to invent new unbinned multivariate g.o.f.\ tests, the high energy physics community would be better served to study the power and applicability of the g.o.f.\ methods published in the statistical literature.  Finally, it would be worthwhile to perform studies similar to this one for other types of high energy physics analyses.  This should be straightforward following the work presented in this paper.

%% file: appendix.tex
\appendix
\section{Goodness-of-Fit from Likelihood Values}

Many high energy physics analyses have attempted to use the m.l.v., $\mathcal{L}_{\rm max}$, as a measure of g.o.f. The method used involves the following steps: the data is fit to obtain $\mathcal{L}_{\rm max}$; the fit p.d.f.\ is used to generate an ensemble of Monte Carlo data sets; the g.o.f.\ is determined using $\mathcal{L}_{\rm max}$ from the data and the distribution of m.l.v.'s obtained from the Monte Carlo.  This method is not published anywhere (that I have been able to find) in the statistical literature.  It is fatally flawed and should not be used.  One can easily see this method is flawed by applying it to the hypothesis $f_0 = {\rm constant}$ (where the likelihood only depends on $n_d$); however, in this Appendix I will follow Ref.~\cite{ref:heinrich} and apply it to a more illustrative example.

Ref.~\cite{ref:heinrich} does an excellent job demonstrating the flaws in this method by applying it to the following simple one-dimensional p.d.f.:
\begin{equation}
  \label{eq:logL-ex-pdf}
  f(x) = \frac{1}{X}e^{-x/X},
\end{equation}
where $X$ is an unknown parameter to be estimated from a fit to the data.  The likelihood for a data set with $n_d$ events is given by
\begin{equation}
  -\log{\mathcal{L}} = \sum\limits_{i=1}^{n_d} \left(x_i/X + \log{X} \right).
\end{equation}
The m.l.v., which occurs when ${\rm d}\log{\mathcal{L}}/{\rm d}X=0$, is
\begin{equation}
  \label{eq:log-L-max}
  -\log{\mathcal{L}_{\rm max}} = n_d \left(1 + \log{\bar{X}} \right), \qquad \bar{X} = \frac{1}{n_d}\sum\limits_{i=1}^{n_d} x_i.
\end{equation}
From Eq.~\ref{eq:log-L-max} one can see that $\mathcal{L}_{\rm max}$ is a simple function of $\bar{X}$; thus, all data sets with the same sample mean, regardless of what their parent p.d.f.\ is, will have the same g.o.f.\ value using this method for the p.d.f.\ defined in Eq.~\ref{eq:logL-ex-pdf}.

\begin{figure}
  \centering
  \includegraphics[width=0.4\textwidth]{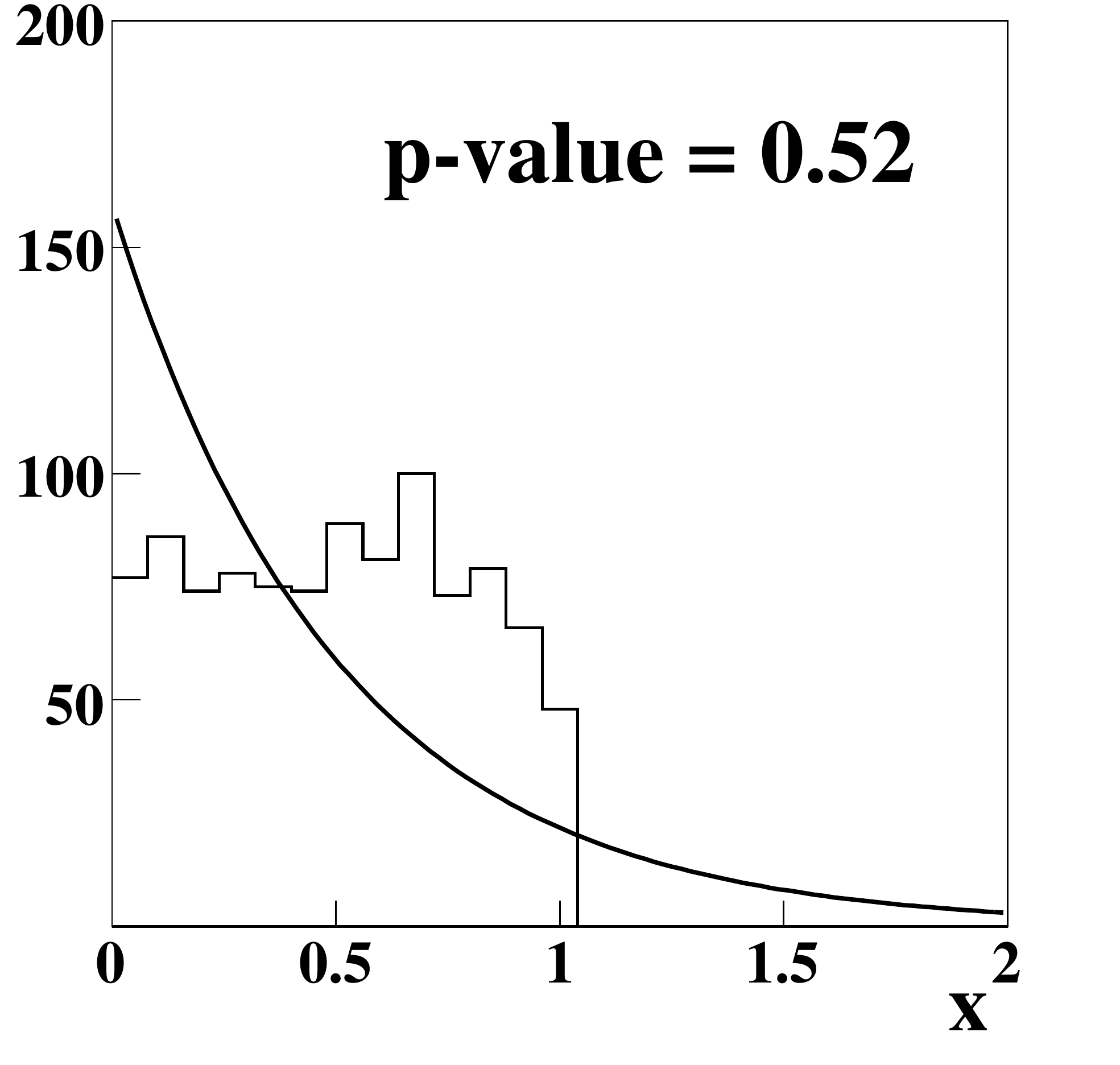}
  \caption[]{\label{fig:logL-ex}
    Data sampled randomly from a uniform distribution on the interval $[0,1)$.  The line represents the results of a fit of the p.d.f.\ given in Eq.~\ref{eq:logL-ex-pdf} to the data.  The $p$-value obtained using the method described in Appendix~A is 0.52.
  }
\end{figure}

To illustrate why this is so disastrous, consider a data set that consists of $n_d = 1000$ events sampled randomly from a uniform distribution on the interval $[0,1)$.  A fit of the p.d.f.\ given in Eq.~\ref{eq:logL-ex-pdf} to this data yields $\bar{X} \approx 1/2$ with the corresponding m.l.v. given by Eq.~\ref{eq:log-L-max}.  Fig.\ref{fig:logL-ex} shows the results of this fit.  Clearly, the fit p.d.f.\ does not reproduce the data; however, applying the g.o.f.\ method described in this Appendix yields a $p$-value of 0.52. What went wrong?  The ensemble of Monte Carlo data sets were generated using the value of $\bar{X}$ obtained from the data.  The sample means of these data sets are then just statistical fluctuations around $\bar{X}$.  Since the m.l.v. is a simple function of the sample mean, one would expect the g.o.f.\ value to always be approximately 0.5 for this p.d.f.\ (regardless of what the true parent p.d.f.\ of the data is).  This method is also not invariant under change of variables and is biased; see Ref.~\cite{ref:heinrich} for more discussion on these topics.

In this example $\mathcal{L}_{\rm max}$ provided no information about the g.o.f.  In general, unless the test-statistic p.d.f.\ is known (which is the case, {\em e.g.}, for the $\chi^2$ test) then the test statistic used to obtain estimators for the unknown parameters in the p.d.f.\ and the one used to determine g.o.f.\ should be weakly correlated.  If the correlation is strong (which is clearly the case when they are the same statistic), then the g.o.f.\ test is redundant.  Of course, the deficiencies in this method are not always this disastrous for more complicated p.d.f.'s.  In some cases this method can expose discrepancies between the fit p.d.f.\ and the data.  It is important to realize that this does not make it a valid g.o.f.\ method; it makes it a cross check.  To be a valid g.o.f.\ method it must (at least) produce a uniform $p$-value distribution if $f = f_0$.  This method, in many cases, does not.

\section{Approximating $\sigma_T^2$ for Mixed-Sample Methods}

The variance of the $T$-statistic distribution used in Ref.~\cite{ref:schil} is difficult to calculate since it depends on $f(\vec{x})$. The limiting value is given by (using the notation from Section~3.1)
\begin{equation}
  \label{eq:schil-var-lim-def}
  \lim_{n \rightarrow \infty} \sigma_T^2 = \frac{1}{n n_k} \left(\frac{n_a n_b}{n^2} + 4\frac{n_a^2 n_b^2}{n^4}n_k \bar{p}_1^{\prime} - \frac{n_a n_b(n_a - n_b)^2}{n^4}n_k (1 - \bar{p}_2^{\prime}) \right),
\end{equation}
where
\begin{equation}
   \bar{p}_i^{\prime} = \frac{1}{n_k^2}\sum\limits_{k=1}^{n_k} \sum\limits_{l=1}^{n_k} \lim_{n \rightarrow \infty} n p_i(k,l),
\end{equation}
and $p_1(k,l)$ and $p_2(k,l)$ are the mutual- and shared-neighbor probabilities, respectively (for a detailed discussion of these quantities, see Refs.~\cite{ref:schil,ref:schil-2}).  Ref.~\cite{ref:schil-2} provides the following very useful limits:
\begin{subequations}
  \label{eq:schil-var-lim}
  \begin{equation}
    \lim_{n_k \to \infty} n_k \bar{p}_1^{\prime} = 1
  \end{equation}
  \begin{equation}
    \lim_{D \to \infty} \bar{p}_2^{\prime} = 1.
  \end{equation}
\end{subequations}
These limits converge very quickly.  Using Eq.~\ref{eq:schil-var-lim} the following limiting values of $\sigma_T$ are obtained:
\begin{subequations}
  \label{eq:schil-var-lim-2}
  \begin{equation}
    \label{eq:schil-var-lim-2a}
    \lim_{n, n_k \to \infty} \sigma_T^2 = \frac{1}{n n_k} \left(\frac{n_a n_b}{n^2} + 4\frac{n_a^2 n_b^2}{n^4}  \right) \qquad {\rm if~} n_a = n_b,
  \end{equation}
  \begin{equation}
    \label{eq:schil-var-lim-2b}
    \lim_{n, n_k, D \to \infty} \sigma_T^2 = \frac{1}{n n_k} \left(\frac{n_a n_b}{n^2} + 4\frac{n_a^2 n_b^2}{n^4}  \right)  \qquad \forall~ n_a, n_b.
  \end{equation}
\end{subequations}
The convergence to these limits is so fast that Eq.~\ref{eq:schil-var-lim-2b} can be used to obtain a good approximation of $\sigma_T$ even for $D=2$ for certain values of $n_a, n_b$ and $n_k$.  

Qualitatively, the constraints required to ensure that Eq.~\ref{eq:schil-var-lim-2b} is a valid approximation of Eq.~\ref{eq:schil-var-lim-def} are not too difficult to see.  Recall that it is the values of $\bar{p}_1^{\prime}$ and $\bar{p}_2^{\prime}$ that must be approximated; however, the limit given in Eq.~\ref{eq:schil-var-lim-2a} converges fast enough that one need not worry about the term containing $\bar{p}_1^{\prime}$ in Eq.~\ref{eq:schil-var-lim}.  The term containing $\bar{p}_2^{\prime}$ is proportional to both $(n_a - n_b)^2$ and $n_k$.  This limits both how much larger one sample can be than the other and how large a value of $n_k$ can be chosen. I have found that the values $n_{a} \lesssim 10~n_b$ and $n_k \lesssim 10$ satisfy all of the relevant constraints.  Changing these values by a factor of two worked fine in the studies I performed; changing them by a factor of 10 did not.  

\section{The Permutation Test}

If the p.d.f.\ of the test statistic, $T$, is not known or is difficult to calculate, then it can often be estimated by performing some kind of re-sampling of the data.  There are many re-sampling techniques described in the statistical literature, {\em e.g.}, bootstrapping, jackknifing, {\em etc.} (see, {\em e.g.}, Ref.~\cite{ref:efron}). The method described in this appendix, referred to as a permutation test, was first proposed by Fisher in 1935~\cite{ref:fisher}.  A detailed discussion on this topic can be found in Ref.~\cite{ref:good}.

The permutation test involves combining the data and Monte Carlo data into a pooled sample of size $n = n_d + n_{mc}$.  The first step towards obtaining an estimate for the $p$-value is to randomly select $n_d$ events from the pooled sample and temporarily label them ``data''; label the remaining $n_{mc}$ events ``Monte Carlo.''   The test statistic, denoted $T_{\rm perm}$, is then calculated with these designations for each event.  This process can be repeated for all of the $n!/(n_d!n_{mc}!)$ possible event combinations; however, if this requires too much processing time, then a random subset of combinations may be used.  This process results in producing the set of $T$ values $\{T^1_{\rm perm}\ldots T^{n_{\rm perm}}_{\rm perm}\}$, where $n_{\rm perm}$ is the number of event combinations used.  The $p$-value is then simply the fraction of times where $T < T_{\rm perm}$.

Why does this technique work? For the case where the test p.d.f.\ and parent p.d.f.\ are the same, the assignment of ``data'' and ``Monte Carlo'' are effectively just labels.  Reassigning these labels should have no effect on the mean value of $T$.  Furthermore, each of the $n!/(n_d!n_{mc}!)$ possible event assignment combinations could have equally well been observed by our experiment.  Thus, we can use them to estimate the p.d.f.\ of $T$ and, in turn, obtain an estimate for the $p$-value.  

For this paper, I chose to use only 100 randomly selected event combinations due to the large number of $p$-values that needed to be calculated (I analyzed ensembles of data sets for multiple p.d.f.'s).  The uncertainty on the $p$-value is obtained from the binomial distribution to be ${\sigma_p = \sqrt{p(1-p)/n_{\rm perm}}}$.  Thus, the number of permutations required depends on the $p$-value obtained.  {\em E.g.}, if after 100 permutations the estimate of the $p$-value is 0.5, then the uncertainty on $p$ is 0.05.  This is sufficient to conclude that the agreement between the fit p.d.f.\ and the data is good.  If, however, the $p$-value estimate is 0.06, then the uncertainty on $p$ is 0.02. More permutations would be required if, {\em e.g.}, one wanted to know whether or not the fit p.d.f.\ is rejected at 95\% confidence level.

\section{Uniformity of the $U$ Statistic}

In this appendix I will prove that the $U$ statistic used in Ref.~\cite{ref:bickel-1983} and defined in Eq.~\ref{eq:bickel-u} is approximately uniform if the parent p.d.f.\ and the test p.d.f.\ are equivalent, {\em i.e.}, if $f = f_0$.  The proof presented here follows the one given in Ref.~\cite{ref:bickel-1983} but includes some additional intermediate steps for illustrative purposes.  

The probability that event $j$ is less than $R$ away from event $i$ is given by
\begin{equation}
  \mathcal{P}(|\vec{x}_i - \vec{x}_j| < R) = \int_{|\vec{x} - \vec{x}_i| < R} f(\vec{x}) d\vec{x},
\end{equation}
which follows from the fact that $\int f(\vec{x}) d\vec{x} = 1$.  The probability that none of the other $n_d - 1$ events falls within $R$ of event $i$ is then
\begin{equation}
  \mathcal{P}(R^{nn}_i \geq R) = \left(1 - \int_{|\vec{x} - \vec{x}_i| < R} f(\vec{x}) d\vec{x} \right)^{n_d - 1}.
\end{equation}
Substituting $y = \int_{|\vec{x} - \vec{x}_i| < R} f(\vec{x}) d\vec{x}$ and using the fact that the value of the integral of the p.d.f.\ is monotonically non-decreasing with $R$ yields
\begin{equation}
  \mathcal{P}(-\frac{1}{n_d}\log{U_i} \geq y) = \left(1 - y \right)^{n_d - 1}
\end{equation}
if $f = f_0$.  Finally, making the substitution $y = -\log{z}/n_d$ yields
\begin{equation}
  \mathcal{P}(U_i \leq z) = \left(1 +\frac{1}{n_d}\log{z} \right)^{n_d - 1},
\end{equation}
for $\log{z} > -n_d$ (which follows from the fact that $y < 1$).  The p.d.f.\ for the $U$'s is then
\begin{equation}
  \label{eq:bickel-u-pdf}
  f_U(z) = \frac{d}{dz} \mathcal{P}(U_i \leq z) = \frac{n_d-1}{n_d z}\left(1 + \frac{1}{n_d}\log{z}\right)^{n_d-2} \approx 1,
\end{equation}
for $e^{-n_d} < z \le 1$.  Thus, for the case $f = f_0$, the $U$ distribution obtained for a data set is approximately uniform. Of course, the $U$ values for each event are not independent (since they include the nearest-neighbor distances) so this is not a true p.d.f.  Because of this, the $U$ distribution is expected to have some deviation from uniformity greater than that given in Eq.~\ref{eq:bickel-u-pdf} (discussion on this is oddly absent from Ref.~\cite{ref:bickel-1983}).

\section{Test Usages in Other Fields}

While the use of unbinned multivariate g.o.f.\ methods in high energy physics is currently very limited, many other scientific fields have been employing these techniques for some time (for decades in some areas).  Below is a (very informal) survey of how the tests studied in this paper have been used in other fields.  Citation counts are taken from {\tt scholar.google.com}.

\begin{description}
\item[Mixed-Sample Methods] \hfill \\
Refs.~\cite{ref:schil,ref:henze} have been cited 39 and 55 times, respectively, including a number of citations in ecology publications.  The most popular ecological application of this method appears to be in the analysis of stable-isotope ratios.  The ratios of the stable isotopes of carbon and nitrogen in the tissue of an animal can be used to determine its dietary composition. The mixed-sample g.o.f.\ method has been used to determine the statistical significance of differences found in carbon-nitrogen isotope space from different biological samples.

\item[Point-to-Point Dissimilarity Methods] \hfill \\
Refs.~\cite{ref:baringhaus,ref:aslan1} have been cited 40 and 11 times, respectively. These publications are both recent (2004), but the list of fields citing them is already diverse; it includes genetics, magnetic resonance imaging, sociology, astronomy, {\em etc.}  This technique appears to be well suited to determining g.o.f.\ in a wide range of multivariate analyses, which is not surprising given how effective it is in a Dalitz-plot analysis.

\item[Distance to Nearest-Neighbor Methods] \hfill \\
Ref.~\cite{ref:bickel-1983} has been cited 71 times, including a number of times in publications that deal with testing the quality of random number generators. It is easy to demonstrate why. Consider the optimally non-random set $\{i/n: i=0,\ldots,n-1\}$.  Many g.o.f.\ tests, including the $\chi^2$ test, would not reject a uniform Poisson hypothesis when applied to this data.  The method presented in Ref.~\cite{ref:bickel-1983}, however, does reject it since the distance from each event to its nearest neighbor is constant.  This results in a $U$ distribution that is a spike (instead of uniform).

\item[Local-Density Methods] \hfill \\
Ref.~\cite{ref:ripley} has a citation count of 985; this includes referencing in a few books that have been cited almost two thousand times each and in a large number of ecology publications.  Ecological processes can be non-Poisson due to factors such as reproduction and competition.  For a Poisson process, the $K$ and $L$ functions take on larger values if $f \ne f_0$ (see Section~3.4).  If, however, the process is not Poisson ({\em i.e.}, if the events are correlated), then the $K$ and $L$ functions can also take on smaller values.  Detecting such correlations is often important in ecology.  It is also important in areas such as public health where the method presented in Ref.~\cite{ref:ripley} has been used to monitor for clusters of disease.  The inhomogeneous extension presented in Ref.~\cite{ref:baddeley} already has 148 citations (more than one per month since its publication).

\item[Kernel-Based Methods] \hfill \\
Ref.~\cite{ref:bickel-1973} has a citation count of 418 including a number of citations in the field of econometrics.  Economic data is highly multi-dimensional.  There is a lot of interest in being able to properly model this data so that future trends and outcomes can be predicted and, in turn, obscene amounts of money made.  The g.o.f.\ of economic models has often been tested using the p.d.e. approach presented in Refs.~\cite{ref:bickel-1973,ref:fan}.

\end{description}
It would appear that many other scientific fields are much more advanced than high energy physics when it comes to unbinned multivariate g.o.f.\ testing. Hopefully this will change in the near future.

%% file: paper.bbl
\begin{thebibliography}{99}

\bibitem{ref:bellman} R.E. Bellman, {\em Adaptive Control Processes}, Princeton University Press, Princeton, NJ (1961).

\bibitem{ref:yates} F. Yates, {\em Contingency tables involving small numbers and the $\chi^2$ test}, Supplement to the J. Roy. Statistical Society {\bf 1}, No. 2 (1934) 217-235. 

\bibitem{ref:dagostino} 
  R.B. D'Agostino and M.A. Stephens, {\em Goodness-of-Fit Techniques}, Marcel Decker, Inc. (1986).

\bibitem{ref:heinrich} J.~Heinrich,
  {\em Pitfalls of goodness-of-fit from likelihood,}
  {\it In the Proceedings of PHYSTAT2003: Statistical Problems in Particle Physics, Astrophysics, and Cosmology, Menlo Park, California, 8-11 Sep 2003, pp MOCT001} [arXiv:physics/0310167].



\bibitem{Blatt}
  J.~Blatt and V.~E.~Weisskopf, {\it Theoretical Nuclear Physics}, J.~Wiley, New York (1952).

\bibitem{ref:zemach}  C.~Zemach, {\em Three pion decays of unstable particles,} Phys.\ Rev.\  {\bf 133} (1964) B1201;  C.~Zemach, {\em Use of angular momentum tensors,} Phys.\ Rev.\  {\bf 140} (1965) B97.


\bibitem{Williams:2008wu}
  M.~Williams,
  {\em Numerical object oriented quantum field theory calculations,}
  Comp. Phys. Comm. {\bf 180}, 1847 (2009) [arXiv:0805.2956 [hep-ph]].

\bibitem{ref:pearson} K. Pearson, {\em On the criterion that a given system of deviations from the probable in the case of a correlated system of variables is such that it can be reasonably supposed to have arisen from random sampling}, The London, Edinburgh and Dublin Philosophical Magazine and Journal of Science {\bf 50}, Ser. 5 (1900) 157-175.

\bibitem{ref:chernoff} H. Chernoff and E.L. Lehmann, {\em The use of maximum likelihood estimates in $\chi^2$ tests for goodness of fit}, Ann. Math. Stat. {\bf 25}, No. 3 (1954) 579-586. 

\bibitem{ref:schil} M.F. Schilling, {\em Multivariate two-sample tests based on nearest neighbors,} J. Amer. Statistical Assoc. {\bf 81}, No. 395 (1986) 799-806.

\bibitem{ref:henze} N. Henze, {\em A multivariate two-sample test based on the number of nearest neighbor type coincidences,} Ann. Stat.{\bf 16}, No. 2 (1988) 772-783.

\bibitem{ref:cuadras} C.M. Cuadras and J. Fortiana, {\em Distance-based multivariate two sample tests} (2003) [{\footnotesize {\tt http://www.imub.ub.es/publications/preprints/pdf/Cuadras-Fortiana.334.pdf}}].

\bibitem{ref:baringhaus} L. Baringhaus and C. Franz, {\em On a new multivariate two-sample test,} J. Multivariate Anal. {\bf 88} (2004) 190-206.

\bibitem{ref:aslan1} B. Aslan and G. Zech, {\em New test for the multivariate two-sample problem based on the concept of minimum energy,} Stat. Comp. Simul. {\bf 75}, Issue 2 (2004) 109-119.

\bibitem{ref:aslan2} B. Aslan and G. Zech, {\em Statistical energy as a tool for binning-free, multivariate goodness-of-fit tests, two-sample comparison and unfolding,} Nucl. Instrum. Methods {\bf A537} (2005) 626-636.

\bibitem{ref:bickel-1983} P.J. Bickel and L. Breimann, {\em Sums of functions of nearest neighbor distances, moment bounds, limit theorems and a goodness of fit test,} Ann. Probab. {\bf 11}, No. 1 (1983) 185-214.

\bibitem{ref:ripley} B.D. Ripley, {\em Modelling spatial patterns,} J. Roy. Stat. Soc. B Met. {\bf 39}, No. 2 (1977) 172-212.

\bibitem{ref:besag} J.E. Besag, Comments in supplement to Ref.~\cite{ref:ripley}.

\bibitem{ref:baddeley} A.J. Baddeley, J. M\o ller and R. Waagepetersen, {\em Non- and semi-parametric estimation of interaction in inhomogeneous point patterns,} Stat. Neerl. {\bf 54}, Issue 3 (2000) 329-350.

\bibitem{ref:bickel-1973} P.J. Bickel and M. Rosenblatt, {\em On some global measures of the deviations of density function estimates,} Ann. Stat. {\bf 1}, No. 6 (1973) 1071-1095.

\bibitem{ref:fan} Y. Fan, {\em Bootstrapping a consistent nonparametric goodness-of-fit test,} Econometric Rev. {\bf 14}, Issue 3 (1995) 367-382.

\bibitem{ref:gourieroux} C. Gourieroux and C. Tenreiro, {\em Local power properties of kernel based goodness of fit tests,} J. Multivariate Anal. {\bf 78}, Issue 2 (2001) 161-190.

\bibitem{ref:efron} B. Efron and R.J. Tibshirani, {\em An Introduction to the Bootstrap}, Chapham and Hall, London (1993).

\bibitem{ref:mise} B.W. Silverman, {\em Density estimation for statistics and data analysis}, Chapham and Hall, London (1986).


\bibitem{ref:schil-2} M.F. Schilling, {\em Mutual and shared neighbor probabilities: finite- and infinite-dimensional results,} Adv. Appl. Probab. {\bf 18}, No. 2 (1986) 388-405. 

\bibitem{ref:fisher} R.A. Fisher, {\em The Design of Experiments}, Oliver and Boyd Ltd., London (1935).

\bibitem{ref:good} P. Good, {\em Permutation Tests: a Practical Guide to Resampling Methods for Testing Hyptheses}, Springer-Verlag, New York (1994).

\end{thebibliography}
